\documentclass[
aps,
pra,
amsmath,
amssymb,
preprint,
floatfix,
longbibliography,
letterpaper,
superscriptaddress,
lengthcheck,
nofootinbib
]{revtex4-1}

\usepackage{graphicx}
\usepackage{rotating}
\usepackage{amsmath}
\usepackage{bbm}
\usepackage{subfigure}
\usepackage{color}
\usepackage{bbold}
\usepackage{mathtools}
\usepackage{mathrsfs}
\usepackage{relsize}
\usepackage{booktabs}
\usepackage{bm}
\usepackage{ulem}
\usepackage{soul}
\usepackage{multirow}
\usepackage{array}
\usepackage{url}
\usepackage[breaklinks=true,colorlinks=true,urlcolor=blue,linkcolor=blue]{hyperref}

\renewcommand{\emph}[1]{{#1}}

\setul{0.5ex}{0.3ex}

\DeclareMathAlphabet{\mathpzc}{OT1}{pzc}{m}{it}

\newcommand{\braket}[2]{\left\langle \, #1 \, \left|\, #2 \, \right. \right\rangle}

\def\bea{\begin{eqnarray}} 
\def\eea{\end{eqnarray}} 
\def\ben{\begin{equation}} 
\def\een{\end{equation}} 
\def\benu{\begin{enumerate}} 
\def\enu{\end{enumerate}}

\newcommand{\vrt}{\vec{r},t}

\newcommand{\mi}{\mathrm{i}}
\newcommand{\md}{\mathrm{d}}
\newcommand{\me}{\mathrm{e}}

\newcommand{\honemm}{\hspace{1mm}} 
\newcommand{\htwomm}{\hspace{2mm}}

\definecolor{cred}{rgb}{0.7,0.0,0.0}
\definecolor{cgreen}{rgb}{0.0,0.7,0.0}
\definecolor{cblue}{rgb}{0.0,0.0,0.7}
\definecolor{corange}{rgb}{0.7,0.7,0.0}

\graphicspath{
  {figures/pdf/}
  {figures/eps/}
}

\begin{document}

\title{Real-time solutions of coupled Ehrenfest-Maxwell-Pauli-Kohn-Sham equations: \\fundamentals, implementation, and nano-optical applications}

\author{Ren\'e Jest{\"a}dt}
\affiliation{Max Planck Institute for the Structure and Dynamics of Matter, Center for Free Electron Laser Science, 22761 Hamburg, Germany}
\author{Michael Ruggenthaler}
\affiliation{Max Planck Institute for the Structure and Dynamics of Matter, Center for Free Electron Laser Science, 22761 Hamburg, Germany}
\author{Micael J.~T.~Oliveira}
\affiliation{Max Planck Institute for the Structure and Dynamics of Matter, Center for Free Electron Laser Science, 22761 Hamburg, Germany}
\author{Angel Rubio}
\affiliation{Max Planck Institute for the Structure and Dynamics of Matter, Center for Free Electron Laser Science, 22761 Hamburg, Germany}
\affiliation{Center for Computational Quantum Physics (CCQ), Flatiron Institute, 162 Fifth Avenue, New York NY 10010, USA}
\author{Heiko Appel}
\affiliation{Max Planck Institute for the Structure and Dynamics of Matter, Center for Free Electron Laser Science, 22761 Hamburg, Germany}
\email[Electronic address:\;]{heiko.appel@mpsd.mpg.de}

\date{\today}

\begin{abstract}
We present the theoretical foundations and the implementation details of a
density-functional approach for coupled photons, electrons, and effective
nuclei in non-relativistic quantum electrodynamics. Starting point of the
formalism is a generalization of the Pauli-Fierz field theory for which we
establish a one-to-one correspondence between external fields and internal
variables. Based on this correspondence, we introduce a Kohn-Sham construction
which provides a computationally feasible approach for ab-initio light-matter
interactions.  In the mean-field limit for the effective nuclei the formalism
reduces to coupled Ehrenfest-Maxwell-Pauli-Kohn-Sham equations.\\
We present an implementation of the approach in the real-space real-time code
Octopus. For the implementation we use the Riemann-Silberstein formulation of
classical electrodynamics and rewrite Maxwell's equations in Schr\"odinger
form. This allows us to use existing time-evolution algorithms developed for
quantum-mechanical systems also for Maxwell's equations. We introduce a
predictor-corrector scheme and show how to couple the Riemann-Silberstein
time-evolution of the electromagnetic fields self-consistently to the
time-evolution of the electrons and nuclei. Furthermore, the Riemann-Silberstein approach
allows to seamlessly combine macroscopic dielectric media with a microscopic
coupling to matter currents.  For an efficient absorption of outgoing
electromagnetic waves, we present a perfectly matched layer for the
Riemann-Silberstein vector. We introduce the concept of electromagnetic
detectors, which allow to measure outgoing radiation in the far field and
provide a direct way to record various spectroscopies.  We present a
multi-scale approach in space and time which allows to deal with the different
length-scales of light and matter for a multitude of applications. We
apply the approach to laser-induced plasmon excitation in a nanoplasmonic dimer
system.  We find that the self-consistent coupling of light and matter leads to
significant local field effects which can not be captured with the conventional
light-matter forward coupling. For our nanoplasmonic example
we show that the self-consistent foward-backward coupling leads to 
changes in observables which are larger than the difference between local density
and gradient corrected approximations for the exchange correlation functional.
In addition, in our example we observe harmonic
generation which appears only beyond the dipole approximation and can be
directly observed in the outgoing electromagnetic waves on the simulation grid.
The self-consistent coupling of the electromagnetic fields to the ion motion
reveals significant energy transfer from the electromagnetic fields to matter
on the scale of a few tens of femtoseconds. \\
Overall, our approach is ideally suited for applications in nano-optics,
nano-plasmonics, (photo) electrocatalysis, light-matter coupling in 2D
materials, cases where laser pulses carry orbital angular momentum, or
light-tailored chemical reactions in optical cavities to name but a few.
\end{abstract}


\date{\today}

\maketitle

\tableofcontents

\section{Introduction}
\label{sec:Intro}

Low-energy quantum physics has been divided traditionally into different
subfields, e.g., quantum chemistry, quantum optics or solid-state physics. Each
of this subfield focuses on a specific part of coupled light-matter systems.
Roughly speaking (see Ref.~\cite{ruggenthaler2018quantum} for details), one either prescribes how the
electromagnetic field looks like and then determines properties of the matter
subsystem, e.g., in quantum chemistry or solid-state physics, or one prescribes
the properties of matter and then determines how the photon subsystem behaves,
as done in, e.g., quantum optics or photonics. This division is reflected also
in the available theoretical methodologies, which either focus on the matter
degrees of freedom (see, e.g.,
Refs~\cite{szabo2012,martin2016,Schollwoeck2011,engel2011}) or on the
electromagnetic field (see, e.g.,
Refs~\cite{loudon1988,grynberg2010,born2013}). This rough division is further
differentiated depending on which part of the matter or photon degrees of
freedom are investigated, e.g., nuclear dynamics that drive chemical
reactions~\cite{tavernelli2015nonadiabatic}.

However, there have been a lot of recent experimental results that question
these traditional distinctions. For instance, when matter and light couple
strongly and neither can be considered a perturbation of the other, then the
properties of matter and light can be strongly modified and novel states of
matter emerge, such as polaritons (light-matter hybrid states). This happens
for single molecules in nanocavities~\cite{chikkaraddy2016} or
microcavities~\cite{wang2017coherent}, where the confined light modes interact
strongly with the matter degrees of freedom. But also in other situations, such
as at interfaces or nanostructures~\cite{sukharev2017}, strong coupling can
change well-established results such as the usual selection rules of quantum
chemistry~\cite{yamamoto2014}. Even for rather bad cavities and ambient
conditions strong coupling can be achieved by, for example, increasing the number of
molecules or atoms~\cite{ebbesen2016}, which in turn provides a novel and very
robust tool to influence and control chemical properties. It has been observed
that by merely coupling to the changed vacuum of the electromagnetic field,
chemical reactions can be modified~\cite{hutchison2012}, well-established
limits for energy transfer can be broken~\cite{zhong2017}, or Raman processes
can be enhanced~\cite{shalabney2015b}. Strong coupling has been achieved for
many different physical systems, e.g., even for living
bacteria~\cite{coles2014}, and it can be used to engineer novel states of
matter such as polariton condensates~\cite{byrnes2014}. Besides these strong
coupling situations many more cases are known where light and matter become
equally important, such as in the case of screening, polarization and
retardation effects as observed, e.g., in the energy transfer induced by
attosecond laser pulses~\cite{sommer2016attosecond} or more traditionally in
optical responses~\cite{maki1991linear}. Furthermore there are many situations
where usually neglected properties of the light field lead to substantial
changes in the matter system, such as in strong-field
physics~\cite{ludwig2014breakdown}, when photons carry a large angular
momentum~\cite{schmiegelow2016transfer,birula2018,Yue2018}, or when the emission spectrum is
investigated in detail~\cite{taminiau2012quantifying}. Indeed, considering the
photon and matter degrees of freedom at the same time can lead to impressive
novel practical applications such as daytime radiative
cooling~\cite{FAN2017264}.

In the above examples the complex interplay between the basic constituents of
coupled light-matter systems -- electrons, nuclei and photons -- are essential.
In most theoretical treatments, however, a strong reduction to only a few
important degrees of freedom is performed \textit{a
priori}~\cite{ruggenthaler2018quantum}, such as in electronic-structure
theory~\cite{szabo2012}, where the nuclei and the photons are treated only as
external perturbations. While many advanced methods to solve the resulting
many-electron equation
exist~\cite{booth2009,Orus2014,schollwoeck2005,kotliar2006,gull2011,Metzner2012,RevModPhys.86.779},
they miss most effects of the correlation with nuclei and photons. Furthermore, many
approaches have been developed that try to tackle the full electron-nucleus problem, such as
exact-factorization approaches~\cite{abedi2010exact,abedi2012correlated} or
trajectory formulations~\cite{kapral2015, tully2012, miller2012}. However,
what the correlation with the light-field
is concerned, there are only a few such approaches available to date. On the
one hand we have coupling with classical light fields, such as presented
in~\cite{lorin2007,lopata2009,PhysRevB.85.045134,Lucchini916,PhysRevE.94.023314,2018arXiv180702733Y,
2018arXiv181006168Y, 2018arXiv181006500U, 2018arXiv181008344Y}, on the other
hand we have also coupling to the full quantized field as discussed
in~\cite{ruggenthaler2011time, tokatly2013, ruggenthaler2014quantum,
ruggenthaler2015,galego2015cavity,flick2018strong,feist2017,ribeiro2018polariton}.
Only very recently also practical formulations that include all three
constituents explicitly have
emerged~\cite{melo2015,flick2017cavity,flick2018cavity,schafer2018ab}. These
coupled matter-photon approaches allow to investigate in detail the change of
chemical structures due to changes in the electromagnetic
vacuum~\cite{flick2018ab}, coupled light-matter observables such as polariton
states and novel potential-energy surfaces~\cite{flick2017atoms}, or changes in
Maxwell's equations due to the interaction with
matter~\cite{flick2018light}. A further advantage of a coupled light-matter
description is that observables that are measured via the light field, such as
absorption or emission spectra, do no longer need to be approximately
determined by matter degrees but are accessible directly by the calculated
photon field~\cite{ruggenthaler2018quantum}.

However, the above presented approaches and techniques are themselves either
simplifications of the full problem, e.g., by only assuming dipole coupling or
neglecting the nuclear degrees of freedom, or have not been made practical for
the general case. In this work we will close this gap and provide the first
\textit{full ab-initio} treatment of electrons, nuclei and photons on equal
footing via a density-functional reformulation of a generalization of the
Pauli-Fierz Hamiltonian of non-relativistic quantum electrodynamics
(QED)~\cite{spohn2004, ruggenthaler2018quantum} together with a numerical
implementation of the resulting Maxwell-Pauli-Kohn-Sham (MPKS) equations. By applying
the resulting multi-scale and multi-species formulation to a nanoplasmonic case
study, we highlight how discarding electromagnetic and/or nuclear degrees of
freedom can alter certain observables as well as how observables differ if they
are computed directly from the Maxwell field as opposed to the usual
approximate treatment. These results provide a completely new perspective on
fundamental low-energy physics, where a disagreement between theory and
experiment is often attributed to missing correlations among only one species
of particles, i.e., electrons, nuclei or photons, but the discarded degrees of
freedom are completely neglected. Furthermore, this unbiased approach allows to
tackle many of the above mentioned experimental results and introduces a novel
tool to employ the complex interplay between light and matter for the design
and control of novel materials.

The paper is structured as follows. In section \ref{sec_fundamentals}, we
introduce the fundamentals that lead to the generalized many-body Pauli-Fierz
Hamiltonian that we consider as starting point of our approach. Next, we provide
in section \ref{sec_QEDFT} a one-to-one correspondence between external and internal
variables for the Pauli-Fierz field theory. Based on this we can establish a
density-functional theory (DFT) of non-relativistic QED for photons, electrons
and effective nuclei. In section \ref{subsec_KS}, we introduce the Kohn-Sham
construction for our generalization of quantum-electrodyamical density-functional theory (QEDFT), which leads in the mean-field approximation for
the nuclei to coupled Ehrenfest-Maxwell-Pauli-Kohn-Sham (EMPKS) equations.
All practical details for the implementation of a solution of these coupled
equations are provided in section \ref{sec_implementation}. In particular,
using the Riemann-Silberstein vector of classical electrodynamics, we rewrite
Maxwell's equations in Schr\"odinger form and introduce the corresponding
time-evolution operators for the homogeneous and for the inhomogeneous cases.
For an efficient absorption of outgoing waves, we introduce absorbing boundary
conditions and a perfectly matched layer for the Riemann-Silberstein vector.
We discuss full-minimal coupling and a multipole expansion and introduce a
predictor-corrector scheme for self-consistent forward-backward coupling of
light and matter.  We place an emphasis on the spatial and temporal multiscale
aspects of light-matter coupling and conclude the section with a validation of
the method and a comparison to the finite-difference-time-domain (FDTD)
approach for solving Maxwell's equations.  Finally, in section
\ref{sec_applications} we illustrate the coupled
EMPKS approach for a nanoplasmonic system. We
investigate electric field enhancements, harmonic generation and analyze
the role of nuclear motion. Since in our approach we also propagate the
electromagnetic fields on a grid, we can define {\it electromagnetic detectors}
in the far field which record all outgoing electromagnetic waves.  We conclude
the paper in section \ref{sec_summary} and provide an outlook in section
\ref{sec_outlook}.


\section{Fundamentals} \label{sec_fundamentals}


  Let us start by defining some notation. We will use the usual vector notation alongside
  the relativistic covariant notation. From a fundamental point of view this is
  convenient since we can easily connect to QED and its relativistic equations,
  i.e., the Dirac and Maxwell equations. We therefore make a difference between
  upper and lower indices, which are connected via the Minkowski metric with
  signature $g \equiv (+,-,-,-)$, 
  \begin{equation}
      g_{\mu \nu} 
    = \begin{pmatrix}
        \hspace{1mm} 1 & \phantom{-} 0 & \phantom{-} 0 & \phantom{-} 0 \hspace{1mm}      \\
        \hspace{1mm} 0 &          -  1 & \phantom{-} 0 & \phantom{-} 0 \hspace{1mm}      \\
        \hspace{1mm} 0 & \phantom{-} 0 &          -  1 & \phantom{-} 0 \hspace{1mm}      \\
        \hspace{1mm} 0 & \phantom{-} 0 & \phantom{-} 0 &          -  1 \hspace{1mm} 
      \end{pmatrix}  .                                                                                                  \label{eq_RN3}
  \end{equation}
  We denote by $x^{0} = c_{0} t$ the (temporal) zero component of the four-component
  vector $x^{\mu}$, where Greek indices go over all space-time dimensions,
  $\mu=\{0,1,2,3\}$, and roman letters in the covariant notation go over only the
  three spatial dimensions, $k=\{1,2,3\}$. Further we denote the speed of light as $c_{0}$, which is related to the vacuum permeability $\mu_0$ and the vacuum permittivity $\epsilon_0$ via $c_{0} = 1/\sqrt{\mu_0 \epsilon_0}$. To make
  switching between notations easier, we provide the following table, where the Einstein
  summation convention over repeated upper and lower indices is implied:
  \begin{equation}
    \begin{alignedat}{2}
           \vec{A}
         & \equiv A^{k} \;,                                                              \\
           \vec{A} \cdot \vec{B}
         & \equiv - A^{k} B_{k} = - A_{k} B^{k} = - A^{k} B^{l} g_{lk} \;,                             \\
           \vec{\nabla} \cdot \vec{A} 
         & \equiv \partial_{k} A^{k} \;,                                                 \\
           A_{k} = g_{kl} A^{l}
         & \equiv - \vec{A} \;,                                                                        \\
           \vec{\nabla} \times \vec{A}
         & \equiv - \epsilon^{klm} \partial_{l} A_{m} \;.                                              
    \end{alignedat}                                                                                \label{eq_relativistic_notation}
  \end{equation}
  Here $g_{kl}$ is the three-dimensional (spatial) submatrix of $g_{\mu \nu}$ and
  $\epsilon^{klm}$ corresponds to the totally anti-symmetric Levi-Civita
  symbol that arises due the Pauli-matrix algebra (see Eq.~\eqref{eq_RN2} below). Accordingly we
  define $\epsilon_{klm} = \epsilon^{a b c}g_{ak}g_{bl} g_{cm}$.

  \subsection{Relativistic wave equations}

    Next, we consider the relativistic wave equations that form the basis of 
    coupled light-matter systems. Here we follow the seminal work of Dirac and
    introduce specific matrix algebras that allow to rewrite a second-order
    partial-differential equation, which usually corresponds to the relativistic
    energy-momentum relation $E^2 = m^2 c_0^4 + p^2 c_0^2$, into a first-order
    partial-differential equation. The
    type of matrix algebra to use depends on the statistics of the particle one wants to
    describe. The most well-known case is of course the Dirac equation,
    where the spin-$1/2$ nature of the electrons dictates the use of  
    \begin{align}
      \gamma^{0} = \begin{pmatrix} \mathbb{1}_2 & 0 \\ 0 & -\mathbb{1}_2 \end{pmatrix}  \hspace{1mm}, \hspace{5mm}
      \gamma^{k} = \begin{pmatrix} 0 & \sigma^{k}   \\ - \sigma^{k} & 0 \end{pmatrix} .                                   \label{eq_RN1}
    \end{align}
    Here $\mathbb{1}_2$ are two-dimensional identity matrices and the Pauli matrices $\sigma^{k}$ obey
    \begin{equation}
      \begin{alignedat}{2}
           \sigma^{k} \sigma^{l}
        &= \frac{1}{2} \left( \left\{ \sigma^{k}, \sigma^{l} \right\} + \left[ \sigma^{k}, \sigma^{l} \right] \right)    \\
        &= \delta^{kl} \mathbb{1}_2 - \mathrm{i} \epsilon^{klj} \sigma_j.
      \end{alignedat}                                                                                                    \label{eq_RN2}
    \end{equation}
    We see here the Levi-Civita symbol appearing and also that the indices of the
    matrices are connected via the Minkowski metric, e.g., $\sigma_{k} = g_{k
    l}\sigma^{l}$. Using these definitions we find that the Dirac operator $\mi
    \hbar c_0 \gamma^{\mu}\partial_\mu$ applied twice to a solution of the Dirac
    equation $\mi \hbar c_0 \gamma^{\mu}\partial_\mu \psi = m c_0^2 \psi$ for a four
    component wave function $\psi$ implies the relativistic energy-momentum
    relation, i.e., the Klein-Gordon equation $\hbar^2 c_0^2 (\nabla^2 -
    \tfrac{1}{c_0^2} \partial_t^2)\psi = m^2 c_0^4 \psi$.\\
    
    A similar procedure can also be applied (to some extent at least~\cite{Gersten-1999}) to particles
    with other spin. Of particular importance among those are photons, massless spin-$1$
    particles. In this case, instead of the spin-$1/2$ Pauli matrices we use the
    corresponding spin-$1$ matrices in a somewhat non-standard
    form~\cite{Gersten-1999}  
    \begin{equation}
      \begin{alignedat}{3}
           S^1
        &= 
           \begin{pmatrix}
           \hspace{1mm} 0 & \phantom{-} 0 & \phantom{-} 0 \hspace{1mm} \\
           \hspace{1mm} 0 & \phantom{-} 0 & - \mi \hspace{1mm} \\
           \hspace{1mm} 0 & \phantom{-} \mi &  \phantom{-}  0 \hspace{1mm}
           \end{pmatrix} \;,
           \\[3mm]
           S^2
        &= 
           \begin{pmatrix}
             \hspace{1mm}         0  &     \phantom{-}    0 &  \phantom{-} \mathrm{i} \hspace{1mm} \\
             \hspace{1mm}  0 &        \phantom{-}  0  &      \phantom{-}    0 \hspace{1mm} \\
             \hspace{1mm}   -   \mathrm{i}       & \phantom{-} 0 &      \phantom{-}   0  \hspace{1mm}
           \end{pmatrix} \;,
           \\[3mm]
           S^3
        &= 
           \begin{pmatrix}
             \hspace{1mm} 0 &  -\mi & \phantom{-} 0 \hspace{1mm} \\
             \hspace{1mm} \mi & \phantom{-} 0 & \phantom{-} 0 \hspace{1mm} \\
             \hspace{1mm} 0 & \phantom{-} 0 & \phantom{-} 0 \hspace{1mm}
           \end{pmatrix} \;.
      \end{alignedat}                                                                                                    \label{eq_RN4}
    \end{equation}
    While the matrices also obey the spin-algebra $[S^{k},S^{l}]= - \mi
    \epsilon^{klj}S_{j}$ and $\vec{S}^2 = \mathbb{1}_3$, they do not fulfill the
    same algebra as the Pauli matrices in Eq.~\eqref{eq_RN2}. Therefore we cannot find
    a similar simple form of the relativistic energy-momentum relation but instead have
    side conditions~\cite{Gersten-1999}. This leads to the famous
    Riemann-Silberstein formulation of electrodynamics~\cite{Silberstein_1907,Oppenheimer_1931,Keller}. To be more
    specific, we find with the above spin-$1$ matrices~\cite{Gersten-1999} 
    \begin{align}
        \left(\frac{E^2}{c_0^2} - p^2\right) \vec{F} &= \left(\frac{E}{c_0} \mathbb{1}_{3}
        - p_k S^k\right)\left(\frac{E}{c_0} \mathbb{1}_{3} + p_k S^k\right) \vec{F} 
        \\
      &-\begin{pmatrix}
          \hspace{1mm} (p_1)^2 & p_1 p_2 & p_1 p_3 \hspace{1mm} \\
          \hspace{1mm} p_{2}p_{1} & (p_{2})^{2} & p_{2}p_{3} \hspace{1mm} \\
          \hspace{1mm} p_{3}p_{1} & p_{3}p_{2} & (p_{3})^2 \hspace{1mm}
      \end{pmatrix} \vec{F} = 0,                                                         \label{eq_Maxwell_field_energy_decomposition}
    \end{align}
    where $E/c_0 \equiv \mi \hbar \partial_0$, $p_{k} \equiv -\mi \hbar \partial_{k}
    $ and $\vec{F}$ is the Riemann-Silberstein vector. This vector is a three-component
    wave function such that we have an entry for each spin state. The above equation
    holds if we equivalently satisfy
    \begin{align}
        \mi \hbar \partial_0 \vec{F} 
     = -\mi \hbar \partial_k S^{k} \vec{F} &\equiv - \hbar \vec{\nabla} \times \vec{F},
      \\
      -\mi \hbar \vec{\nabla}\cdot \vec{F} &= 0,                                         \label{eq_RS_vector_definition}
    \end{align}
    where we have expressed the momentum mix-term as a side condition on $\vec{F}$.
    Taking into account that also the complex conjugate of the above equations
    leads to the right energy-momentum relation and by defining the positive $(+)$
    and negative $(-)$ Riemann-Silberstein helicity states
    \begin{align}\label{eq:RSvectors}
      F^{k}_{\pm} = \sqrt{\frac{\epsilon_0}{2}}\left( E^k \pm \mi c_0 B^k\right),
    \end{align}
    where $E^{k}$ is the electric field and $B^{k}$ is the magnetic field, the above equations can be
    recast as the usual homogeneous Maxwell equations in vacuum~\cite{Gersten-1999}
    \begin{align}
      \vec{\nabla} \times \vec{E} &= - \partial_t \vec{B},                     \label{eq_Mx_equation_Faraday_homogenous}
      \\
      \vec{\nabla} \times \vec{B} &= - \frac{1}{c_0^2}\partial_t \vec{E},        \label{eq_Mx_equation_Ampere_homogenous}
      \\
      \vec{\nabla}\cdot \vec{E}& = 0,                                          \label{eq_Mx_equation_Gauss_E_homogeneous}
      \\
      \vec{\nabla}\cdot \vec{B}&= 0.                                           \label{eq_Mx_equation_Gauss_B_homogeneous}
    \end{align}
    We emphasize here that the above side condition translates into the Gauss laws and
    that the $\hbar$ has canceled out. Only upon quantizing the classical fields
    (see Sec.~\ref{subsec_photons}) will Planck's constant reappear in the electromagnetic field
    equations. In analogy to the Dirac equation we can consider $\vec{F}_{\pm}$ as the single-photon wave
    function in real space~\cite{Keller}. This analogy will become of practical importance
    when we actually want to solve the equations of motion for the electromagnetic
    field numerically (see Sec.~\ref{subsec_Maxwell_equation_RS}). Furthermore, in the case of the photons we refer to the spin
    as helicity. In the following, in order to make explicit that we only have two physically allowed
    independent (circular) polarizations, we will always employ the Coulomb
    gauge when using the vector-potential formulation to couple to matter.   
    In this case the above homogeneous Maxwell equations can be compactly
    reformulated by introducing vector potentials $A_{\mu}$ that obey the Coulomb
    gauge condition $\partial_k A^{k} = 0$ and the second-order relativistic wave
    equation
    \begin{align*}
      \left(\partial_0^2 + \partial_l \partial^{l}\right) A^{k} = 0.
    \end{align*}
    To connect to the previous versions of the homogeneous Maxwell equations we
    only need the relation between the vector potential and the physical fields,
    i.e., $E^{k} = - \partial_0 A^{k}$ and $B^k = -\frac{1}{c_0} \epsilon^{klm} \partial_{l} A_{m}$.\\
    
    One can extend the above considerations to arbitrary spins (also for particles
    with mass), which leads to the Bargmann-Wigner equations. However, finding simple and physical side conditions, as Gauss' law in the above photon case, is not always possible. This is one reason why in the following 
    we will use the non-relativistic limit of the Dirac equation (also for the
    electrons) and then replace the spin-$1/2$ matrices by different spin matrices
    for each species of nuclei. In this way we only keep the most important degrees
    of freedom of the nuclei in our considerations, i.e., quantized translations,
    rotations and vibrations, and do not describe the protons and neutrons
    explicitly, which themselves are effective spin-$1/2$ particles. The second,
    more relevant reason why we will not use relativistic equations to describe the
    matter degrees of freedom (while the photons are treated fully
    relativistically) is that we would like to have a stable ground state and a
    mathematically well-defined non-perturbative theory~\cite{spohn2004}. Without
    further restrictions the Dirac equation does not have a ground state, as can be
    seen from the fact that the spectrum is in general unbounded from
    below~\cite{thaller2013}. Taking the non-relativistic limit of the Dirac
    equation minimally coupled to a classical \textit{external} vector potential $a_{\mu}$,
    i.e., $\partial_\mu \rightarrow \partial_\mu + \mi q a_{\mu}/\hbar c_0$ with $q$
    being the charge of the particle species, leads in first order of $1/M c_0^2$ to the Pauli-equation
    \begin{align}
        \hat{h}
      = \frac{1}{2 M} \left( - \mathrm{i} \hbar \vec{\nabla} - \frac{q}{c_0} \vec{a}(\vec{r},t) \right)^2
        + q a^{0}(\vec{r},t) - \frac{q \hbar}{2 M} \vec{S} \cdot \vec{b}(\vec{r},t).                                        \label{eq_RN6}
    \end{align}
    Here we have already replaced the Pauli matrices with general spin matrices
    $S^{k}$, and $M$ is the mass of the particle and we used $\vec{b}(\vec{r},t) = \frac{1}{c_0} \vec{\nabla} \times \vec{a}(\vec{r},t)$. This generalized Pauli equation
    for arbitrary spin
    will be, together with the homogeneous Maxwell equations for the photons, our
    mathematical representation of the basic building blocks of our theory for
    electrons, nuclei and photons. Furthermore, we use the notation convention that an \textit{external} field, i.e., a field that is not part of the modelled light-matter system such as a classical pump or probe pulse, is denoted with a lower-case letter. We will next combine all these ingredients into one
    general framework, which will be a generalized form the Pauli-Fierz
    Hamiltonian~\cite{spohn2004}. For completeness, we note that semi-relativistic
    extensions of the Pauli-Fierz Hamiltonian
    exist~\cite{miyao2009,konenberg2011,hiroshima2014}, but as starting point we stay
    within the non-relativistic limit for the matter subsystem. 
    This limit is already enough for a vast set of applications.

  \subsection{Free matter Hamiltonians}

    Let us start by merging the different mathematical representations of our
    fundamental building blocks. We first consider the matter subsystems,
    i.e., electrons and effective nuclei. Since at this point we are not yet coupling the matter
    subsystems to the quantized electromagnetic field, i.e., the photons, these
    particles are not interacting. In the end the photons are the gauge bosons that
    make charged particles interact. We therefore follow the usual construction of
    quantum field theory to establish interacting theories~\cite{greiner1996}.\\
    
    In order to consider many non-interacting particles we will lift our
    single-particle description of Eq.~\eqref{eq_RN6} to arbitrarily but finitely
    many particles. Formally this is done most easily by working in Fock space. We
    therefore introduce Fock-space creation and annihilation field
    operators that obey
    \begin{align}
        \left[ \hat{\Phi}(\vec{r},s), \hat{\Phi}^{\dagger}(\vec{r} \hspace{0.5mm}',s \hspace{0.5mm}') \right]_{\pm}
      = \delta_{ss \hspace{0.5mm}'} \delta^3(\vec{r}-\vec{r} \hspace{0.5mm}') ,                                         \label{eq_RN7}
    \end{align}
    where $s$ corresponds to the different possible spins of the particles, and
    $\pm$ refers to anti-commutation (fermions) or commutations (bosons) relations,
    respectively. 
    Mathematically these objects are somewhat
    inconvenient~\cite{thirring2013} but they allow for very efficient formal
    manipulations. Thus we introduce the ``second-quantized'' notation for
    computational convenience. However, since we work with number-conserving
    Hamiltonians, we can always switch back to the usual ``first quantized'' form
    in which every object can be made well-defined. Only for the photons the Fock
    space is necessary. In this case, however, the mathematical problems can be
    kept in check~\cite{spohn2004}. 
    If we then introduce the conventions
    \begin{align}
      \hat{\Phi}^{\dagger} \hat{\Phi} &\equiv \sum \limits_s \hat{\Phi}^{\dagger}(\vec{r},s) \hat{\Phi}(\vec{r},s), 
      \\
      \hat{\Phi}^{\dagger} S_k \hat{\Phi} &\equiv  \sum \limits_{s,s'} \hat{\Phi}^{\dagger}(\vec{r},s) (S_k)_{s,s'} \hat{\Phi}(\vec{r},s'),     \label{eq_RN9}
    \end{align}
    we can lift the single-particle Pauli equation to the Fock space via
    \begin{equation}
      \begin{alignedat}{2}
           \hat{H}
        &= \sum \limits_s \int \mathrm{d}^3 r \; \hat{\Phi}^{\dagger} (\vec{r},s) \hat{h} \hat{\Phi}(\vec{r},s)        \\
        &= -  \int \md^3 r \frac{1}{2 M} \hat{\Phi}^{\dagger} 
           ( - \mathrm{i} \hbar \partial_k + \frac{q}{c_0} a_{k} )
           (- \mathrm{i} \hbar \partial^{k}  + \frac{q}{c_0} a^{k} ) \hat{\Phi}                                           \\
        &\quad +  \int \md^3 r q a^{0} \hat{\Phi}^{\dagger} \hat{\Phi}
           -  \int \md^3 r \frac{q \hbar}{2 M}
           \hat{\Phi}^{\dagger} S_k \hat{\Phi} \left( \frac{1}{c_0} \epsilon^{klm} \partial_l a_m \right) .
      \end{alignedat}                                                                                                  \label{eq_RN8}
    \end{equation}
    We can do this now for every species of particles. If we have $N$ different
    species of particles, i.e., electrons and effective nuclei, we then have a
    direct sum of Fock spaces. The resulting Hamiltonian on this sum of Fock spaces
    is given with the definition of the respective field operators, masses, charges
    and spin matrices
    \begin{align}
      \left\{
        \hat{\Phi}_{(n)} \hspace{1mm} ; \hspace{1mm} \hat{\Phi}_{(n)}^{\dagger} \hspace{1mm} ; \hspace{1mm}
        M_{(n)} \hspace{1mm}; \hspace{1mm} q_{(n)} \hspace{1mm} ; \hspace{1mm} S_{(n)}
      \right\}                                                                                                         \label{eq_RN10}
    \end{align}
    as
    \begin{equation}
      \begin{alignedat}{2}
           \hat{H}^{(0)}
        &= \hspace{-0.7mm} \sum \limits_{n=1}^{N} \hat{H}_{(n)}                                                         \\
        &  \hspace{-8mm}
         = \hspace{-1.0mm} \sum \limits_{n=1}^{N}
           \hspace{-0.7mm} - \frac{1}{2 M_{\hspace{-0.5mm}(n)}} \hspace{-1mm}
           \int \hspace{-1.5mm} \md^3 r \hat{\Phi}_{\hspace{-0.5mm}(n)}^{\dagger}
           \hspace{-1.4mm} \left( \hspace{-1.1mm} - \mathrm{i} \hbar \partial_{k}
           \hspace{-0.6mm} + \hspace{-0.6mm} \frac{q_{(n)}}{c_0} a_{k} \hspace{-1mm} \right) \hspace{-1mm}
           \hspace{-0.7mm} \left( \hspace{-1.1mm} - \mathrm{i} \hbar \partial^{k}
           \hspace{-0.9mm} + \hspace{-0.6mm} \frac{q_{(n)}}{c_0} a^{k} \hspace{-1mm} \right) \hspace{-1mm}
           \hat{\Phi}_{\hspace{-0.5mm}(n)}                                                                              \\
        &  \hspace{-5mm}
           + \sum \limits_{n=1}^{N} \int \hspace{-1.5mm} \md^3 r \; q_{(n)} a^{0}
           \hat{\Phi}_{(n)}^{\dagger} \hat{\Phi}_{(n)}                                                                  \\
        &  \hspace{-5mm} + \sum \limits_{n=1}^{N} 
           \hspace{-0.7mm} - \frac{1}{2 M_{\hspace{-0.5mm}(n)}} \hspace{-1mm} \int \hspace{-1.5mm} \md^3 r \;
           \hat{\Phi}_{(n)}^{\dagger} S_{k}^{(n)} \hat{\Phi}
           \left( \frac{1}{c_0} \epsilon^{klm} \partial_{l} a_{m} \right)   \honemm .
      \end{alignedat}                                                                                                   \label{eq_RN11}
    \end{equation}
    We note here that while all of the different particles do not interact with
    each other, we still assume that they all see the same external field $a_{\mu}$
    in Coulomb gauge. By lifting this external classical field to a quantum field
    we will make the particles interact.

  \subsection{Free photon Hamiltonian} \label{subsec_photons}

    Before we do so, we will first quantize the electromagnetic field. We will follow the standard 
    procedure~\cite{greiner1996}, but will also connect to the Riemann-Silberstein formulation of QED~\cite{Keller}. 
    Since we have chosen the Coulomb gauge, the canonical quantization procedure only affects the
    transversal fields~\cite{greiner1996} and the canonical commutation relations read
    \begin{align}
        \left[ \hat{A}_{k}(\vec{r}) ; \epsilon_{0} \hat{E}_l^{\perp} (\vec{r} \hspace{0.5mm}') \right]
      = - \mathrm{i} \hbar c_0 \delta_{kl}^{\perp}(\vec{r} - \vec{r} \hspace{0.5mm}'),                                     \nonumber
    \end{align}
    where we employed the transversal delta distribution
    \begin{align}
        \delta_{ij}^{\perp}(\vec{r}-\vec{r} \hspace{0.5mm} ')
      = \left( \delta_{ij} \partial_i \frac{1}{\Delta} \partial_j \right) \delta^{3} (\vec{r} - \vec{r} \hspace{0.5mm}').
    \end{align}
    Here $1/\Delta$ is the inverse of the Laplacian $\Delta \equiv \vec{\nabla}^2$. Due to this quantization
    procedure in Coulomb gauge, the longitudinal part of the electromagnetic field stays classical and does not
    influence the quantized degrees of freedom. This can be seen most easily if we construct the Hamiltonian of
    the free Maxwell field. To do so we introduce the vector-potential operator in terms of creation and annihilation field operators in momentum space  
    \begin{equation}
      \begin{alignedat}{2}
          \hspace{1mm}
          \hat{A}^{k}(\vec{r} \,) \hspace{-0.5mm} &=                                                                           \\
        & \hspace{-8mm} \sqrt{ \frac{\hbar c_0^2}{\epsilon_{0} (2 \pi)^3 } } \hspace{-1mm}
          \int \hspace{-1mm} \frac{\mathrm{d}^3 k}{\sqrt{2 \omega_{k}}} \hspace{-1mm}
          \sum \limits_{\lambda=1}^2 \hspace{-0.5mm} \vec{\epsilon} \, (\vec{k},\lambda) \hspace{-1mm}
          \left[ 
          \hat{a}(\vec{k},\lambda) \mathrm{e}^{ \mathrm{i} \vec{k} \cdot \vec{r}} \hspace{-1mm}
          + \hspace{-0.5mm} \hat{a}^{\dagger}(\vec{k},\lambda) \mathrm{e}^{-\mathrm{i} \vec{k} \cdot \vec{r}}
          \right]    ,                                                                                                      \label{eq_RN19}
      \end{alignedat}
    \end{equation}
    where $\omega_k = c_0|\vec{k}|$ and $\vec{\epsilon} \, (\vec{k}, \lambda)$ is the transversal polarization vector
    that obeys $\vec{k}\cdot \vec{\epsilon} \, (\vec{k}, \lambda) = \vec{\epsilon} \, (\vec{k}, 1)\cdot \vec{\epsilon} \, (\vec{k}, 2)=0$~\cite{greiner1996}.
    The momentum-space annihilation and creation field operators obey the usual commutation relations.
    The purely transversal electric field is then given in accordance to the classical case by $\partial_0 \hat{A}^{k} = - \hat{E}^{k}_{\perp}$
    as
    \begin{equation}
      \begin{alignedat}{2}
          \hspace{1mm}
          \hat{E}^{k}_{\perp}(\vec{r} \,) &=                                                                              \\
        & \hspace{-8mm} \sqrt{ \frac{\hbar c_0^2}{\epsilon_{0} (2 \pi)^3 } } \hspace{-1mm}
          \int \hspace{-1mm} \frac{\mathrm{d}^3 k \mathrm{i} \omega_{k}}{\sqrt{2 \omega_{k}}} \hspace{-1mm}
          \sum \limits_{\lambda=1}^2 \hspace{-0.5mm} \vec{\epsilon} \, (\vec{k},\lambda) \hspace{-1mm}
          \left[ 
          \hat{a}(\vec{k},\lambda) \mathrm{e}^{\mathrm{i} \vec{k} \cdot \vec{r}} \hspace{-1mm}
          - \hspace{-0.5mm} \hat{a}^{\dagger}(\vec{k},\lambda) \mathrm{e}^{-\mathrm{i} \vec{k} \cdot \vec{r}}
          \right]    ,     
      \end{alignedat}      
    \end{equation}
    and the magnetic field via $\hat{B}^{k} = - \tfrac{1}{c_0}\epsilon^{klm}\partial_l \hat{A}_{m}$ as
    \begin{equation}
      \begin{alignedat}{2}
          \hspace{1mm}
          \hat{B}^{k}(\vec{r}\,) &=                                                                                      \\
        & \hspace{-8mm} \sqrt{ \frac{\hbar c_0^2}{\epsilon_{0} (2 \pi)^3 } } \hspace{-1mm}
          \int \hspace{-1mm} \frac{\mathrm{d}^3 k}{\sqrt{2 \omega_{k}}} \hspace{-1mm}
          \sum \limits_{\lambda=1}^2 \hspace{-0.5mm} \mathrm{i} \vec{k}
          \hspace{-0.8mm} \times \hspace{-0.8mm} \vec{\epsilon} \, (\vec{k},\lambda) \hspace{-1mm}
          \left[ 
          \hat{a}(\vec{k},\lambda) \mathrm{e}^{\mathrm{i} \vec{k} \cdot \vec{r}} \hspace{-1mm}
          - \hspace{-0.5mm} \hat{a}^{\dagger}(\vec{k},\lambda) \mathrm{e}^{-\mathrm{i} \vec{k} \cdot \vec{r}}
          \right]    . 
      \end{alignedat}
    \end{equation}
    Again, following the classical definition of the energy of the electromagnetic field, we find
    \begin{equation}
      \begin{alignedat}{2}
           \hat{H}_{\mathrm{P}}
        &= \underbrace{\frac{\epsilon_{0}}{2} \int \mathrm{d}^3 r
           : \left( \hat{E}_{\perp}^2(\vec{r}) + c_0^2 \hat{B}^2(\vec{r}) \right) :}_{= \sum_{\lambda} \int \mathrm{d}^3 k  \;
           \hbar \omega_k \hat{a}^{\dagger}(\vec{k},\lambda) \hat{a}(\vec{k},\lambda)} - \frac{\epsilon_{0}}{2}\int \mathrm{d}^3 r  \vec{E}_{\parallel}^{2} (\vec{r},t)         
           \\
        & \quad + \frac{1}{c_0} \int \mathrm{d}^3 r j^{k}(\vec{r},t) \hat{A}_{k}(\vec{r}) + \frac{1}{c_0} \int \mathrm{d}^3 r j^{0}(\vec{r},t) A_{0}(\vec{r},t), 
      \end{alignedat}  \label{eq_Hp_1}                                                                                                 
    \end{equation}
    where we used normal ordering, denoted as $::$, to discard the constant energy shift~\cite{greiner1996} and included
    the coupling to a classical external charge current $j^{\mu}$. Using the Green's function of the Laplacian in real-space representation, i.e.,
    $G(\vec{r}, \vec{r}') = \braket{\vec{r}}{(\Delta^{-1}) \vec{r}\hspace{0.5mm}'}$, we can further express the zero component of the
    field as
    \begin{equation}
      \begin{alignedat}{2}
           & \quad - \vec{\nabla}^2 A^0
          && = \frac{j^0}{\epsilon_{0} c_0}                                                                                    \\[5mm]
           & \Rightarrow \hspace{0.7cm} A^0 
          && = \frac{1}{\epsilon_{0} c_0} \int \limits_{\Omega} \mathrm{d}^3 r' 
             \left( - G (\vec{r},\vec{r} \hspace{0.5mm}') \right) j^0(\vec{r} \hspace{0.5mm}',t)                             \\
        & && \underset{\underset{\Omega = \mathbb{R}^3}{|}}{=} \frac{1}{c_0} \int \limits_{\mathbb{R}^3} \mathrm{d}^3 r' 
             \frac{j^0(\vec{r}\hspace{0.5mm}',t)}{4 \pi \epsilon_{0} | \vec{r} - \vec{r} \hspace{0.5mm}' |}.
      \end{alignedat}                                                                                                   \label{eq_RN12}
    \end{equation}
    This, together with $\vec{E}_{\parallel}(\vec{r},t) =- \vec{\nabla}A_{0}(\vec{r},t)$, 
    allows us, after partial integration, to rewrite Eq.~\eqref{eq_Hp_1} as
    \begin{equation} 
      \begin{alignedat}{2}
           \hat{H}_{\mathrm{P}}
        &= \sum_{\lambda} \int \mathrm{d}^3 k \hbar \omega_k \hat{a}^{\dagger}(\vec{k},\lambda) \hat{a}(\vec{k},\lambda) + \frac{1}{c_0} \int \mathrm{d}^3 r j^{k}(\vec{r},t) \hat{A}_{k}(\vec{r})
           \\
        & \quad + \frac{1}{2c_0^2} \int \int \mathrm{d}^3 r \mathrm{d}^3 r' \; w(\vec{r}, \vec{r} \,')j^0(\vec{r} \hspace{0.5mm}',t) j^0(\vec{r},t ). \label{eq_Hp_2}
      \end{alignedat}
    \end{equation}
    Here we have defined
    \begin{align}
      w(\vec{r},\vec{r} \hspace{0.5mm}')
      = - \frac{1}{\epsilon_{0}} G (\vec{r},\vec{r} \hspace{0.5mm}') 
    	  =  \frac{1}{4 \pi \epsilon_{0} | \vec{r} - \vec{r} \hspace{0.5mm}' |}.              \label{eq_RN13}
    \end{align}
    However, the last term in Eq.~\eqref{eq_Hp_2}, which corresponds to the longitudinal degrees of freedom of
    the field, is merely a constant (not an operator) and can thus be discarded in the current case.
    It commutes with all observables. When we couple to the quantum particles, the external current will
    correspond instead to an operator-valued field and thus will no longer vanish. Indeed, it is this part
    that will lead to the longitudinal Coulomb interaction among the charged particles (see Sec.~\ref{sec:CoulombInteraction}).\\
    
    Before we move on and derive the inhomogeneous Maxwell equation from the above Hamiltonian,
    let us rewrite the previously introduced quantum field in a form such that we can easily connect it
    to the Riemann-Silberstein formulation. In accordance to the classical case we can introduce
    the Riemann-Silberstein operators
    \begin{align}
        \hat{F}^{k}_{\pm}(\vec{r})
      = \sqrt{\frac{\epsilon_0}{2}} \left( \hat{E}^{k}_{\perp}(\vec{r}) \pm \mi c_0 \hat{B}^{k}(\vec{r}) \right).    \label{eq_RS_vector_pm}
    \end{align}
    Thus, we see that we can use the expectation value of the Riemann-Silberstein vectors $\vec{F}_{\pm}$
    to re-express the transversal electric and magnetic fields. Some further analysis~\cite{Keller} shows
    that one can furthermore decompose the operators and the expectation values in positive and
    negative frequency parts, which then give rise to helicity creation and annihilation field operators
    and helicity single-photon wave functions.

    Irrespective thereof, from the Hamiltonian of Eq.~\eqref{eq_Hp_2} describing a photon field coupled to a
    classical external current, we can derive the operator
    form of the inhomogeneous Maxwell equation in Coulomb gauge by applying the Heisenberg equation of motion twice, i.e.,
    \begin{align}
      &  (\partial_0^2 + \partial_k \partial^k) \hat{A}^{i}(\vec{r}) \nonumber
         \\
      &= \mu_{0} c_0 j^{i} (\vec{r},t) + \partial^i \partial^0 \frac{1}{c_0}
         \int \mathrm{d}^3 r' w(\vec{r},\vec{r} \hspace{0.5mm}) j^{0}(\vec{r} \hspace{0.5mm},t) \nonumber
         \\
      &= \mu_{0} c_0 j^{i}_{ \perp} (\vec{r},t).                                                              \label{eq_RN20}
    \end{align}
    Here we have assumed in the last step that the external classical charge current $j^{\mu}$ obeys
    the continuity equation $\partial_{\mu}j^{\mu}=0$, such that by the Helmholtz decomposition only
    the transversal part of the charge current $j^{i}_{\perp}$ couples to the purely transversal photon field.
    As pointed out before, the longitudinal component of the photon Hamiltonian does not influence the quantized degrees
    of freedom since it commutes with the field operators. The classical component is determined by Eq.~\eqref{eq_RN12}.

  \subsection{Interaction Hamiltonians}

    After providing the basic uncoupled Hamiltonians of non-interacting particles of different species in Eq.~\eqref{eq_RN11}
    and of uncoupled photons in Eq.~\eqref{eq_Hp_2}, we now join them. Following the minimal-coupling prescription of QED~\cite{greiner1996},
    we know that we merely have to use the conserved total charge current of all species and, in accordance to
    classical electrodynamics, couple it linearly to the vector-potential operator~\cite{Ruggenthaler_2014}. This can be done by
    promoting the above external classical fields to operator-valued fields. Since we work in Coulomb gauge we can do
    the coupling conveniently in two consecutive steps: first only due to the longitudinal and then due to the
    transversal electromagnetic field.

    \subsubsection{Longitudinal interactions}
    \label{sec:CoulombInteraction}

      The zero component of the charge-current operator for the multi-species case is 
      \begin{align}
          \hat{J}^{0}(\vec{r}) 
        = \sum \limits_{n=1}^{N} \underbrace{ q_{(n)} c_{0}
          \hat{\Phi}_{(n)}^{\dagger} \hat{\Phi}_{(n)} }_{= \hat{J}^{0}_{(n)}(\vec{r}) }.                                        \label{eq_RN14}
      \end{align}
      In accordance to the units of a charge current we have multiplied the usual densities $\hat{\Phi}^{\dagger}_{(n)}\hat{\Phi}_{(n)}$
      not only by the respective charge of the species $q_{(n)}$ but also by the velocity of light $c_0$. The longitudinal part
      of the photon field now gets an operator-valued contribution and thus can no longer be discarded. The corresponding term then reads
      \begin{equation}
        \begin{alignedat}{2}
             \hat{W}
          &= \frac{1}{2 c_0^2} \int \int \mathrm{d}^3 r \mathrm{d}^3 r' w(\vec{r},\vec{r} \hspace{0.5mm}')
             : \hat{J}^{0}(\vec{r}) \hat{J}^{0}(\vec{r} \hspace{0.5mm}') :                                                        \\
          &= \frac{1}{2 c_0^2} \int \int \mathrm{d}^3 r \mathrm{d}^3 r' w(\vec{r},\vec{r} \hspace{0.5mm}')
             \sum_{n,n'} : \hat{J}^{0}_{(n)}(\vec{r}) \hat{J}^{0}_{(n')}(\vec{r} \hspace{0.5mm}') :                               \\
          &= \frac{1}{c_0^2} \int \int \mathrm{d}^3 r \mathrm{d}^3 r' w(\vec{r},\vec{r} \hspace{0.5mm}')
             \sum \limits_{n>n'}  \hat{J}^{0}_{(n)}(\vec{r}) \hat{J}^{0}_{(n')}(\vec{r} \hspace{0.5mm}')    \\
          &\quad + \frac{1}{2 c_0^2} \int \int \mathrm{d}^3 r \mathrm{d}^3 r' w(\vec{r},\vec{r} \hspace{0.5mm}')
             \sum_{n} : \hat{J}^{0}_{(n)}(\vec{r})  \hat{J}^{0}_{(n)}(\vec{r} \hspace{0.5mm}')  : \hspace{1mm} .
        \end{alignedat}                                                                                                   \label{eq_RN15}
      \end{equation}
      Here we used normal ordering to bring the interaction in the usual Coulomb form. We note that in the third line,
      since we consider different species, the normal ordering does not affect the expression as the operators commute.
      This term we refer to as the \textit{inter-species} Coulomb interaction, i.e., how the different effective nuclei and
      electrons act on each other via longitudinal photons. The last line describes how the particles of each individual
      species interact longitudinally with each other. We call this term the \textit{intra-species} Coulomb interaction.
      For later reference and also to highlight the nature of the somewhat unfamiliar term of the inter-species
      Coulomb interaction, we note that $w(\vec{r}, \vec{r}\hspace{0.5mm}')$ is the real-space Green's function
      of the Poisson equation. This allows us to define the operator-valued scalar field that one specific species feels
      due to all the other species, i.e.,
      \begin{align}
          \hat{A}_{(n)}^{0}(\vec{r})
        = \sum \limits_{ \underset{n' \neq n}{n'} } \frac{1}{c_0}
          \int \mathrm{d}^3 r' w(\vec{r},\vec{r}') \hat{J}^{0}_{(n')}(\vec{r} \hspace{0.5mm}').                                  \label{eq_RN16}
      \end{align}
%
      Though now operator-valued, this term still only affects the matter subsystem and commutes with the photon
      field observables. Since we have expressed the longitudinal interaction purely in matter degrees of freedom
      we get an interaction among the particles but not with the transversal photon field. The resulting equations
      would be the interacting Pauli equation for a multi-species and multi-particle problem and uncoupled
      (transversal) photons. However, to make the resulting Pauli equation physically reasonable we still
      need to change the masses of the particles. Indeed, to agree with the observed spectrum, we need to use
      the renormalized physical masses that take into account the effect of the transversal photon degrees 
      instead of the bare (unobservable) masses with which we built the coupled problem~\cite{ryder1996quantum, spohn2004}.
      The physical mass is always a sum of the bare plus the electromagnetic mass. The electromagnetic mass
      just subsumes the energy that is stored in the transversal photon field , which is always non-zero when coupled
      to charges. In this way even when we discard the coupling to the transversal photon field, it implicitly
      shows up in our physical masses of electrons and effective nuclei.

    \subsubsection{Transverse interactions}

      Let us next consider the coupling between the transversal degrees of freedom of matter and light.
      To be as general as possible at this point we will not only consider the internal degrees of freedom
      but also couple to classical external fields. To do so we introduce total fields that have an operator-valued
      and a classical contribution, i.e.,
      \begin{equation}
        \begin{alignedat}{2}
             \hat{A}_{\mathrm{tot}}^{k}(\vec{r},t) 
          &= a^{k}(\vec{r},t) + \hat{A}^{k}(\vec{r}),                                                                      \\
             A_{\mathrm{tot}}^{0}(\vec{r},t)
          &= a^{0}(\vec{r},t) + \frac{1}{c_0} \int \mathrm{d}^3 r' w(\vec{r},\vec{r} \,') j^{0}(\vec{r} \,',t). \label{eq_RN17}                     
        \end{alignedat}   
      \end{equation}
      Here, due to the fact that the Coulomb gauge only quantizes the physical (transversal) photon degrees of freedom,
      an external charge density $j^{0}$ couples via the Coulomb kernel directly to the charged particles. Therefore, physically an external classical density becomes equivalent to an external scalar potential $a^{0}$.
      Since the external charge density effectively only changes the external classical
      scalar potential, we need to avoid this sort of ``double counting'' when we want to establish a Runge-Gross-type mapping. We then couple these total fields to the three spatial components of the total conserved charge current of the matter system  
      \begin{equation}
        \begin{alignedat}{2}     
             \hat{J}^{k}(\vec{r},t)
          &= \hat{J}_{\mathrm{pmc}}^{k}(\vec{r}) + \hat{J}_{\mathrm{mc}}^{k}(\vec{r})
             - \sum \limits_{n=1}^{N} \frac{q_{(n)}}{M_{(n)} c_0^2}
             \hat{J}^{0}_{(n)}(\vec{r}) \hat{A}_{\mathrm{tot}}^{k}(\vec{r},t).
        \end{alignedat}                                                                                                   \label{eq_RN25}
      \end{equation}
      The first term is the total paramagnetic current density given by
      \begin{align}
           \hat{J}_{\rm{pmc}}^{k}(\vec{r})
         = \sum_{n=1}^{N} \frac{\hbar q_{(n)}}{2 M_{(n)} \mi} 
           \left[ 
             \left( \partial^{k} \hat{\Phi}^{\dagger}_{(n)}\right) \hat{\Phi}_{(n)}
                    + \hat{\Phi}^{\dagger}_{(n)} \partial^{k} \hat{\Phi}_{(n)} 
             \right],
      \end{align}
      the second is the magnetization current due to the Stern-Gerlach (Pauli) term
      \begin{align}
          \hat{J}_{\rm{mc}}^{k}(\vec{r}) 
        = \sum_{n=1}^{N} - \epsilon^{klm} \partial_l \hat{\Phi}_{(n)}^{\dagger}
          \left(\frac{q_{(n)} \hbar}{2 M_{(n)}} S^{(n)}_{m}  \right) \hat{\Phi}_{(n)},
      \end{align}
      and the last term is the diamagnetic term. That the photon field becomes part of the charge current is due to
      the quadratic part in the Pauli minimal coupling, which in turn is due to expressing the anti-particle (positronic)
      degrees of freedom by the particle degrees of freedom in the non-relativistic limit of the Dirac equation~\cite{Ruggenthaler_2014}.
      Since we only couple to the transversal part of the photon field it is also only the transversal part of the total conserved
      charge current $\hat{J}_{\perp}^{k}$ that is needed. Therefore, the zero component of the charge current, i.e., $\hat{J}^{0}$, does
      not couple to the transversal photons but only to the previously studied longitudinal part of the electromagnetic field.
      The resulting fully coupled generalized non-relativistic QED Hamiltonian is given in the next section. For completeness
      we note that in order to have a well-defined self-adjoint operator, a square-integrable mode mask function should be used~\cite{spohn2004}.
      In its simplest form this is just a cut-off such that the integrals over the photon modes stop at some highest allowed frequency.
      Physically, since we treat non-relativistic particles, a sensible choice is the rest-mass energy of the electrons at about $\hbar \omega_{k} \approx 0.5110$ MeV.
      For such high energies the Pauli equation becomes unreliable. Note that this cut-off implies that also the choice of the bare mass
      depends on this cut-off. For instance, if the cut-off is chosen to be zero, i.e., we only have longitudinal coupling to the Maxwell field,
      the masses are the physical masses.

\section{QEDFT for multi-species}\label{sec_QEDFT}

  Let us now collect all the previous results of the last section and give the generalized Pauli-Fierz Hamiltonian of
  interacting multi-species matter systems coupled to photons and to external classical electromagnetic fields and charge currents
  \begin{equation}
    \begin{alignedat}{2}
         \hat{H}
      &= -\sum \limits_{n=1}^{N} \frac{1}{2 M_{(n)}} \hspace{-0.5mm}
         \int \hat{\Phi}^{\dagger}_{(n)} \hspace{-1mm} \mathpzc{P}_{(n),k} \mathpzc{P}_{(n)}^{k}
         \hat{\Phi}_{(n)}                                                                                              \\
      & \quad + \sum \limits_{n=1}^{N} \int
        q_{(n)} A_{\mathrm{tot}}^0
        \hat{\Phi}_{(n)}^{\dagger} \hat{\Phi}_{(n)}                                                                    \\
      & \quad - \sum \limits_{n=1}^{N} \int \hspace{3mm}
        \frac{q_{(n)} \hbar}{2 M_{(n)}}
        \hat{\Phi}_{(n)}^{\dagger} S_{k,(n)} \hat{\Phi}_{(n)}
        \left( \frac{1}{c_0} \epsilon^{klm} \partial_l \hat{A}_{m,\mathrm{tot}} \right)                                  \\
      & \quad - \frac{\epsilon_0}{2} \int 
        : \left( \hat{E}^{k} \hat{E}_{k} + c^2 \hat{B}^{k} \hat{B}_{k} \right) :   + \frac{1}{c_0} \; \int  j^{k} \hat{A}_k                                      \\
      & \quad + \sum \limits_{n,n'}
        \left\{ 
          \frac{1}{2 c^2_0} \int \int  
          w : \hat{J}^{0}_{(n)} \hat{J}^{0}_{(n')}  :
        \right\}   \honemm ,
    \end{alignedat}                                                                                                    \label{eq_RN21}
  \end{equation}
with the following definition of the canonical momentum
  \begin{align*}
      \mathpzc{P}_{(n)}^{k}
    = \left( - \mathrm{i} \hbar \partial^k + \frac{q_{(n)}}{c_0} \hat{A}^k_{\mathrm{tot}} \right)   \honemm .
  \end{align*}
  Here and in the following we have assumed that also the external electromagnetic field is given in Coulomb gauge
  \begin{align}
    \partial_k a^k (\vec{r},t) = 0 \;.                                                                                    \label{eq_RN23}
  \end{align}
  We also assumed that we have only an external classical scalar potential $a^{0}$ and no external classical charge density $j^{0}$,
  which due to Eq.~\eqref{eq_RN17} would only modify the scalar potential. Thus different configurations of
  $a^{0}$ and $j^{0}$ can lead to the same physics. Consequently, since we assumed that the external classical current
  obeys the continuity equation (see Subsec.~\ref{subsec_photons}), we also find
  \begin{align}
    \partial_k j^k(\vec{r},t) = 0.                                                                                      \label{eq_RN22}
  \end{align}
  We note how enforcing that we do not ``double count'' physically equivalent situations leads to an equivalent ``gauge''
  condition on the classical charge current. This reduction to inequivalent external fields is a prerequisite for
  a Runge-Gross-type mapping. Furthermore, one can check for the consistency of our construction by calculating
  the operator-valued continuity equation. By determining the Heisenberg equation of motion of the zero component
  of the charge current one readily finds
  \begin{align}
      \partial_0 
       \sum \limits_{n=1}^{N} \hat{J}^{0}_{(n)}(\vec{r}) 
    = - \partial_k \hat{J}^{k}(\vec{r},t),                                                                              \label{eq_RN24}
  \end{align}
  i.e., the total conserved current is the afore defined operator of Eq.~\eqref{eq_RN25}. Here it is important
  to note a common pitfall. Since the magnetization current $\hat{J}_{\rm{mc}}^{k}$ is given in terms of a curl, by construction it
  does not appear in the continuity equation. Only upon careful consideration of how to express
  the energy in terms of the linear coupling between current and vector potential or via the Gordon decomposition
  of the relativistic charge current~\cite{engel2011,Ruggenthaler_2014} does the full charge current appear.
  This is the reason why the magnetization current is often overlooked.\\ 

  In a next step we want to establish a bijective mapping between the external fields $(a_{\mu},j_{k})$,
  also called the \textit{external pair}, and the \textit{internal pair} $(J^{\mu},A^{k})$, which are given by the expectation
  value of the corresponding operators for the wave function $\Psi(t)$ that is determined by propagating a fixed
  initial state $\Psi_0$ with the Hamiltonian of Eq.~\eqref{eq_RN21}. Hereby we indicate the dependence of
  the Hamiltonian on the chosen external pair by $\hat{H}[a_{\mu},j_{k}]$ and, correspondingly, the propagated
  wave function also becomes dependent on the external pair, i.e., $\Psi([a_{\mu},j_{k}],t)$. Establishing
  a one-to-one correspondence between $(a_{\mu},j_{k})$ and $(J^{\mu},A^{k})$ would then allow to re-express
  the dependence of the wave function of the fully coupled system in terms of the internal pair, i.e., $\Psi([J^{\mu},A^{k}],t)$.
  This makes all observables expressible in terms of the internal pair only and allows to re-express the full Hamiltonian equation
  as an exact quantum-fluid equation~\cite{Ruggenthaler_2014}. This in turn leads to the possibility of a Kohn-Sham-type
  construction for the coupled matter-photon problem at hand (see Subsec.~\ref{subsec_KS}). In order to establish such
  a one-to-one correspondence we will follow the approach of van Leeuwen for the purely electronic case~\cite{leeuwen-1999}
  and combine it with the derivations of Ref.~\cite{Ruggenthaler_2014}, where electrons are coupled to photons in a similar setting
  as the one considered here. As a first step we therefore establish the fundamental equations of motion for the (still operator-valued)
  internal pair. Since the first time-derivative of $\hat{A}^{k}$ merely leads to the conjugate field $\hat{E}_{\perp}^{k}$,
  we need to go to the second time-derivative, which leads to the inhomogeneous Maxwell equation in Coulomb gauge
  \begin{equation}
    \begin{alignedat}{2}
         (\partial_0^2 + \partial_k \partial^k) \hat{A}^{i}(\vec{r})
      &= \mu_{0} c_0
         \left(
           j^{i}(\vec{r},t) + \hat{J}^{i}(\vec{r},t)
         \right)                                                                                                       \\
      & \quad - \mu_{0} c_0
         \partial^{i} \int \mathrm{d}^3 r'
         \frac{ \partial_{k}' \hat{J}^{k}(\vec{r} \hspace{0.5mm}',t) }{ 4 \pi | \vec{r} - \vec{r} \hspace{0.5mm}' |}.
    \end{alignedat}                                                                                                    \label{eq_RN26}
  \end{equation}
  Here the last term on the right-hand side merely takes care to only count the transversal part of the current operator,
  i.e., it subtracts the longitudinal component with an operator-equivalent of the Helmholtz decomposition.
  Alternatively, we could also write $\hat{J}^{k}_{\perp}$, which implies we only take the transversal component of $\hat{J}^{k}$.
  Again we note that the external current is chosen to be purely transversal, i.e., it obeys Eq.~\eqref{eq_RN22}.
  For the current we can use directly the first time derivative. Instead of summing over all contributions directly let us give
  the equation of motion for the species current, which after lengthy derivations reads as
  \begin{equation}
    \begin{alignedat}{2}
         \partial_t \hat{J}_{(n)}^{k}
      &= - \partial_{l} \hat{T}_{(n)}^{kl} + \hat{W}_{(n)}^{k}                                                         \\
      & \hspace{-8mm} + \frac{q_{(n)}}{M_{(n)} c_0} 
         \left[
           \partial^{k} (a_0 + \hat{A}_{0,(n)}) 
          - \partial_0 \underbrace{(a^{k} + \hat{A}^{k})}_{\hat{A}_{\mathrm{tot}}^{k}}
         \right] \hat{J}^{0}_{(n)}                                                                                                      \\
      & \hspace{-8mm} + \frac{q_{(n)}}{M_{(n)} c_0}
        \partial_{l}
        \left(
          \hat{A}_{\mathrm{tot}}^{k} \hat{J}_{(n)}^{l} + \hat{A}_{\mathrm{tot}}^{l} \hat{J}_{{\mathrm{pmc}},(n)}^{k}
        \right)                                                                                                        \\
      & \hspace{-8mm} + \frac{q_{(n)}}{M_{(n)} c_0}
        \hat{J}_{(n)}^{l}
        \left(
          \partial^{k} \hat{A}_{l,\mathrm{tot}} - \partial_{l} \hat{A}_{\mathrm{tot}}^k
        \right)                                                                                                        \\
      & \hspace{-8mm} + \frac{q_{(n)}}{M_{(n)} c_0}
        \epsilon^{lmn} \partial_{m} ( \partial^{k} \hat{A}_{l,\mathrm{tot}} ) \hat{M}_{n,(n)}                          \\
      & \hspace{-8mm} + \epsilon^{klm} \partial_{l}
        \left\{
          \partial_{l'}
          \left[
            \frac{\hbar}{\mathrm{i} 2 M_{(n)}} \hat{\Phi}_{(n)}^{\dagger}
            \frac{q_{(n)}\hbar}{2 M_{(n)}}
            S_{(n)} \partial^{l'} \hat{\Phi}_{(n)}
          \right]
        \right\}                                                                                                        \\
      & \hspace{-8mm} + \epsilon^{klm} \partial_{l}
        \left\{
          \partial_{l'}
          \left[
            \left( \partial^{l'} \hat{\Phi}_{(n)}^{\dagger} \right)
            \frac{q_{(n)} \hbar}{2 M_{(n)}} S_{m,(n)} \hat{\Phi}_{(n)}
          \right]
        \right\}                                                                                                        \\
      & \hspace{-8mm} - \epsilon^{klm} \partial_{l}
        \left\{
          \partial_{l'}
          \left[
            \frac{q_{(n)}}{M_{(n)}c_0} \hat{A}_{\mathrm{tot}}^{l'} \hat{M}_{m,(n)}
          \right]
        \right\}                                                                                                        \\
      & \hspace{-8mm} + \epsilon^{klm} \partial_{l}
        \left\{
          \partial_{l'}
          \left[
            \frac{q_{(n)}}{M_{(n)} c_0}
            \left( \partial_{l'} \hat{A}_{m,\mathrm{tot}} - \partial_{m} \hat{A}_{l',\mathrm{tot}} \right)
            \hat{M}_{(n)}^{l'}
          \right]
        \right\}
    \end{alignedat}                                                                                                    \label{eq_RN27}
  \end{equation}
  In accordance to the usual electronic case, we have defined here the momentum stress tensor
  \begin{align}
       \hat{T}_{(n)}^{kl} 
    &= \frac{q_{(n)}\hbar^2}{2 M_{(n)}^{2}}
       \left[
         \left(\partial^{k}\hat{\Phi}_{(n)}^{\dagger}\right) \left(\partial^{l} \hat{\Phi}_{(n)} \right)
         + \left(\partial^{k}\hat{\Phi}_{(n)}^{\dagger}\right) \left(\partial^{l} \hat{\Phi}_{(n)} \right) \right. \nonumber
       \\
    & \left. - \frac{1}{2} \partial^{k} \partial^{l} \hat{\Phi}^{\dagger}_{(n)}\hat{\Phi}_{(n)} \right]
  \end{align}
  and the interaction stress tensor
  \begin{align}
     \hat{W}^{k}_{(n)} =& \frac{q^{3}_{(n)}}{M_{(n)}}\sum_{s,s'}\int \rm{d}^3 r' \left[ \hat{\Phi}^{\dagger}_{(n)}(\vec{r},s)\hat{\Phi}^{\dagger}_{(n)}(\vec{r} \;', s') \right.
     \\
    &\left. \left(\frac{\partial}{\partial r'_{k}} w(\vec{r}\;', \vec{r})  \right)\hat{\Phi}_{(n)}(\vec{r}\;',s\;')\hat{\Phi}^{\dagger}_{(n)}(\vec{r}, s) \right]. \nonumber
  \end{align}
  Furthermore, in correspondence with the current density, we have also defined a magnetization density of the
  $n$-th particle species
  \begin{align}
    \hat{M}^{k}_{(n)}= \hat{\Phi}^{\dagger}_{(n)}\left( \frac{\hbar q_{(n)}}{2 M_{(n)}} S^{k}_{(n)} \right) \hat{\Phi}_{(n)}
  \end{align}
  as well as the corresponding expressions for the $n$-th paramagnetic current. In this somewhat complicated expression,
  which is consistent with previous electron-photon~\cite{Ruggenthaler_2014} and electron-only results~\cite{stefanucci2013},
  the term of main interest in the following will be  
  \begin{equation}
    \begin{alignedat}{2}
         \sum_{n} \frac{q_{(n)}}{M_{(n)} c_0} \left( \partial_0 \hat{A}_{\mathrm{tot}}^{k} \right) \hat{J}^{0}_{(n)}
      &= (\partial_{0} a^{k})
         \left(
           \sum \limits_{n} \frac{q_{(n)}^2}{M_{(n)}} \hat{\Phi}_{(n)}^{\dagger} \hat{\Phi}_{(n)}
         \right)                                                                                                       \\
      & \quad + \sum \limits_{n} \frac{q_{(n)}^2}{M_{(n)}}
        (\partial_{0} \hat{A}^{k}) \hat{\Phi}_{(n)}^{\dagger} \hat{\Phi}_{(n)}.
    \end{alignedat}                                                                                                    \label{eq_RN28}
  \end{equation}
  Here the first term on the right-hand side will be the reason why we are able to establish a one-to-one correspondence,
  i.e., a Runge-Gross-type result. Due to the fact that we have a $q^{2}_{(n)}$ factor, all different particles species add up
  positively and the expectation value for a physical wave function will be usually strictly positive in all of space, i.e., non-zero everywhere.
  Specifically, in the following we will choose an initial state $\Psi_0$ that obeys
  $\sum_{n}\braket{\Psi_0}{\hat{\Phi}_{(n)}^{\dagger} \hat{\Phi}_{(n)} \Psi_0} > 0 $ in all of space.
  While we already need to assume such a strict positivity for electrons only, we now have a positive contribution
  from each particle species leaving this assumption rather trivial to be fulfilled.\\
  
  Before proving a bijective mapping (under certain assumptions) we introduce some further convenient notation. We rewrite the equation of motion for the charge current in the following compact form
  \begin{equation}
    \begin{alignedat}{2}
         \partial_t \hat{J}^{k}
      &= \left( \partial^{k} a_{0} - \partial_{0} a^{k} \right) 
         \underbrace{ \sum \limits_{n=1}^{N} \frac{q_{n}^2}{M_{(n)}}
         \hat{\Phi}_{(n)} \hat{\Phi}_{(n)} }_{=\hat{\rho}}
         + \; \hat{Q}^{k},
    \end{alignedat}                                                                                                     \label{eq_RN29}
  \end{equation}
  where we have subsumed all the complex expressions of how particles and photons couple with each other in
  \begin{align}\label{eq_Qdef}
      \hat{Q}^{k}
    = \sum \limits_{n=1}^{N} 
      \left[ \frac{q_{(n)}}{M_{(n)} c_0}
        \left( \partial^{k} \hat{A}_0^{(n)} - \partial_0 \hat{A}^{k} \right) \hat{J}^{0}_{(n)}
        - \partial_{l} \hat{T}_{(n)}^{kl} + \hat{W}_{(n)}^{k} + ...
      \right].                                                                                                           \nonumber
  \end{align}
  Here it is important to note that no time-derivatives of the external potentials appear in $\hat{Q}^{k}$. In a next step we define higher-order time derivatives of expectation values of operators at $t=0$ by
  \begin{align}
    J^{k, \{\alpha\}}(\vec{r}) = \partial_{t}^{{\alpha}} \left. \braket{\Psi(t)}{\hat{J}^{k}(\vec{r},t) \Psi(t)}\right|_{t=0}
   \end{align}
  and accordingly for the classical external fields by
  \begin{align}
    a^{k, \{\alpha\}}(\vec{r}) = \partial_{t}^{{\alpha}} \left. a^{k}(\vec{r},t) \right|_{t=0},
  \end{align}
  where we put the time-derivative in brackets, i.e., $\{ \alpha \}$, to distinguish them from other components. Here $\alpha \in \mathbb{N}_{0}$ and we assume from now on that all external fields and expectation values are real analytic in time,
  i.e., that their Taylor series in time has a finite convergence radius. In general one can get away with fewer
  conditions~\cite{Ruggenthaler_review} but for the sake of simplicity we stick with these rather stringent ones.
  Having assumed that all time-derivatives at zero exist (which implies a rather well-behaved initial state) we can,
  based on the Eq.~\eqref{eq_RN29}, deduce how all the different time-derivatives at $t=0$ are connected.
  Following van Leeuwen~\cite{leeuwen-1999} we find
  \begin{equation}
    \begin{alignedat}{2}
         J^{k,\{\alpha+1\}}
      &= Q^{k,\{\alpha\}}                                                                  \\
      & \quad + \sum \limits_{\beta=0}^{\alpha} \begin{pmatrix} \alpha \\ \beta \end{pmatrix} \rho^{\{\alpha-\beta\}}
         \left[ \partial^{k} a^{0,\{\beta\}} - \frac{1}{c_0} a^{k,\{\alpha+1\}} \right].
    \end{alignedat}                                                                                                    \label{eq_RN30}
  \end{equation}
  Accordingly we can define a similar equation based on Eq.~\eqref{eq_RN26} for the vector-potential expectation
  value and the external classical charge current
  \begin{equation}
    \begin{alignedat}{2}
         A^{k,\{\alpha+2\}}
      &= -c^2_0 \partial_{k} \partial^{l} A^{k,\{\alpha\}}                    \\
      &\quad + \mu_{0} c^3_0 \left( j^{k,\{\alpha\}} + J_{\perp}^{k,\{\alpha\}} \right)\;.
    \end{alignedat}                                                                                                    \label{eq_RN33}
  \end{equation}
  We can now show constructively that there is a bijective mapping $(a_{\mu},j_{k}) \leftrightarrow (J^{\mu}, A^{k})$
  by prescribing the Taylor series of an internal pair and then constructing a unique external pair that generates
  $(J^{\mu}, A^{k})$ by propagating the given initial state $\Psi_0$ with the corresponding Hamiltonian $\hat{H}[a_{\mu},j_{k}]$.\\

  To do so we first of all need to guarantee that the prescribed internal current and vector potential at $t=0$ agrees
  with the fixed initial state $\Psi_0$. Therefore it needs to hold that
  \begin{align} \label{eq_inistate}
       A^{k,\{0\}} 
    &= \braket{\Psi_0}{\hat{A}^{k}\Psi_0}, \; A^{k,\{1\}} = -c_0\braket{\Psi_0}{\hat{E}^{k}_{\perp} \Psi_0}, \nonumber
       \\
       J^{0,\{0\}}
    &= \braket{\Psi_0}{ \hat{J}^{0} \Psi_0}.
  \end{align}
  The last condition guarantees that also $J^{0}$ is exactly reproduced via the continuity equation as we go along in time. We note that
  due to the fact that the current at the initial time also contains $a^{k,\{0\}}$ we get a first condition on
  the chosen current as well as the $\alpha=0$ component of the external field by 
  \begin{align}\label{eq_akalpa0}
       a^{k,\{0\}}
    &= \frac{1}{\rho^{\{0\}}}
       \left[ 
         J^{k,\{0\}} - J_{\mathrm{pmc}}^{k,\{0\}} 
        \right. \nonumber \\
     & + \left. \sum \limits_{n} \frac{q_{(n)}^2}{M_{(n)}} \braket{\Psi_0}{ \hat{n}_{(n)} \hat{A}^{k} \Psi_0}
         - J_{\rm{mc}}^{k,\{0\}}
       \right], 
  \end{align}
  where all other quantities are given via the initial state. Due to the fact that we have chosen the Coulomb gauge, the left hand side should obey $\partial_k a^{k,\{0\}}=0$. This in turn implies that also the right-hand side should be purely transversal. If this is not the case,
  and since we know that the charge current is gauge-independent, we would need to perform a gauge transformation
  (which we now know)
  \begin{align}
    a'^{0} = a^{0}-\partial^{0} \chi  \hspace{1mm}, \hspace{4mm} a'^{k} = a^{k} - \partial^{k} \chi
    \label{eq_RN32}
  \end{align}                                                                                %
  on the initial state such that it complies with the Coulomb gauge or we use a different gauge choice. For the other
  $\alpha=0$ components we merely employ the equations of motions from Eqs.~\eqref{eq_RN25} and \eqref{eq_RN30},
  which leads to
  \begin{equation}
    \begin{alignedat}{2}
         j^{k,\{0\}}
      &= \frac{1}{\mu_{0} c_0^3} 
         \left[A^{k,\{2\}}+
           c^2_0 \partial_{l} \partial^{l} A^{k,\{0\}} - \mu_{0} c_0^3 J_{\perp}^{k,\{0\}}
         \right]         ,                                                                                              \\
         - \partial_{k} \partial^{k} a^{0,\{0\}}
       &= \partial_k \frac{1}{\rho^{\{0\}}} \left( Q^{k,\{0\}} - J^{k, \{1\}} \right) .
    \end{alignedat}                                                                                                    \label{eq_RN37}
  \end{equation}
  Here $Q^{k,\{0\}}$ contains only the previously determined components of the external vector potential.
  For the higher orders of the external pair we now only use the equations of motion, i.e., 
  \begin{equation}
    \begin{alignedat}{2}
         a^{k,\{\alpha+1\}}
      &= \frac{c_0}{\rho^{\{0\}}} 
         \Bigg[
           Q^{k,\{\alpha\}} - J^{k,\{\alpha+1\}} 
           +\rho_{\{0\}} \partial^{k} a^{0,\{\alpha\}}                                              \\
      &\qquad + \sum_{\beta=0}^{\alpha-1} \begin{pmatrix} \alpha \\ \beta \end{pmatrix} 
           \left[ \partial^{k} a^{0, \{\beta \}} - \frac{1}{c_0} a^{k,\{\beta+1\}} \right] \rho^{\{\alpha-\beta\}}
         \Bigg]_{\perp}                                                                                                \\
         - \partial_{k} \partial^{k}
      &  a^{0,\{ \alpha\}} = \partial_{k} \frac{1}{\rho^{\{0\}}}
         \Bigg[
           Q^{k,\{\alpha\}} - J^{k,\{\alpha+1\}}                                                                                     \\
      &  \hspace{9mm} + \sum \limits_{\beta=0}^{\alpha-1} \begin{pmatrix} \alpha \\ \beta \end{pmatrix}
           \left( \partial^{k} a^{0,\{\beta\}} - \frac{1}{c_0} a^{k,\{\beta + 1\}} \right) \rho^{\{\alpha-\beta\}}
         \Bigg]                                                                                                        \\
         j^{k,\{\alpha\}}
      &= \frac{1}{\mu_{0} c_0^3}
         \Bigg[ 
           A^{k,\{\alpha+2\}} + c_0^2 \partial_{k'} \partial^{k'} A^{k,\{ \alpha\}}
           - \mu_{0} c_0^3 J_{\perp}^{k,\{\alpha\}}
         \Bigg]
    \end{alignedat}                                                                                                    \label{eq_RN38}
  \end{equation}
  The $a^{k,\{1\}}$-term we get again in two steps. From the first equation above (in the case $\alpha=0$ the sum over $\beta$ is to be discarded) we determine an $\tilde{a}^{k,\{1\}}$, and from the condition that the resulting field is purely transversal we determine the corresponding gauge function in next order, i.e., 
  \begin{align}
    \left. \partial_{t}\partial_{k} \chi(\vec{r},t)\right|_{t=0},
  \end{align}
  which then leads to $a^{k,\{1\}}$. We indicate this procedure with a $\perp$ at the right-hand side of the equation. This procedure is possible since 
  only the previously determined orders appear in $Q^{k,\{1\}}$.
  The first-order for the external current is simply determined from plugging in the prescribed internal pair.
  We can now repeat this order by order and with this construct the corresponding external pair
  \begin{align}
    \left (a^{\mu} = \sum_{\alpha=0}^{\infty} \frac{a^{\mu, \{ \alpha \}}}{\alpha!} t^{\alpha},j^{k} = \sum_{\alpha=0}^{\infty} \frac{j^{k, \{ \alpha \}}}{\alpha!} t^{\alpha}  \right).
  \end{align}
  In this procedure we see the necessity of the afore-mentioned (after Eq.~\eqref{eq_RN28}) strict positivity of
  $\rho^{\{0\}}$. If $\rho^{\{0\}}$ would be zero somewhere, the external field would be undetermined by this construction.
  Let us finally point out that within a given gauge choice only one such pair can exist. For the $j^{k}$ this is trivial
  due to the linearity of the Maxwell equation and the ``gauge'' condition $\partial_k j^k =0$. For the vector potential
  in zeroth order, i.e., Eq.~\eqref{eq_akalpa0}, we immediately see that the difference can only be due to a gauge
  transformation. Fixing the gauge leaves us with only one choice. This in turn fixes the zero component of the
  scalar potential, i.e., Eq.~\eqref{eq_RN37}. In the next order, i.e., Eq.~\eqref{eq_RN30} for $\alpha=1$, we are then
  again left with only a gauge choice and by fixing $\partial_k a^{k,\{1\}}=0$ we find the unique representative.
  In this way we can go through all orders and find one and only one $a^{\mu}$.

\section{Maxwell-Pauli-Kohn-Sham coupling} \label{subsec_KS}

  Although we could have shown the one-to-one correspondence more easily by following the original Runge-Gross
  approach~\cite{runge-1984}, the constructive approach by van Leeuwen~\cite{leeuwen-1999} is beneficial
  for showing the existence of a Kohn-Sham-type system. But before, let us briefly explain how the above
  one-to-one correspondence becomes relevant for practical applications.\\

  In the previous section we established that in principle the full wave function of the generalized Pauli-Fierz Hamiltonian of
  Eq.~\eqref{eq_RN21} is determined uniquely by the given initial state $\Psi_0$ and the internal pair
  $(J^{\mu}, A^k)$. From this we realize that instead of solving the wave-function-based Hamiltonian formulation of the problem
  (that due to the infinitely many degrees of freedom of the photon field is impossible even for only one particle)
  we can alternatively solve two coupled non-linear fluid equations
  \begin{align}
        \partial_t J^{k}(\vec{r},t)
     &= \left( \partial^{k} a_{0}(\vec{r},t) - \partial_{0} a^{k}(\vec{r},t) \right) \rho([J^{\nu},A^{l}];\vec{r},t) \nonumber
        \\
     &+ Q^{k}([J^{\nu},A^{l}; a^{\tau}];\vec{r},t), \label{eq:fluid1}
        \\
        (\partial_0^2 +  \partial_k& \partial^k) A^{l}(\vec{r},t)
      = \mu_{0} c_0
         \left(
           j^{l}(\vec{r},t) + J^{l}_{\perp}(\vec{r},t).
         \right)           \label{eq:fluid2}                             
  \end{align}
  Here $\rho([J^{\nu},A^{l}];\vec{r},t)$ is a functional of the internal pair and initial state respectively
  (we do not indicate the dependence on the initial state for simplicity) and $Q^{k}([J^{\nu},A^{l}; a^{\tau}];\vec{r},t)$
  does also depend on the chosen external vector potential (see the partial definition in Eq.~\eqref{eq_Qdef} and
  Eq.\eqref{eq_RN27}, respectively). It is easy to rewrite the term $Q^{k}$ in such a way as to make the dependence on
  $a^{\mu}$ explicit, but for the sake of brevity we refrain from this here. The above coupled equation are the exact
  quantum Navier-Stokes equations~\cite{tokatly2005quantum, Ruggenthaler_2014, Ruggenthaler_review} and make calculations
  of particles coupled to the photon field practical. In principle, for a given external pair the solution of these
  two coupled equations would lead to the exact internal pair, from which we could construct all physical observables.
  The only severe drawback is that we do not know the expressions of $\rho$ and $Q^k$ in terms of the currents and fields,
  but only in terms of the intractable wave function, so we have to apply some approximations in practice. Instead of trying
  to find approximations in terms of $J^{\mu}$ and $A^{k}$ directly, we will follow the well-established approach to employ
  a numerically simpler quantum system that shares as much similarities with the full generalized Pauli-Fierz problem and
  then build approximations on top of this ersatz system. To put it differently, we will use a still numerically solvable system
  to build approximations to $\rho$ and $Q^k$ in terms of the resulting auxiliary wave functions. This procedure is called
  a Kohn-Sham construction~\cite{kohn-1965,tokatly2005quantum, Ruggenthaler_2014, Ruggenthaler_review}.\\

  While we have many possibilities, the simplest choice for an auxiliary Kohn-Sham-type system is a non-interacting
  (and consequently also uncoupled) system as already introduced for the electron-photon case~\cite{ruggenthaler2011time, Ruggenthaler_2014}.
  Similarly to the fully interacting case, we can find a one-to-one correspondence between external and internal pairs
  \begin{align}
    (a^{\mu}_{(\rm{s})}, j^{k}_{(\rm{s})}) \leftrightarrow (J_{\mu}, A_{k} ).
  \end{align}
  Here we follow the usual convention to denote the external fields for a non-interacting system with a subindex ``$\rm{s}$''.
  By following the construction of Sec.~\ref{sec_QEDFT}, we can also provide a one-to-one correspondence for the non-interacting case and thus generate two maps
  \begin{equation}
    \begin{alignedat}{2}
      \left( J_{\mu}, A_{k} \right) \mapsto \left( a^{\mu} , j^{k}  \right)            \\
      \left( J_{\mu}, A_{k} \right) \mapsto \left( a^{\mu}_{(\rm{s})} , j^{k}_{(\rm{s})}  \right)  
    \end{alignedat}                                                                                                    \label{eq_RN39}
  \end{equation}
  of a given internal to different external pairs for given initial states $\Psi_0$ (of the interacting) and $\Phi_0$
  (of the non-interacting), respectively. We therefore see that the above extended Runge-Gross-type approach helps us in
  showing that a fully interacting problem is representable by a non-interacting problem. Since the Kohn-Sham photon field is uncoupled,
  instead of solving for the Kohn-Sham photon wave function for infinitely many degrees, we can just solve the corresponding
  classical inhomogeneous Maxwell equation~\cite{Ruggenthaler_2014}. Both the classical as well as the fully quantized yet
  uncoupled Kohn-Sham photon equation lead, by construction, to the same field $A^{k}$. This allows us to use a matter wave function $\Phi_0$
  of non-interacting particles only, which we assume to obey
  \begin{align}
      \braket{\Psi_0}{ \hat{J}^{0} \Psi_0} 
    &=\braket{\Phi_0}{ \hat{J}^{0} \Phi_0}.
  \end{align}
  The initial state of the fully coupled wave function $\Psi_0$ then further provides the initial condition for the
  classical photon field, i.e., Eqs.~\eqref{eq_inistate}. For the external current, due to the linearity of
  the Maxwell equation, the mappings are the same and hence lead to
  \begin{align}
       j^{k} \left[ J_{\nu}, A_{l} \right]
    =  j^{k}_{(\rm{s})} \left[ J_{\nu}, A_{l} \right].
  \end{align}
  Following the usual Kohn-Sham construction~\cite{Ruggenthaler_review} we merely need to introduce
  a mean-field exchange-correlation (Mxc) potential
  \begin{align}
      a^{\mu}_{\mathrm{Mxc}} \left[ J_{\nu}, A_{l} \right] 
    = a_{(\rm{s})}^{\mu} \left[ J_{\nu}, A_{l} \right] - a^{\mu} \left[ J_{\nu}, A_{l} \right],                                  \label{eq_RN40}
  \end{align}
  to take care of the difference between the physical and the auxiliary system. Alternatively we can directly construct the Kohn-Sham vector potential $a_{\rm{KS}}^{\mu} = a^{\mu} + a^{\mu}_{\rm{Mxc}}$ by employing the Eqs.~\eqref{eq_RN38} for the non-interacting case together with Eq.~\eqref{eq_RN30} for the interacting case to express the unknown $J^{k,\{\alpha+1\}}$ in terms of the basic variables.
%
%
  With these definitions in place,
  assuming that the initial state is a tensor product of Slater determinants (fermions) and permanents (bosons),
  we find the auxiliary Maxwell-Pauli-Kohn-Sham (MPKS) equations  
  \begin{equation}
    \begin{alignedat}{2}
         \mathrm{i} \hbar \partial_t \phi_{(n,i)} (\vec{r} s_{(n)}, t) &=                                              \\
      & \hspace{-27mm} \Bigg\{ - \frac{1}{2 M_{(n)}} \tilde{\mathpzc{P}}_{(n)}^k \tilde{\mathpzc{P}}_{(n),k}
        + q_{(n)} \left( a^{0} + a_{\mathrm{Mxc}}^{0} \right)                                                          \\
      & \hspace{-25.5mm} - \frac{q_{(n)} \hbar}{2 M_{(n)} c_0} 
        S_{k}^{(n)} \left[\epsilon^{klm} \partial_{l}
        \left( a_{m} + a^{\mathrm{Mxc}}_{m} \right)\right] \Bigg\}
        \phi_{(n,i)}(\vec{r} s_{(n)}, t)   \honemm ,
    \end{alignedat}                                                                                                    \label{eq_RN26_a}
  \end{equation}
  \begin{align*}
      \tilde{\mathpzc{P}}_{(n)}^k
    = - \mathrm{i} \hbar \partial^{k} + \frac{q_{(n)}}{c_0}
      \left( a^{k} + a^{k}_{\mathrm{Mxc}} \right)   \honemm ,
  \end{align*}
  \begin{equation}
    \begin{alignedat}{2}
         \left( \partial_{0}^{2} + \partial_{k} \partial^{k} \right) A^{l}(\vec{r},t)
      &= \mu_{0} c_0 
         \left(
           j^{l}(\vec{r},t) + J^{l}_{\perp}(\vec{r},t)
         \right)    \honemm ,
    \end{alignedat}                                                                                                    \label{eq_RN26_b}
  \end{equation}
  where the internal charge current is determined by the auxiliary wave function $\Phi(t)$. 
  This wave function consists of auxiliary single-particle spin-orbitals $\phi_{(n,i)}(t)$, where we have a corresponding spin-degree of freedom $s_{(n)}$ for each species $n$. 
  To distinguish the indices $n$ and $i$ from covariant indices we put them in parenthesis. We stress that, provided we have the exact Mxc potential, the coupled auxiliary MPKS problem \textit{predicts}
  the exact internal pair $(J^{\mu}, A^{k})$ for a given generalized Pauli-Fierz Hamiltonian
  $\hat{H}[a^{\mu}, j^{k}]$. The infinite number of degrees of freedom of the quantized photon field
  has been \textit{exactly} subsumed in the classical Maxwell equation and in the non-linear coupling 
  to the matter subsystem. In this way the MPKS equation takes into account the full photon and phonon
  bath in an exact manner.

  \subsection{Classical limit for Nuclei}    \label{subsec_classical_nuclei}

    At this point we note that we have quite some freedom in establishing the
    mappings as well as the MPKS systems. For instance, we can use instead of the uncoupled particles with their respective bare masses just the uncoupled particles with their physical masses. That is, we already subsume as usually done in quantum mechanics, the fluctuations of the bare electromagnetic vacuum in the physical (dressed) masses~\cite{ruggenthaler2018quantum}. This we will do in the following, i.e., the MPKS systems is build in practice by using the physical masses of the particles. Furthermore, we could look at each
    individual particle-species' internal current and find that also each
    species' internal current can be used to establish a mapping individually.
    We could then (unphysically) assume that each species sees a different
    external field and then establish purpose built $a^{\mu}_{\rm{Mxc},(n)}$.
    Such an approach might be easier in establishing more accurate
    approximations to the ultimately unknown Mxc potentials. Here we will not
    follow this route but instead try to find a first approximation that
    simplifies the MPKS construction slightly. Even though we have rewritten
    the coupled generalized Pauli-Fierz problem in terms of single-particle
    quantum equations, the solution of a large amount of these non-linear
    equations is still numerically demanding.  Furthermore, for the initial
    states, which can be determined from a ground-state reformulation of the
    generalized Pauli-Fierz problem following Ref.~\cite{ruggenthaler2015}, it
    is often beneficial to make the Born-Oppenheimer approximation and treat
    the nuclei semi-classically. We point out, that due to the
    coupling of the photons to the electron-nucleus system the problem without
    external fields is now invariant with respect to the total momentum and
    total angular momentum of the {\it coupled} matter-photon
    system~\cite{spohn2004}, not the coupled matter system only.  Thus when
    translating or rotating only the matter-system in real space, the photon
    field is changed, which breaks the usual real-space translational and
    rotational symmetry.  This can be beneficial to overcome the usual drawback
    of electron-nucleus quantum mechanics (and the corresponding DFT
    formulation) that the densities of such matter systems are homogeneous and
    no direct molecular structure is apparent. The physical rationale to then perform the classical limit for the nuclei subsystem is
    that the nuclei are much heavier than the electrons and hence they behave
    more classically.  In the following we will therefore simplify the MPKS
    construction slightly and describe the nuclei classically.  We note that
    more advanced alternatives exists that, e.g., are based on the exact
    factorization of electron-nuclei wave
    functions~\cite{abedi2010exact,abedi2012correlated}.

    In order to see whether this is possible in a simple manner let us consider
    the MPKS equations for the nuclear orbitals. A first approximation will be
    to discard the Stern-Gerlach term. Next we rewrite the spatial orbitals in
    polar representation
    \begin{align}
      \phi_{(n,i)}(\vec{r},t) = | \phi_{(n,i)}(\vec{r},t) | \mathrm{e}^{(\mathrm{i}/\hbar)S_{(n,i)}(\vec{r},t)}.          \label{eq_CT_MQC_a}
    \end{align}
    If we then plug this into the MPKS equation without the Stern-Gerlach term (which makes the solution spin-independent)
    we find a Hamilton-Jacobi-type equation for the phase~\cite{min2017ab} 
    \begin{equation}
      \begin{alignedat}{2}
           \partial_t S_{(n,i)}(\vec{r},t)
        &= -
           \frac{ \left( \vec{\nabla} S_{(n,i)}(\vec{r},t) - \frac{q_{(n)}}{c_0}\vec{a}_{\rm{KS}}(\vec{r},t) \right)^2 }{ 2 M_{(n)} }  \\
        & \quad  - q_{(n)} a^{0}_{\rm{KS}}(\vec{r},t) + \frac{\hbar^2 }{2 M_{(n)}} 
           \frac{ \vec{\nabla}^2 | \phi_{(n,i)}(\vec{r},t) | }{ | \phi_{(n,i)}(\vec{r},t) | },
      \end{alignedat}                                                                                                    \label{eq_CT_MQC_S2}
    \end{equation}
    where we denote $a^{\mu}_{\rm{KS}} = a^{\mu} + a^{\mu}_{\rm{Mxc}}$. Using
    the above equation for the phase we will turn the quantum evolution
    equation in a classical equation in the following. The physical argument is
    based on the fact that the $\hbar^2/2M_{(n)}$ factor becomes very small for
    heavy particles such as nuclei, and can thus be discarded in a first
    approximation. But before we do so, let us connect the relevant quantum
    observables to their classical counter parts. To do so we first consider
    the conserved current for Eq.~\eqref{eq_RN26_a} without the Stern-Gerlach
    term, which is $\vec{J}_{(n,i)}(\vec{r},t) =
    \vec{J}_{\rm{pmc}}^{(n,i)}(\vec{r},t) +
    \vec{J}_{\rm{dmc}}^{(n,i)}(\vec{r},t) =
    \tfrac{q_{(n)}}{M_{(n)}}|\phi_{(n,i)}(\vec{r},t)| \vec{\nabla}
    S_{(n,i)}(\vec{r},t) - \tfrac{q_{(n)}^2}{M_{(n)c_{0}}}
    \vec{a}_{\mathrm{KS}}(\vec{r},t) |\phi_{(n,i)}(\vec{r},t)|$. With this the
    total velocity field becomes
    \begin{align}
    \vec{v}_{(n,i)}(\vec{r},t) & = \frac{\vec{J}_{(n,i)}(\vec{r},t)}{q_{(n)}|\phi_{(n,i)}(\vec{r},t)|}  \nonumber \\
                                 & = \frac{1}{M_{(n)}} \left( \vec{\nabla} S_{(n,i)}(\vec{r},t) - \frac{q_{(n)}}{c_{0}} \vec{a}_{\mathrm{KS}}(\vec{r},t)\right).
    \end{align}
    And accordingly we can define the total momentum field $\vec{p}_{(n,i)} = \vec{v}_{(n,i)}/M_{(n)}$.

    If we discard the quantum-potential part in Eq.~\eqref{eq_CT_MQC_S2} assuming the classical limit $\hbar \rightarrow 0$ for the slowly moving nuclei, then the resulting equation becomes
    \begin{equation}
    \begin{alignedat}{2}
    \partial_t S_{(n,i)}(\vec{r},t)
    &= 
    - \left( \vec{v}_{(n,i)}(\vec{r},t) \cdot \vec{\nabla} \right) \vec{p}_{(n,i)}(\vec{r},t) - q_{(n)} a^{0}_{\mathrm{KS}}(\vec{r},t)              \\
    & \quad + \frac{q_{(n)}}{M_{(n)}} \vec{p}_{(n,i)}(\vec{r},t) \times \vec{b}_{\mathrm{KS}} (\vec{r},t),
    \end{alignedat}                                                                     
    \end{equation}
    where $\vec{b}_{\mathrm{KS}}(\vec{r},t) = \tfrac{1}{c_{0}} \vec{\nabla} \times \vec{a}_{\mathrm{KS}}(\vec{r},t)$. If we then add to both sides the partial time-derivative of the Kohn-Sham vector potential and define the total derivative for a co-moving reference frame that moves with the velocity field $\vec{v}_{(n,i)}$, i.e., 
    \begin{equation}
    \begin{alignedat}{2}
    \dot{\vec{p}}_{(n,i)}(\vec{r},t) = \partial_t \vec{p}_{(n,i)}(\vec{r},t) + \left( \vec{v}_{(n,i)}(\vec{r},t) \cdot \vec{\nabla} \right) \vec{p}_{(n,i)}(\vec{r},t),
    \end{alignedat}                                                                     
    \end{equation}
    then the above equation becomes
    \begin{equation}
    \begin{alignedat}{2}
    \dot{\vec{p}}_{(n,i)}(\vec{r},t) 
    &= q_{(n)} \vec{v}_{(n,i)}(\vec{r},t) \times \vec{b}_{\mathrm{KS}} (\vec{r},t) + q_{(n)} \vec{E}_{\mathrm{KS}}(\vec{r},t).
    \end{alignedat}                                                                     
    \end{equation}
    Here we used that $-\partial_0 \vec{a}_{\mathrm{KS}} = \vec{E}_{\perp}^{\mathrm{KS}}$ gives the transversal electric field and $- \vec{\nabla} a^{0}_{\mathrm{KS}} = \vec{E}_{\parallel}^{\mathrm{KS}}$ gives the longitudinal electric field such that $\vec{E}_{\mathrm{KS}} = \vec{E}_{\perp}^{\mathrm{KS}} + \vec{E}_{\parallel}^{\mathrm{KS}}$. This is  just the classical Lorentz equation which can be solved by the method of characteristics, i.e., we can follow a specific classical trajectory that starts at $\vec{r}_{(n,i)}$ and $\vec{p}_{(n,i)}$. The initial wave function then gives us the initial distribution of these trajectories. Using this classical approximation we can determine the charge current of the nuclei that contribute to the total current and thus the effective Kohn-Sham field $a^{\mu}_{\rm{KS}}$. In the case that we use classical trajectories for the nuclei in our MPKS approach, we call the resulting simplification in analogy to matter-only quantum dynamics the Ehrenfest-Maxwell-Pauli-Kohn-Sham (EMPKS) approach.

\subsection{MPKS-Outlook}

As is obvious from the extent of the physics included in our considerations -- from matter-only quantum mechanics to quantum optics and beyond -- the unknown exact effective fields $a^{\mu}_{\rm{Mxc}}$ need to contain all knowledge of the correlated quantum dynamics. Already for electron-only quantum mechanics the quest for such exact expressions is a herculean effort. And this will not become simpler for the full matter-photon problem of electrons, nuclei and photons. Specifically challenging is that the description of the matter subsystem is based on the current density and not on the density as is the usual case in density-functional theory~\cite{gross2013,ullrich2011,engel2011}. Hence most approximations developed for the effective fields only cover the zero component and little is known besides linear-response kernels~\cite{vignale1996current,ullrich2002time} for the more advanced current-density-functional theory~\cite{vignale1988current, vignale2004mapping}. Therefore a necessary step to develop more accurate functionals beyond treating the zero component via density-functional approximations and the spatial components on the mean-field level includes also more advanced current-density functionals for electron-only theory. Of course we also need to account for new terms that are due to the transversal photon interaction and that are not covered by matter-only theories. It is helpful in this regard that for processes in free space where many photons are involved, e.g., when the system is perturbed by a weak laser pulse, the mean-field limit is very accurate~\cite{ruggenthaler2018quantum, flick2015kohn}. This changes if we consider situations where it is the fluctuations of the photon field that are important, such as changes in the ground-state of the combined light-matter system~\cite{flick2018ab}, and few-photon strong coupling situations~\cite{flick2018light}. But for such specific cases a slight simplification of the presented theory to dipole coupling~\cite{tokatly2013,Ruggenthaler_2014,ruggenthaler2015} becomes a sensible alternative. In this case it is only the density of the matter-subsystem that couples to the photon field~\cite{schaefer2018} and standard density-functional approximation schemes become applicable. The currently most sophisticated is an extension of the optimized-effective potential approach to matter-photon systems~\cite{pellegrini2015optimized}, which has been already applied to real systems~\cite{flick2018ab}. In this work an exact-exchange-type of approximation for the transversal and longitudinal matter-photon interactions is employed. This means for the novel transversal part that, e.g., leads to the Lamb shift of the matter states, a single-photon approximation is used. Thus multi-photon processes of the fluctuating photon field are not included. An interesting alternative to include all orders of field-fluctuations are trajectory-based methods. In this case instead of one Maxwell field many realization are propagated at the same time, which is akin to a field-quantization procedure in the Riemann-Silberstein formulation of electrodynamics~\cite{Bialynicki-Birula_1996,bialynicki2013role}. Furthermore, the present formulation allows to inculude many different species of particles. This suggests to model quantum dynamics in a complex environment by, e.g., treating certain particle types with a crude yet efficient orbital-free quantum Navier-Stokes description based directly on Eqs.~\eqref{eq:fluid1} and \eqref{eq:fluid2}, while the system of interest is treated with a more accurate Kohn-Sham description.


\section{Real-space implementation} \label{sec_implementation}

  After presenting the theoretical fundamentals, we
  introduce in the following sections our first implementation of the
  coupled EMPKS equations in real-time. Our implementation is based on
  finite-difference discretizations and real-space grid representations for
  both the matter wave functions and the electromagnetic fields. While not the
  only possible choice, this representation has the advantage to allow for a uniform and
  unbiased description of the combined light-matter system. More importantly,
  this choice also simplifies the description of coupling between matter and radiation,
  since also the QED couplings are prescribed in real space. Moreover,
  the real-space representation is suited for the multi-scale aspects of the coupling,
  and allows to reuse simulation techniques and algorithms for the matter and the radiation subsystems. 
  We have integrated our EMPKS implementation in the real-space real-time code
  Octopus \cite{Octopus_2015, doi:10.1002/pssb.200642067}, an open source simulation
  package for quantum-mechanical ab-initio calculations based on time-dependent
  density-functional theory.

  We start by going into more detail of the Maxwell's equations in
  Riemann-Silberstein representation in section
  \ref{subsec_Maxwell_equation_RS}.  In section
  \ref{subsec_maxwell_time_evolution_equation}, we introduce a time-evolution
  operator for the Riemann-Silberstein vector, while section
  \ref{subsec_maxwell_boundary_conditions} is focussing on different boundary
  conditions. We continue our discussion in section
  \ref{subsec_time_evolution_Kohn_Sham_orbitals} with details on the time
  evolution of the matter degrees of freedom before we turn our attention in
  sections \ref{subsec_multipole_expansion}, \ref{subsec_multi_scale_implementation}, and 
  \ref{subsec_predictor_corrector_scheme}
  on the coupling of radiation and matter which involves a multipole expansion for
  the coupling Hamiltonian, a multi-scale implementation, and a predictor-corrector scheme 
  for a self-consistent coupling of radiation and matter.
  We conclude the discussion of our implementation in
  section \ref{subsec_validation_and_comparison_to_FDTD} with a validation of
  the method and a comparison to FDTD methods.

  \subsection{Maxwell's equations in Riemann-Silberstein representation}
  \label{subsec_Maxwell_equation_RS}

  Most common electromagnetic simulations are based on the FDTD method, which
  was already introduced in 1966 by Yee \cite{Yee_1966,
  taflove2005computational}. FDTD uses two grids, shifted by half of the chosen
  grid spacing. One grid represents the electric field and the other one
  the magnetic field.  Based on this description, the time evolution in FDTD is
  given by two update equations one for each electromagnetic field.  In
  contrast, in the Octopus code several methods are implemented to approximate
  the quantum-mechanical time-evolution operator \cite{Castro2004} for the
  propagation of wave functions in real-time. By transforming Maxwell's
  equations into a six-component Riemann-Silberstein representation~\cite{MOHR_2010}, the
  underlying equation of motion for the Maxwell fields can be expressed as a
  Schr\"odinger-like equation. This fact gives us the opportunity to propagate
  also the electromagnetic field with quantum-mechanical time-evolution methods
  modified for Maxwell fields.  Besides the fact that we can immediately reuse
  with Octopus an existing time-evolution engine that is very efficient and
  highly parallelized, the reformulation of Maxwell's equations in terms of the
  six-dimensional Riemann-Silberstein vector also has the advantage over FDTD that higher order
  discretizations for the spatial derivatives can be used which in turn allow
  for much larger grid spacings and time steps. Furthermore, the very same grid
  can be used for the electric and for the magnetic field which simplifies the
  coupling to microscopic matter charge currents. Also, from our experience it
  allows for more stable time-stepping of coupled radiation and matter, and
  improves the maintainability of the implementation.

    In Eq.~(\ref{eq:RSvectors}) we have introduced already the complex
    Riemann-Silberstein vector which can also be written in the form 
    \begin{equation}
      \begin{alignedat}{2}
           \vec{F}_{\pm}(\vec{r},t)
        &= \sqrt{\frac{\epsilon_{0}}{2}} \vec{E}(\vec{r},t)
           \pm \mathrm{i} \sqrt{ \frac{1}{2 \mu_{0}} } \vec{B}(\vec{r},t)                  \label{eq_RS_vector_pm_2}.
      \end{alignedat}
    \end{equation} 
    As already stated before, the sign of the imaginary part of the
    Riemann-Silberstein vector corresponds to different helicities. To convert
    Maxwell's equations to Riemann-Silberstein form, we have to be able to keep
    track of superposition states of different helicities which correspond to a
    linear combination of $ \vec{F}_{+} $ and $ \vec{F}_{-} $. Hence, it is
    useful to combine both vectors in a six-dimensional vector
    \begin{align}
        \mathcal{F}(\vec{r},t)
      = \begin{pmatrix}
          \vec{F}_{+}(\vec{r},t)   \\
          \vec{F}_{-}(\vec{r},t) 
        \end{pmatrix}    \hspace{1mm} .                                                                      \label{eq_RS_vector_6_dim}
    \end{align}
    The inverse transformation for given Riemann Silberstein vectors $ \vec{F}_{+}(\vec{r},t) $ and
    $ \vec{F}_{-}(\vec{r},t)$ then takes simply the form

    \begin{align}
        \vec{E}(\vec{r},t)
      = \sqrt{ \frac{1}{2 \epsilon_{0}} } \left( \vec{F}_{+}(\vec{r},t) + \vec{F}_{-}(\vec{r},t) \right),        \label{eq_E_field_from_RS}
    \end{align}

    \begin{align}
        \vec{B}(\vec{r},t)
      = -\mi\sqrt{ \frac{\mu_{0}}{2} } \left( \vec{F}_{+}(\vec{r},t) - \vec{F}_{-}(\vec{r},t) \right),               \label{eq_B_field_from_RS}
    \end{align}

     which allows always to reconstruct the electric and magnetic fields from the
     Riemann-Silberstein vector.

    \subsubsection{Maxwell's equations in Schr\"odinger form}

    In this section, we generalize the Riemann-Silberstein description of the
    homogeneous Maxwell's equations 
    Eqs.~(\ref{eq_Mx_equation_Faraday_homogenous}-\ref{eq_Mx_equation_Gauss_B_homogeneous})
    to the inhomogeneous case.
    While here we consider classical external charge and current densities, later, for a fully microscopic description, we use
    the expectation values of the previously derived quantum-mechanical charge density Eq.~(\ref{eq_RN14})
    and current density Eq.~(\ref{eq_RN25}) which will lead to a modification of the usual Maxwell's equations.

    We start by considering the two
    Gauss' laws for the electric and the magnetic field
    \begin{align}
      \partial_{k} E^{k}(\vec{r},t)& = \frac{\rho(\vec{r},t)}{\epsilon_0},                              \label{eq_Mx_equation_Gauss_E_inhomogeneous}
      \\
      \partial_{k} B^{k}(\vec{r},t)&= 0,                                                     \label{eq_Mx_equation_Gauss_B_inhomogeneous}
    \end{align}
    where we introduced as the external charge density
    $\rho(\vec{r},t) = \tfrac{1}{c_0} j^{0}(\vec{r},t)$. Later we will then use also the internal
    charge density $n(\vec{r},t) = \tfrac{1}{c_{0}} J^{0}(\vec{r},t)$. 
    Both Gauss' laws can be combined into one equation 
    \begin{equation}
      \begin{alignedat}{2}
           \begin{pmatrix}
             \partial_{k} F_{+}^{k}(\vrt)   \\[2mm]
             \partial_{k'} F_{-}^{k'}(\vrt)
           \end{pmatrix}
         = \frac{1}{\sqrt{2 \epsilon_0}}
           \begin{pmatrix}
             \rho(\vec{r},t)   \\[2mm]
             \rho(\vec{r},t) 
           \end{pmatrix}    \honemm .                      \label{eq_RS_Gauss_6_dim}
      \end{alignedat}
    \end{equation}
    The remaining two Maxwell's equations, the Amp\`ere's and Faraday's law
    \begin{align}
      \partial_t E^{k}(\vec{r},t)
      &= c_0^2 \epsilon^{klm} \partial_{l} B_{m}(\vec{r},t) + c_0^2 \mu_0 j^{k}(\vec{r},t),        \label{eq_Mx_equation_Ampere_inhomogeneous}
      \\
      \partial_t B^{k}(\vec{r},t)
      &= \epsilon^{klm} \partial_{l} E_{m}(\vec{r},t),                                             \label{eq_Mx_equation_Faraday_inhomogenous}
    \end{align}
    can also be combined into one evolution equation for the Riemann-Silberstein
    vectors as
    \begin{equation}
      \begin{alignedat}{2}
          \mi \hbar
           \begin{pmatrix}
             \partial_{t} F_{+}^{k}(\vrt)        \\
             \partial_{t} F_{-}^{k'}(\vrt)
           \end{pmatrix}
        &= \hbar c_{0}
           \begin{pmatrix}
                      -  \epsilon^{klm} \partial_{l} F_{+,m}(\vrt)              \\
             \phantom{-} \epsilon^{k'l'm'} \partial_{l'} F_{-,m'}(\vrt)
           \end{pmatrix}                                                                   \\
        & \quad - \frac{\mi \hbar}{\sqrt{2 \epsilon_{0}}}
           \begin{pmatrix}
             j^{k}(\vrt)   \\
             j^{k'}(\vrt)
           \end{pmatrix}.                                                                   \label{eq_RS_Ampere_Faraday_curl}
      \end{alignedat}
    \end{equation}
    The first term on the right-hand side of
    Eq.~(\ref{eq_RS_Ampere_Faraday_curl}) describes the curl operation and can be
    represented by the already previously introduced spin-1 matrices for the
    photons in Eq. (\ref{eq_RN4}). Taking them into account, Amp\`ere's and
    Faraday's laws in Riemann-Silberstein form can be written as 
    \begin{equation}
      \begin{alignedat}{2}
          \mi \hbar \hspace{-1.2mm}
           \begin{pmatrix}
             \partial_{t} \vec{F}_{+}(\vrt)                \\[2mm]
             \partial_{t} \vec{F}_{-}(\vrt)
           \end{pmatrix}
        &= \mi \hbar c_{0} \hspace{-1.3mm}
           \begin{pmatrix}
             \hspace{-0.5mm} \phantom{-} S^{l} \partial_{l}    \vec{F}_{+}(\vrt)      \\[2mm]
             \hspace{-0.5mm} - S^{l'} \partial_{l'}  \vec{F}_{-}(\vrt)
           \end{pmatrix}                                                                           \\[2mm]
        & \quad - \frac{\mi \hbar}{\sqrt{2 \epsilon_{0}}} \hspace{-1mm}
           \begin{pmatrix}
             \vec{j}(\vrt)             \\[2mm]
             \vec{j}(\vrt)
           \end{pmatrix}   \honemm .                                                               \label{eq_RS_Ampere_Faraday_spin_matrices}
      \end{alignedat}
    \end{equation}
    This equation has the form of an inhomogeneous Schr{\"o}dinger equation with
    single particle "photon" Hamiltonian
    \begin{align}
       H_{\mathrm{Mx}} = - \mi \hbar c_{0}   S^{l} \partial_{l}  .
       \label{eq_single_particle_photon_Hamiltonian}
    \end{align}
    As consequence, Maxwell's equations in Riemann-Silberstein form can be interpreted as the first quantized
    wave equation for a single photon in real-space. Nevertheless, we are dealing
    still with a classical equation, since $\hbar$ can be cancelled in all terms 
    in Eq. (\ref{eq_RS_Ampere_Faraday_spin_matrices}).

    \subsubsection{Current densities and integral kernels}

    As already noted, to reach a microscopic description of the electromagnetic fields, 
    the current and charge densities that
    appear in our Maxwell's equations in Riemann-Silberstein form have to
    correspond to the multi-species currents that we introduced in Eq.
    (\ref{eq_RN25}). The current densities consist of three terms, the
    paramagnetic current term $J_{\mathrm{pmc}}^{k} $, the diamagnetic current
    term $J_{\mathrm{dmc}}^{k} $, and the magnetization current term $
    J_{\mathrm{mc}}^{k} $. These contributions to the total current can be expressed
    in terms of $ I_{(n)} $ auxiliary one-body Kohn-Sham orbitals $
    \phi_{(n,i)}(\vec{r}s_{(n)},t) $,  and the Kohn-Sham densities $
    n_{(n,i)}(\vec{r}s_{(n)},t) = |\phi_{(n,i)}(\vec{r}s_{(n)},t)|^2$ for the corresponding
    species $ n $, which gives rise to the charge density
    $n(\vec{r},t) = \tfrac{1}{c_0}J^{0}(\vec{r},t)
    = \sum_{n=1}^{N} q_{(n)}\sum_{i,s_{(n)}}^{I_{(n)}} n_{(n,i)}(\vec{r}s_{(n)},t)$ in the form

    \begin{equation}
      \begin{alignedat}{2}
          \hspace{1mm}
          J_{\mathrm{pmc}}^{k}(\vec{r},t) &=                                                                \\
        & \hspace{-7mm} \sum \limits_{n=1}^{N} \hspace{-1mm}
          \frac{\hbar q_{(n)}}{\mathrm{i} 2 M_{(n)}} \hspace{-1mm}
          \sum \limits_{i,s_{(n)}}^{I_{(n)}} \hspace{-0.5mm}
          \bigg[ \hspace{-0.5mm}
            \Big( 
              \hspace{-0.5mm} \partial^{k} \phi_{(n,i)}^{\dagger}(\vec{r} s_{(n)},t) \hspace{-0.5mm}
            \Big) \hspace{0mm} \phi_{(n,i)}(\vec{r} s_{(n)},t)                                              \\
        & \hspace{-5mm}  \hspace{20mm} - \hspace{-0.5mm} \phi_{(n,i)}^{\dagger}(\vec{r} s_{(n)},t) \hspace{0mm}
            \Big( \hspace{-0.5mm} \partial^k \phi_{(n,i)}(\vec{r} s_{(n)},t) \hspace{-0.5mm} \Big)
          \bigg]   \honemm ,                                                                                \label{eq_paramagnetic_current}
      \end{alignedat}
    \end{equation}
   
    \begin{equation}
      \begin{alignedat}{2}
          \hspace{1mm}
          J_{\mathrm{dmc}}^{k}(\vec{r},t) &=                                                                 \\
        & \hspace{-1mm} - \sum \limits_{n=1}^{N} \frac{q_{(n)}^2}{M_{(n)} c_{0}}
          \left[ 
            \sum \limits_{i,s_{(n)}}^{I_{(n)}} n_{(n,i)}(\vec{r} s_{(n)},t)
          \right] a_{\mathrm{KS}}^{k}(\vrt)
      \end{alignedat}   \honemm,                                                                             \label{eq_diamagnetic_current}
    \end{equation}

    \begin{equation}
      \begin{alignedat}{2}
          \hspace{1mm}
          J_{\mathrm{mc}}^{k}(\vec{r},t) &=                                                                  \\
        & \hspace{-13mm} - \hspace{-1mm} \sum \limits_{n=1}^{N} \hspace{-1mm}
          \left[ \sum \limits_{i,s_{(n)}}^{I_{(n)}} \hspace{-2mm}
            \epsilon^{klm} \partial_l \phi_{(n,i)}^{\dagger} \hspace{-0.5mm} (\vec{r} s_{(n)},t) \hspace{-1mm}
            \left(
              \frac{ q_{(n)} \hbar}{2 M_{(n)}} S_{m}^{(n)}
            \right) \hspace{-1mm} \phi_{(n,i)} \hspace{-0.5mm} (\vec{r} s_{(n)},t)
          \right]   \honemm .                                                                                \label{eq_magnetization_current}
      \end{alignedat}
    \end{equation}
    where the summations go over all
    spin-states $s_{(n)}$, Kohn-Sham orbitals $i$ and species types $n$.
    All three current terms as well as the total charge density depend
    on the particle charge $ q_{(n)} $, the particle mass $ M_{(n)} $, and the
    single-particle wave functions.  While the paramagnetic current and magnetization
    current do not depend explicitly on the Maxwell fields, the diamagnetic
    current depends on the vector potential $ a_{\mathrm{KS}}^{k} = a^{k} + a^{k}_{\rm{xc}} + A^{k}$, which is
    implicitly determined by the Riemann-Silberstein vectors $ F_{\pm}^{k} $.
    The mean-field vector potential $ A^{k} $ can
    be expressed in terms of the magnetic field via a Poisson equation for each
    magnetic field vector component as
    \begin{equation}
      \begin{alignedat}{2}
           A^{k}(\vec{r},t)
        &= - c_0
             \int \mathrm{d}^3 r' 
             \frac{ \epsilon^{klm} \partial'_{l} B_{m}(\vec{r} \hspace{0.5mm}' \hspace{-1mm} ,t) }{ 4 \pi | \vec{r} - \vec{r}\hspace{0.5mm}' | }    \\
        & \hspace{-7mm} = \mathrm{i} \sqrt{\frac{c_0^2\mu_0}{2}} 
           \hspace{-1mm}
           \underbrace{
             \int \hspace{-1mm} \mathrm{d}^3 r' \frac{\epsilon^{klm} \partial'_{l}}{4 \pi |\vec{r} - \vec{r}\hspace{0.5mm}'|} \hspace{-0.8mm}
             \left(
               F_{+}^{k}(\vec{r} \hspace{0.5mm}' \hspace{-1mm} ,t)
               - F_{-}^{k}(\vec{r} \hspace{0.5mm}' \hspace{-1mm} ,t)
             \right)
           }_{\mathrm{Poisson \hspace{1mm} equation}}   \honemm .
      \end{alignedat}                                                                              \label{eq_vector_potential_via_spatial_integral}
    \end{equation}
    But we can also use the transversal part of the electric field such that the vector potential
     $ A^{k} $ results from an integral
    over time starting at the initial time $ t = t_0 $ and ending at the
    current time $ t $
    \begin{equation}
      \begin{alignedat}{2}
           A^{k}(\vec{r},t) &=                                                                              \\
        & \hspace{-5mm} - \sqrt{ \frac{c_0^2}{2 \epsilon_0} } \int \limits_{t_0}^{t} \md t'
           \left(
             F^{+}_{m}(\vec{r} ,t')
             + F^{-}_{m}(\vec{r} ,t')
           \right)_{\perp} + A^{k}(\vec{r},t_0)  \honemm .
      \end{alignedat}                                                                                        \label{eq_vector_potential_via_time_integral}
    \end{equation}
    Here $\perp$ again indicates that only the transversal degrees are to be considered.
    This can be made explicit with the help of the Helmholtz decomposition, as also shown 
    in Eq.~\eqref{eq_RN26}. Also, the initial value $A^{k}(\vec{r},t_0)$ is in principle determined by the initial wave function of the interacting system. Alternatively, provided the external field obeys itself the homogeneous Maxwell's equation $(\partial_0^2 + \partial_l \partial^{l})a^{k} =0$, we can also use instead the total field $A^{k}_{\rm{tot}} = a^{k} + A^{k}$ in the above considerations. This is the standard case and in the implementation we usually work directly with the total field and the corresponding Riemann-Silberstein vector. Since the diamagnetic current term is obtained by one of the two
    different equivalent vector potential expressions from above, the charge current $
    J_{\mathrm{dmc}}^{k} $ carries non-local
    contributions, which originate from the electromagnetic fields. Here we make the distinction between an internal part that depends directly on $A^{k}$ and an external part of the diamagnetic current that comes from the external field $a^{k}$ and from the missing exchange-correlation forces $a^{k}_{\rm{xc}}$. Furthermore, as the
    internal diamagnetic current depends explicitly on the Riemann-Silberstein vectors, we can
    combine it with the curl operation of
    Eq.~(\ref{eq_RS_Ampere_Faraday_spin_matrices}) to define a single integral
    operator. The Hamiltonian-like integral operator
    $ \bm{\hat{\mathcal{H}}} $ then acts on the Riemann-Silberstein vector,
    which in the following we denote by
    \begin{align}
        \bm{\hat{\mathcal{H}}}(\vec{r},t) \bm{\mathcal{F}}(\vec{r},t) 
      := \int  \md^3 r' \int \limits_{t_0}^{t} \md t' 
        \hat{\bar{\mathcal{H}}}(\vec{r}, \vec{r} \hspace{0.5mm}' \hspace{-1mm}, t, t') 
        \mathcal{F}(\vec{r} \hspace{0.5mm}' \hspace{-1mm}, t')   \honemm .                                   \label{eq_RS_Hamiltonian_operation}
    \end{align}

    The operator $ \bm{\hat{\mathcal{H}}} $ is determined by its corresponding integral kernel $ \hat{\bar{\mathcal{H}}} $ and here given by 
   
    \begin{align}
        \hat{\bar{\mathcal{H}}}(\vec{r},\vec{r} \hspace{0.5mm}' \hspace{-1mm},t,t')
      = \hat{\bar{\mathcal{H}}}_{(0)}(\vec{r},\vec{r} \hspace{0.5mm}' \hspace{-1mm},t,t')
        + \hat{\bar{\mathcal{K}}}(\vec{r},\vec{r} \hspace{0.5mm}' \hspace{-1mm}, t,t')   \honemm .       \label{eq_RS_Hamiltonian}
    \end{align}

    The matrix $ \hat{\bar{\mathcal{H}}}_{(0)} $ represents the curl operation with the spin-1 matrices in equation (\ref{eq_RS_Ampere_Faraday_spin_matrices})
    without any diamagnetic current 
   
    \begin{equation}
      \begin{alignedat}{2}
          \hat{\bar{\mathcal{H}}}_{(0)}(\vec{r}, \vec{r} \hspace{0.5mm}' \hspace{-1mm}, t, t') =                 \\[2mm]
        & \hspace{-15mm}
          \begin{pmatrix}
            \begin{array}{rr}
              \hspace{-1mm} 1 & \hspace{1mm}          0 \hspace{-1mm} \\
              \hspace{-1mm} 0 & \hspace{1mm} \text{-} 1 \hspace{-1mm}
           \end{array}
          \end{pmatrix}_{\hspace{-1mm} 2 \times 2}
          \hspace{-1mm} \otimes \hspace{-1mm}
          \bigg(
            \hspace{-1mm} - \mathrm{i} \hbar c_0 
            \delta(\vec{r} \hspace{-0.5mm} - \hspace{-0.5mm} \vec{r} \hspace{0.5mm}' ) 
            \delta(t \hspace{-0.5mm} - \hspace{-0.5mm} t') \hspace{-0.3mm}
             S^{m} \partial_{m} \hspace{-2mm}' \hspace{1.0mm}
          \bigg)_{\hspace{-1mm} 3 \times 3}   \honemm .                                                      \label{eq_RS_Hamiltonian_0_kernel}
      \end{alignedat}
    \end{equation}

    Here and in the following, we describe the 6x6 matrices which act on the
    six-dimensional Riemann-Silberstein state $ \mathcal{F} $ as a Kronecker
    product of a 2x2 and a 3x3 matrix. The first 2x2 matrix which we call
    "coupling" matrix  shows whether the two Riemann-Silberstein vectors $
    \vec{F}_{\pm} $ couple to each other. If the off-diagonal of this 2x2 matrix is zero, the
    resulting 6x6 matrix does not couple the two vectors. 
    Furthermore, the second 3x3 kernel
    matrix, contains all necessary physical variables and operations to satisfy
    the underlying Maxwell's equations.  Due to the delta functions in $
    \hat{\bar{\mathcal{H}}}_{(0)}$, the application of $\bm{\hat{\mathcal{H}}}_{(0)}$
    then results in a local linear operator acting on the Riemann-Silberstein vector, i.e.,
    \begin{align}
        \bm{\hat{\mathcal{H}}}_{(0)}(\vec{r},t) \bm{\mathcal{F}}(\vec{r},t) 
      = \hat{\mathcal{H}}_{(0)} \mathcal{F}(\vrt)   \honemm ,  
    \end{align}
    where
    \begin{align}
        \hat{\mathcal{H}}_{(0)}
      = \Bigg[
          \begin{pmatrix}
            \begin{array}{rr}
              \hspace{-1mm} 1 & \hspace{1mm} 0 \hspace{-1mm} \\ \hspace{-1mm} 0 & \hspace{1mm} \text{-} 1 \hspace{-1mm}
            \end{array}
          \end{pmatrix}_{\hspace{-1mm} 2 \times 2}
          \hspace{-1mm} \otimes \hspace{-1mm}
          \bigg(
            \hspace{-1mm} - \mathrm{i} \hbar c_0 
             S^{m} \partial_{m}
          \bigg)_{\hspace{-1mm} 3 \times 3}
        \Bigg]    \honemm .                                                                                                      \label{eq_RS_Hamiltonian_0}
    \end{align}
    In contrast to this, the integral kernel originating from the diamagnetic current contribution
    in equation (\ref{eq_RS_Hamiltonian}) carries a non-locality in space due to
    \begin{equation}
      \begin{alignedat}{2}
          \hat{\bar{\mathcal{K}}}(\vec{r},\vec{r} \hspace{0.5mm}' \hspace{-1mm},t,t') =                                                \\
        & \hspace{-20mm}
          \begin{pmatrix}
             \begin{array}{rr}
               \hspace{-1mm} 1 & \honemm \text{-} 1 \hspace{-1mm} \\[2mm] \hspace{-1mm} 1 & \honemm \text{-} 1 \hspace{-1mm}
             \end{array}
           \end{pmatrix}_{\hspace{-1mm}2 \times 2} \hspace{-2.5mm}
           \otimes \hspace{-1mm}
           \left(
             \hspace{-1mm} - \hbar \sqrt{\frac{\mu_0}{2}} \delta(t-t') \kappa (\vrt)
               \frac{S^{m} \partial_{m}'}{4 \pi |\vec{r}-\vec{r} \hspace{0.5mm}'|}
           \right)_{\hspace{-1mm} 3 \times 3}   .                                                   \label{eq_RS_Hamiltonian_dia_current_nls}
      \end{alignedat}
    \end{equation}
    The integrals over the kernel have to be kept explicitly. Note, that the 2x2 coupling matrix
    contains off-diagonal entries in this case.
    The prefactor $ \kappa$ is given by
    \begin{equation}
      \begin{alignedat}{2}
          \kappa(\vrt)
        = \sum \limits_{n=1}^{N} \frac{q_{(n)}^2}{M_{(n)}}
          \left[ \sum \limits_{i=1,s_{(n)}}^{I_{(n)}} n_{(n,i)}(\vec{r} s_{(n)},t) \right]   \honemm ,
      \end{alignedat}
    \end{equation}
%
    %
    This non-local term that arises due to the microscopic treatment of
    the matter system therefore leads already for a classical treatment of the
    photon field to effective "photon-photon" interactions and changes the usual
    Maxwell's equation. This is discussed in more detail in~\cite{flick2018light}. The remaining current densities in Eqs.~(\ref{eq_paramagnetic_current})
    and (\ref{eq_magnetization_current}) as well as the external diamagnetic currents can be added to the
    external current such that we have a inhomogeneous current
    $ J_{\mathrm{inh}}^{k} = J^k_{\mathrm{pmc}} + J^k_{\mathrm{mc}} - \sum_{n} \tfrac{q^2_{(n)}}{M_{(n)c_0}}\bigl[\sum_{i,s_{(n)}} n_{(n,i)}\bigr](a^{k} + a^{k}_{\rm{xc}})+ j^{k}$,
    which constitutes a six-dimensional vector for the Riemann-Silberstein Maxwell's
    equation given by
   
    \begin{equation}
      \begin{alignedat}{2}
           \mathcal{J}(\vec{r},t)
        &= \begin{pmatrix} 1 \\[1mm] 1 \end{pmatrix}_{2 \times 1} \hspace{-4mm} \otimes
           \left(
           \frac{1}{\sqrt{2 \epsilon_{0}}}
             \vec{J}_{\mathrm{inh}}(\vrt)
           \right)_{3 \times 1}.
      \end{alignedat}                                                                                                    \label{eq_RS_current}
    \end{equation}
    Combining the previous considerations, allows us finally to formulate
    Amp\`ere's and Faraday's law in Riemann-Silberstein representation 
   
    \begin{align}
        \mathrm{i} \hbar \frac{\partial}{\partial t} \mathcal{F}(\vec{r},t)
      = \bm{\hat{\mathcal{H}}}(\vec{r},t)
        \bm{\mathcal{F}}(\vec{r},t) 
        - \mathrm{i} \hbar \mathcal{J}(\vec{r},t)   \honemm .                                                            \label{eq_RS_Faraday_Ampere_schroedinger}
    \end{align}
    The Maxwell fields that follow this equation are due to
    microscopic quantum-mechanical currents from Eq.~(\ref{eq_RS_current}). We
    call these fields in the following \textit{microscopic} Maxwell fields. We can,
    however, also include other types of contributions in our implementation.
    In the following Sec.\ref{subsec:linmedium}, we will discuss how we
    can in a similar manner include also \textit{macroscopic} Maxwell fields
    due to, e.g., lenses or surfaces. Further note that in contrast to the
    quantum mechanical second-order derivative operator for the kinetic energy, the
    integral kernel $ \hat{\mathcal{H}} $ contains only first-order derivatives.
    Without inhomogeneity this reflects the linear dispersion relation for
    photons. On the other hand, including the inhomogeneity introduces a
    non-linear dispersion for the photons.
   
    In the discussion so far, we have seen from the matrix operator $ \hat{\mathcal{H}} $ in
    Eq.~(\ref{eq_RS_Hamiltonian}) and its underlying matrix operators $
    \hat{\mathcal{H}}_{(0)} $, and $ \hat{\mathcal{K}} $ in
    Eqs.~(\ref{eq_RS_Hamiltonian_0}) and
    (\ref{eq_RS_Hamiltonian_dia_current_nls}) that the
    two different Riemann-Silberstein vectors $ F_{+}^{k} $ and $ F_{-}^{k} $
    only couple in the presence of a diamagnetic current $ J_{\mathrm{dmc}}^{k}
    $.  Without this contribution, the combined Amp\`ere's and
    Faraday's law in Riemann-Silberstein form in
    (\ref{eq_RS_Faraday_Ampere_schroedinger}) decouple into two
    three-component equations
   
    \begin{align}
         \mathrm{i} \hbar \frac{\partial}{\partial t} F_{+}^{k}(\vrt)
      &= c_{0} \hspace{-0.5mm} \big[ S^{l} \partial_{l} \big]^{\hspace{-0.5mm}k}_{m} \hspace{-0.5mm} F_{+}^m(\vrt)
         \hspace{-0.5mm} - \hspace{-0.5mm} \frac{\mi \hbar}{\sqrt{2 \epsilon_{0}}} J_{\mathrm{inh}}^{k}(\vrt)
    \end{align}

    for positive helicity fields $ F_{+}^{k} $, and
   
    \begin{align}
         \mathrm{i} \hbar \frac{\partial}{\partial t} F_{-}^{k}(\vrt)
      &= - c_{0} \hspace{-0.5mm} \big[ S^{l} \partial_{l} \big]^{\hspace{-0.5mm}k}_{m} \hspace{-0.5mm} F_{-}^m(\vrt)
         \hspace{-0.5mm} - \hspace{-0.5mm} \frac{\mi \hbar}{\sqrt{2 \epsilon_{0}}} J_{\mathrm{inh}}^{k}(\vrt)
    \end{align}
   
    for negative helicity fields $ F_{-}^{k} $.

    \subsubsection{Helicity coupling in a linear medium}
    \label{subsec:linmedium}

    So far, we have only focused on the microscopic Maxwell's equations and
    have obtained an explicit coupling between $ F_{+}^{k} $ and $ F_{-}^{k} $ caused by
    the diamagnetic current. The formulation in terms of integral kernels,
    however, allows also to seamlessly combine a microscopic propagation with
    macroscopic electrodynamics for linear media. This has the merit that
    macroscopic optical elements like lenses or mirrors can easily be
    incorporated in the Riemann-Silberstein propagation. In the following, we
    briefly discuss this relation.\\ The coupling matrix $ \mathcal{K} $ bears a
    direct similarity to the Riemann-Silberstein Maxwell's equation in a linear medium
    \cite{Sameen_Ahmed_Khan_2005,Bialynicki-Birula_1996}, where the
    electromagnetic fields are described by the electric displacement field $
    \vec{D}(\vrt) $ and the magnetic $ \vec{H}(\vrt) $ field. In linear optics,
    it is assumed that $ \vec{D} $ and $ \vec{B} $ depend approximately
    linearly on $ \vec{E} $ and $ \vec{H} $
    \begin{align}
      \vec{D}(\vrt) = \epsilon(\vrt) \vec{E}(\vrt) \htwomm , \quad
      \vec{B}(\vrt) = \mu(\vrt) \vec{H}(\vrt)  \honemm.
    \end{align}
    Here, $ \epsilon(\vrt) $ and $ \mu(\vrt) $ denote the electric permittivity
    and magnetic permeability in the medium which in general depend on space
    and time.  Referring to the Riemann-Silberstein definition in vacuum in
    Eq.~(\ref{eq_RS_vector_pm}) with constant vacuum electric permittivity and
    magnetic permeability, the Riemann-Silberstein vector in a linear medium $
    \vec{F}_{\pm,\mathrm{lm}} $ has to be adapted according to
    \begin{align}
        F_{\pm,\mathrm{lm}}^{k}(\vrt)
      = \sqrt{\frac{\epsilon(\vrt)}{2}} E^{k}(\vrt)
        \pm \mi \sqrt{\frac{1}{2 \mu(\vrt)}} B^{k}(\vrt)   \honemm .           \label{eq_RS_vector_medium_pm}
    \end{align}
    Consequently, the speed of light in the medium also depends now on space and time with
    \begin{align}
      c(\vrt) = \frac{1}{\sqrt{\epsilon(\vrt) \mu(\vrt)}}   \honemm .
    \end{align}
    Based on this new definition for the Riemann-Silberstein vectors in
    Eq.~(\ref{eq_RS_vector_medium_pm}), Amp\`ere's and Faraday's law for a linear
    medium \cite{jackson1998classical} are then given by
    \cite{Sameen_Ahmed_Khan_2005,Bialynicki-Birula_1996}
    \begin{equation}
      \begin{alignedat}{2}
          \mathrm{i} \hbar \frac{\partial}{\partial t} \mathcal{F}(\vec{r},t)
        = \hat{\mathcal{H}}_{\mathrm{lm}}(\vec{r},t)
          \mathcal{F}(\vec{r},t)
          - \mathrm{i} \hbar \mathcal{J}_{\mathrm{lm}}(\vec{r},t),
      \end{alignedat}                                                                              \label{eq_RS_Faraday_Ampere_schroedinger_medium}
    \end{equation}
    where the Hamiltonian $ \hat{\mathcal{H}}_{\mathrm{lm}} $ is given by
    \begin{align}
        \hat{\mathcal{H}}_{\mathrm{lm}}(\vec{r}, t)
      = \hat{\mathcal{H}}_{\mathrm{lm},(0)}(\vec{r}, t) 
        + \hat{\mathcal{K}}_{\mathrm{lm}}(\vec{r}, t)   \honemm .
    \end{align}
    The first term describes the free evolution in the linear medium 
    \begin{equation}
      \begin{alignedat}{2}
          \hat{\mathcal{H}}_{\mathrm{lm},(0)}(\vec{r}, t) =
          \begin{pmatrix}
            \begin{array}{rr}
              \hspace{-1mm} 1 & \hspace{1mm} 0 \hspace{-1mm} \\ \hspace{-1mm} 0 & \hspace{1mm} \text{-} 1 \hspace{-1mm}
           \end{array}
          \end{pmatrix}_{\hspace{-1mm} 2 \times 2}
          \hspace{-1mm} \otimes \hspace{-1mm}
          \bigg(
            \hspace{-1mm} - \mathrm{i} \hbar c(\vrt) 
             S^{m} \partial_{m}  \hspace{-1mm}
          \bigg)_{\hspace{-1mm} 3 \times 3}   \honemm .                                            \label{eq_RS_Hamiltonian_medium_0}
      \end{alignedat}
    \end{equation}
    Note that, due to $c(\vrt)$ this term contains now an explicit spatial and
    temporal dependence. The remaining term $ \hat{\mathcal{K}}_{\mathrm{lm}} $
    describes the properties of the linear medium and takes the form \cite{Bialynicki-Birula_1996,Sameen_Ahmed_Khan_2005}
    \begin{equation}
      \begin{alignedat}{2}
           \hat{\mathcal{K}}_{\mathrm{lm}}(\vec{r}, t) =                                                                               \\[2mm]
        &  \hspace{-10mm} \hspace{4mm}
           \hspace{-1mm} 
           \begin{pmatrix}
             \begin{array}{rr}
                \hspace{-1mm} \text{-} 1 & \text{-} 1 \hspace{-1mm} \\ \hspace{-1mm} 1 & 1 \hspace{-1mm}
             \end{array}
           \end{pmatrix}_{\hspace{-1mm} 2 \times 2}
           \hspace{-3.5mm} \otimes \hspace{-0.5mm} 
           \bigg( \hspace{-1mm}
             - \mi \frac{ \hbar c(\vec{r}, t) }{4 \epsilon(\vec{r}, t)}
             S^{m} \big( \partial_{m} \epsilon(\vec{r}, t) \big)
           \bigg)_{\hspace{-1mm} 3 \times 3}                                                                                     \\[2mm]
        &  \hspace{-10mm} + 
           \hspace{-1mm}
           \begin{pmatrix}
             \begin{array}{rr}
               \hspace{-1mm} \text{-} 1 & \hspace{1mm} 1 \hspace{-1mm} \\ \hspace{-1mm} \text{-} 1 & \hspace{1mm} 1 \hspace{-1mm}
             \end{array}
           \end{pmatrix}_{\hspace{-1mm} 2 \times 2}
           \hspace{-3.5mm} \otimes \hspace{-0.5mm} 
           \bigg( \hspace{-1mm}
             - \mi \frac{\hbar c(\vec{r}, t)}{4 \mu(\vec{r}, t)}
             S^{m} \big( \partial_{m} \mu(\vec{r}, t) \big) 
           \bigg)_{\hspace{-1mm} 3 \times 3}                                                                                     \\[2mm]
        &  \hspace{-10mm} + 
           \hspace{-1mm}
           \begin{pmatrix}
             \begin{array}{rr}
               \hspace{-1mm} \text{-} 1 & \text{-} 1 \hspace{-1mm} \\ \hspace{-1mm} \text{-} 1 & \text{-} 1 \hspace{-1mm}
             \end{array}
           \end{pmatrix}_{\hspace{-1mm} 2 \times 2}
           \hspace{-3.5mm} \otimes \mi \hbar
           \frac{\dot{\epsilon}(\vec{r}, t)}{4 \epsilon(\vec{r}, t)}
           \mathlarger{{\mathbb{1}}}_{3 \times 3}                                                                                \\[2mm]
        &  \hspace{-10mm} +
           \hspace{-1mm}
           \begin{pmatrix}
             \begin{array}{rr}
               \hspace{-1mm} \text{-} 1 & 1 \hspace{-1mm} \\ \hspace{-1mm} 1 & \text{-} 1 \hspace{-1mm}
             \end{array}
           \end{pmatrix}_{\hspace{-1mm} 2 \times 2}
           \hspace{-3.5mm} \otimes \mi \hbar
           \frac{\dot{\mu}(\vec{r}, t)}{4 \mu(\vec{r}, t)}
           \mathlarger{{\mathbb{1}}}_{3 \times 3} \hspace{1mm} , 
      \end{alignedat}                                                                              \label{eq_RS_Hamiltonian_coupling_medium}
    \end{equation}
    where $ \dot{\epsilon}(\vrt) $ and $ \dot{\mu}(\vrt) $ represent the total
    time derivative of $ \epsilon(\vrt) $ and $ \mu(\vrt) $.  From the 2x2
    coupling matrices contained in  $ \hat{\mathcal{K}}_{\mathrm{lm}} $ it becomes
    evident that the medium is capable to couple $ F_{+}^{k} $ with $ F_{-}^{k} $, as
    expected from linear optics.\\ 
    Based on the present formulation, it is instructive to further analyze the relation of
    the coupling that is induced by the diamagnetic contribution of the current
    density and the coupling that appears in linear and non-linear media. This
    provides a route to directly connect microscopic electrodynamics with a macroscopic
    formulation and will be considered in an upcoming publication. 

  \subsection{Time-evolution operator for the Riemann-Silberstein vector}  \label{subsec_maxwell_time_evolution_equation}
    
    In the previous section, we have illustrated how to express Amp\`ere's law and Faraday's law
    as an inhomogeneous Schr\"odinger-like equation in Riemann-Silberstein form. In the following, we
    construct the time-evolution operators for this equation in the homogeneous and inhomogeneous case.
    We start with the homogeneous case, i.e.,
    \begin{align}
        \frac{\partial}{\partial t} \mathcal{F}(\vec{r}, t)
      = - \frac{\mi}{\hbar} \bm{\hat{\mathcal{H}}}(\vec{r},t) \bm{\mathcal{F}}(\vec{r},t) \honemm .        \label{eq_RS_time_derivative_homogeneous}
    \end{align}
    The goal is to find an expression for a time-evolution operator that is
    acting on the Riemann-Silberstein vector $\mathcal{F}(\vec{r},t)$ in
    analogy to the time-evolution operator used in quantum mechanics for matter
    wave functions. Since the right hand-side of
    Eq.~(\ref{eq_RS_time_derivative_homogeneous}) contains in general an integral operator
    with integral kernel $ \hat{\mathcal{H}}$, we use as ansatz for the
    time-evolution operator $ \bm{\hat{\mathcal{U}}}(t,t_0) $ also an
    integral operator. This operator should obey the evolution equation
    \begin{align}
        \frac{\partial}{\partial t} 
        \bm{\hat{\mathcal{U}}}(t, t_0)
        \bm{\mathcal{F}}(\vec{r}, t_0)
      = - \frac{\mi}{\hbar} 
        \bm{\hat{\mathcal{H}}}(\vec{r},t) 
        \bm{\hat{\mathcal{U}}}(t, t_0) 
        \bm{\mathcal{F}}(\vec{r}, t_0) \honemm ,                     \label{eq_RS_time_derivative_U_homogeneous}
    \end{align}
    with the following boundary condition and group composition laws
    \begin{equation}
      \begin{alignedat}{4}
        & 1.) \hspace{5mm} && \bm{\hat{\mathcal{U}}}^{\dagger} \bm{\hat{\mathcal{U}}}           && = && \hspace{5mm} \mathbb{1}                  \honemm , \\
        & 2.) \hspace{5mm} && \hspace{-1mm} \lim\limits_{t \to t_0} \bm{\hat{\mathcal{U}}}(t,t_0) && = && \hspace{5mm} \mathbb{1}                  \honemm , \\
        & 3.) \hspace{5mm} && \bm{\hat{\mathcal{U}}}(t_2,t_0) 
                           && = && \hspace{5mm} \bm{\hat{\mathcal{U}}}(t_2,t_1) \bm{\hat{\mathcal{U}}}(t_1,t_0)                                  \honemm .
      \end{alignedat}                                                                                        \label{eq_RS_time_evolution_properties}
    \end{equation}
    
    From Eqs.~\eqref{eq_RS_time_derivative_U_homogeneous} we then find the implicit definition
    \begin{equation}
      \begin{alignedat}{2}
          \hspace{-5mm} \bm{\hat{\mathcal{U}}}(t,t_0) \bm{\mathcal{F}}(\vec{r},t_0)  - \mathcal{F}(\vec{r}, t_0) =                                                              \\
        & \hspace{-35mm} - \frac{\mi}{\hbar} \int \limits_{t_0}^{t} \hspace{-1mm} \md t'
          \bm{\hat{\mathcal{H}}}(\vec{r},t')
          \bm{\hat{\mathcal{U}}}(t',t_0) \bm{\mathcal{F}}(\vec{r} \hspace{-1mm}, t_0)  \honemm ,
      \end{alignedat}
    \end{equation}
    which fulfills the conditions in Eq.~(\ref{eq_RS_time_evolution_properties}).
    Iterating this equation allows to construct a series expansion for the time
    evolution-operator 
    %
    \begin{equation}
      \begin{alignedat}{2}
          \hspace{-1mm} \bm{\hat{\mathcal{U}}}(t,t_0) \bm{\mathcal{F}}(\vec{r}, & t_0) = \mathcal{F}(\vec{r},t_0)               \\
        & \hspace{-23mm} + \hspace{-1mm} \sum_{k=1}^{\infty} \hspace{-0.5mm} \frac{1}{k!} \hspace{-0.5mm}
          \bigg( \hspace{-1mm} \text{-} \frac{\mi}{\hbar} \bigg)^{\hspace{-1mm}k} \hspace{-1.5mm}
          \hspace{0.5mm} \mathcal{\hat{T}}
          \prod\limits_{p=1}^{k} \hspace{1mm}
          \hspace{-1mm} \int\limits_{t_0}^{t} \hspace{-1.5mm} \md \tau_{p}
          \hspace{-1.5mm} \int\limits \hspace{-1.5mm} \md^3 r_{p}
          \hspace{-1.5mm} \int\limits_{t_0}^{\tau_p} \hspace{-1.5mm} \md t_{p}
          \hat{\bar{\mathcal{H}}}(\vec{r}_{p-1} \hspace{-0.5mm}, \vec{r}_{p}, t_p, t_0) 
          \mathcal{F}(\vec{r}_{p}, t_{0})   ,
      \end{alignedat}                                                                              \label{eq_RS_time_evolution_homogen_diamagnetic_current}
    \end{equation}
    %
    %
    where we used \eqref{eq_RS_Hamiltonian_operation}, $\vec{r}_0 = \vec{r}$ and $\mathcal{\hat{T}}$ 
    is the time-ordering operator such that earlier times go to the right.
    In the homogeneous microscopic case, when the diamagnetic current and $ \mathcal{J} $
    are zero, the time evolution equation in
    Eq.~(\ref{eq_RS_time_evolution_homogen_diamagnetic_current}) simplifies further to
    \begin{align}
      \mi \hbar \frac{\partial}{\partial t} \mathcal{F}(\vrt) = \hat{\mathcal{H}}_{(0)} \mathcal{F}(\vrt)    \honemm .
    \end{align}
    In this limit the Hamiltonian $ \hat{H}_{\mathrm{Mx},(0)} $ is time-independent
    which allows to write the time-evolution operator for the
    Riemann-Silberstein vector formally as exponential
    \begin{align}
      \mathcal{U}_{(0)}(t,t_0) = \mathrm{exp} \left[ - \frac{\mi}{\hbar} (t - t_0) \hat{\mathcal{H}}_{(0)} \right],       \label{eq_RS_time_evolution_homogen}
    \end{align}
    which has the familiar form of evolution operators for matter wave functions.
    In the inhomogeneous case, but with $ j_{\mathrm{mdc}} $ equal to zero, we start with the inhomogeneous
    Schr{\"o}dinger-like equation
    \begin{align}
        \frac{\partial}{\partial t} \mathcal{F}(\vec{r},t)
      = - \frac{\mathrm{i}}{\hbar} \hat{\mathcal{H}}_{(0)}
        \bm{\mathcal{F}}(\vec{r}, t) - \mathcal{J}(\vec{r},t)   \honemm .                          \label{eq_RS_Faraday_Ampere_schroedinger_inhomogen}
    \end{align}
    Such inhomogeneous time evolution equations are also found in quantum
    mechanics \cite{Ioana_Serban_2005, Mamadou_Ndong_2009}, which can be
    readily transferred to our Riemann-Silberstein case. By multiplying the
    time-evolution operator $ \hat{\mathcal{U}}_{(0)}(t, t_0) $ with an
    additional time-dependent operator $ \hat{\mathcal{Z}}(t) $, we
    propose
    \begin{align}
        \mathcal{F}(\vec{r},t) 
      = \hat{\mathcal{U}}_{(0)}(t, t_0)
        \hat{\mathcal{Z}}(t) 
        \mathcal{F}(\vec{r}, t_0)   \honemm           \label{eq_RS_time_evolution_inhomogeneous_ansatz}
    \end{align}
    as an ansatz for the solution of the inhomogeneous equation. The time derivative of this expression leads to the time-evolution equation in
    terms of $ \hat{\mathcal{U}}_{(0)}(t, t_0) $ and $
    \bm{\hat{\mathcal{Z}}}(t) $.  Taking 
    Eq.~(\ref{eq_RS_time_derivative_U_homogeneous}) into account leads to
    \begin{equation}
      \begin{alignedat}{2}
          \frac{\partial}{\partial t} \hat{\mathcal{U}}(t, t_0)  
          \hat{\mathcal{Z}}(t) \mathcal{F}&(\vec{r}, t_0) =                                        \\
        & \hspace{-30mm} - \frac{\mi}{\hbar} \hat{\mathcal{H}}_{(0)}(\vec{r},t) 
          \hat{\mathcal{U}}_{(0)}(t, t_0) \hat{\mathcal{Z}}(t)\mathcal{F}(\vec{r}, t_0)
          + \hat{\mathcal{U}}_{(0)}(t,t_0) 
          \Big( \partial_t \hat{\mathcal{Z}}(t) \Big) \mathcal{F}(\vec{r}, t_0)   \honemm .        \label{eq_RS_time_derivative_inhomogeneous}
      \end{alignedat}
    \end{equation}
    This equation has to be equivalent to Eq.~(\ref{eq_RS_Faraday_Ampere_schroedinger_inhomogen})
    such that the last term on the right-hand side of
    Eq.~(\ref{eq_RS_time_derivative_inhomogeneous}) is equal to $ - \mathcal{J} $.
    This condition gives us that the $ \hat{\mathcal{Z}}(t)
    \mathcal{F}(\vec{r},t_0) $ operation has to be equal to
    \begin{align}
        \hat{\mathcal{Z}}(t) \mathcal{F}(\vec{r},t_0) 
      = \mathcal{F}(\vec{r}, t_0) - \int \limits_{t_0}^{t} \md \tau \hat{\mathcal{U}}_{(0)}(t_0, \tau) \mathcal{J}(\vec{r}, \tau)   \honemm .
    \end{align}
    Substituting this result into the ansatz in
    Eq.~(\ref{eq_RS_time_evolution_inhomogeneous_ansatz}) yields the final
    time-evolution equation for the inhomogeneous Maxwell's
    equation in Riemann-Silberstein form
    \begin{align}
           \mathcal{F}(\vec{r},t)
        = \hat{\mathcal{U}}_{(0)}(t, t_0) \mathcal{F}(\vec{r}, t_0)
           - \int\limits_{t_0}^{t} \mathrm{d} \tau \hat{\mathcal{U}}_{(0)}(t, \tau)
           \mathcal{J}(\vec{r},\tau)   \honemm .                                              \label{eq_RS_time_evolution_inhomogeneous_no_dia}
    \end{align}
    For a non-zero diamagnetic current $ j_{\mathrm{mdc}} $, we have to replace the
    first term on the right-hand side of Eq.~(\ref{eq_RS_time_evolution_inhomogeneous})
    by the corresponding Eq.~(\ref{eq_RS_time_evolution_homogen_diamagnetic_current}). The
    final equation provides the correct propagation including the diamagnetic current
    and reads
    \begin{align}
           \mathcal{F}(\vec{r},t)
        = \bm{\hat{\mathcal{U}}}(t, t_0) \bm{\mathcal{F}}(\vec{r}, t_0)
           - \int\limits_{t_0}^{t} \mathrm{d} \tau \hat{\mathcal{U}}_{(0)}(t, \tau)
           \mathcal{J}(\vec{r},\tau)   \honemm .                                              \label{eq_RS_time_evolution_inhomogeneous}
    \end{align}
    We note, that due to the continuity equation, the time-evolution
    Eq.~(\ref{eq_RS_time_evolution_inhomogeneous}) conserves the Gau\ss{} side 
    condition of the electromagnetic field during the propagation.

    In general, the time-evolution of the Riemann-Silberstein vector can not be
    expressed analytically in closed form. However, the form of
    Eq.~(\ref{eq_RS_time_evolution_inhomogeneous}) is already suitable for a
    numerical time-stepping scheme. Iterating the time-evolution $ (m+1 ) $
    times for a time-step $\Delta t$ yields
    \begin{equation}
      \begin{alignedat}{2}
             \mathcal{F}(\vec{r},(m+1) \Delta t)
          &= \bm{\hat{\mathcal{U}}}((m+1) \Delta t, m \Delta t)
             \bm{\mathcal{F}}(\vec{r}, m \Delta t)                                                 \\
          & \quad - \hspace{-3mm} \int\limits_{m \Delta t}^{(m+1) \Delta t} \hspace{-2mm} 
             \mathrm{d} \tau \hat{\mathcal{U}}_{(0)}((m+1) \Delta t, \tau)
             \mathcal{J}(\vec{r}, \tau)    \honemm .                                          \label{eq_RS_time_evolution_equation_recursive}
      \end{alignedat}
    \end{equation}
    For convergence, the length of $ \Delta t $ is chosen such that the
    propagation stays stable and results in an accurate evolution of the
    Riemann-Silberstein vector with negligible error.  With this, we have now reached a
    numerical time-stepping scheme for the Maxwell's equations coupled to
    microscopic as well as macroscopic matter.

    \subsubsection{Forward-backward coupling and time-reversal symmetry}

    The time-evolution of the Pauli-Fierz Hamiltonian in Eq. (\ref{eq_RN21}) is
    given by the time-dependent Schr{\"o}dinger equation of the fully coupled
    matter-photon system. If there are no time-dependent external fields acting
    on the coupled light-matter system it is by construction time reversal
    symmetric when $ t \rightarrow -t $. This symmetry is only strictly given,
    if we consider the whole coupled light-matter system where both subsystems
    influence each other. Breaking the forward and backward coupling between
    matter and electromagnetic fields also destroys the time-reversal symmetry.\\
    As consequence of the time-reversal symmetry for the full photon-matter coupling, the corresponding
    Maxwell time stepping has to obey the property that a reverse time step
    from $ \mathcal{F}(\vec{r}, m \Delta t) $ leads again to the previous
    result $ \mathcal{F}(\vec{r}, (m-1) \Delta t) $. 
    We can construct a numerical
    time-evolution equation based on the enforced time-reversal symmetry
    (ETRS)~\cite{Castro2004} propagator to also numerically ensure time-reversal symmetry for
    the fully coupled case.  The underlying condition of the ETRS algorithm
    requires that a propagation forward by one step $ \Delta t $ starting from
    $ \mathcal F(\vec{r},m \Delta t) $ and then half a step backwards with $
    \Delta t / 2 $ is equivalent to propagating half a step forward.  As
    result, a modified numerical time-evolution equation for ETRS based on
    Eq. (\ref{eq_RS_time_evolution_equation_recursive}) takes the form
    \begin{equation}
      \begin{alignedat}{2}
           \mathcal{F}(\vec{r},(m+1) \Delta t)
        &= \bm{\hat{\mathcal{U}}}((m+1) \Delta t, m \Delta t) \bm{\mathcal{F}}(\vec{r}, m \Delta t)                                      \\
        &  \quad - \hspace{-4mm} \int \limits_{m \Delta t}^{(m + \frac{1}{2}) \Delta t} \hspace{-4mm} \md \tau 
           \hat{\mathcal{U}}_{(0)}(m \Delta t, \tau) \mathcal{J}(\vec{r}, \tau)                                                      \\
        &  \quad + \hspace{-4mm} \int \limits_{(m+1) \Delta t}^{(m + \frac{1}{2}) \Delta t} \hspace{-4mm}
                   \md \tau \hat{\mathcal{U}}_{(0)}((m+1) \Delta t,\tau)
                   \mathcal{J}(\vec{r},\tau)   \honemm .                            \label{eq_RS_time_evolution_equation_recursive_etrs}
      \end{alignedat}
    \end{equation}
    The integrals in Eq. (\ref{eq_RS_time_evolution_equation_recursive}) and
    (\ref{eq_RS_time_evolution_equation_recursive_etrs}) can be approximated by the
    trapezoidal rule so that the numerical
    time-evolution equations take the explicit form 
    \begin{equation}
      \begin{alignedat}{2}
          \mathcal{F}(\vec{r},(m \hspace{-0.5mm} + \hspace{-0.5mm} 1) \Delta t)
        &\approx \bm{\hat{\mathcal{U}}}((m \hspace{-0.5mm} + \hspace{-0.5mm} 1) \Delta t, m \Delta t)
          \bm{\mathcal{F}}(\vec{r} \hspace{0.5mm} \hspace{-1mm} ,m \Delta t)                                                                   \\
        & - \frac{\Delta t}{2}
          \hat{\mathcal{U}}_{(0)}((m \hspace{-0.5mm} + \hspace{-0.5mm} 1) \Delta t, m \Delta t)
          \mathcal{J}(\vec{r} \hspace{0.5mm} \hspace{-1mm}, m \Delta t)                                                             \\
        & - \frac{\Delta t}{2} \mathcal{J}(\vec{r}, (m+1) \Delta t)   \honemm ,     \label{eq_RS_time_evolution_equation_recursive_numerical}
      \end{alignedat}
    \end{equation}
    for the simple direct propagation, and
    \begin{equation}
      \begin{alignedat}{2}
          \mathcal{F}(\vec{r}, \hspace{-0.5mm} (m \hspace{-0.5mm} + \hspace{-0.8mm} 1) \Delta t)
        & \approx \bm{\hat{\mathcal{U}}}((m \hspace{-0.5mm} + \hspace{-0.8mm} 1) \Delta t, m \Delta t)
          \bm{\mathcal{F}}(\vec{r}, m \Delta t)                                                                                                                \\
        & \hspace{-15mm} - \frac{\Delta t}{4} 
          \hat{\mathcal{U}}_{(0)}((m \hspace{-0.5mm} + \hspace{-0.8mm} 1) \Delta t, m \Delta t)
          \mathcal{J}(\vec{r}, m \Delta t)                                                                                                                \\
        & \hspace{-15mm} - \frac{\Delta t}{4} 
          \hat{\mathcal{U}}_{(0)}((m \hspace{-0.5mm} + \hspace{-0.8mm} 1) \Delta t, \hspace{-0.5mm} (m \hspace{-0.5mm} + \hspace{-0.8mm} 1/2) \Delta t)
          \mathcal{J}(\vec{r}, \hspace{-0.5mm} (m \hspace{-0.5mm} + \hspace{-0.8mm} 1/2) \Delta t)                                                        \\
        & \hspace{-15mm} - \frac{\Delta t}{4} 
          \hat{\mathcal{U}}_{(0)}( m \Delta t, \hspace{-0.5mm} (m \hspace{-0.5mm} + \hspace{-0.8mm} 1/2) \Delta t)
          \mathcal{J}(\vec{r}, \hspace{-0.5mm} (m \hspace{-0.5mm} + \hspace{-0.8mm} 1/2) \Delta t)                                                        \\
        & \hspace{-15mm} - \frac{\Delta t}{4} 
          \mathcal{J}(\vec{r}, \hspace{-0.5mm} (m \hspace{-0.5mm} + \hspace{-0.8mm} 1) \Delta t)
      \end{alignedat}                                                                              \label{eq_RS_time_evolution_equation_recursive_numerical_etrs}
    \end{equation}
    for the ETRS propagation.
 
    Note, that a strict time-reversal symmetry can only be obtained if
    the propagation of the matter wave function is considered simultaneously
    with an ETRS scheme and if both, matter and electromagnetic fields
    are coupled self-consistently in every time-step. We provide more details for such
    a self-consistent coupling in section \ref{subsec_predictor_corrector_scheme}.\\
    The time step parameter $ \Delta t $ has to be chosen such that the
    propagation remains stable and accurate.  Both criteria are
    not defined strictly, but it is possible to reduce the value $ \Delta t $
    until convergence is reached. A well-known
    criterion for the time step $ \Delta t $ is given by the
    Courant-Friedrichs-Lewy (CFL) condition \cite{Courant_1928,Courant_1967}
    \begin{align}
        \Delta t_{\mathrm{Mx,CFL}}
      \le \frac{ S_{\mathrm{CFL,max}} }{ c \sqrt{ \frac{1}{\Delta x^2} 
          + \frac{1}{\Delta y^2} + \frac{1}{\Delta z^2} } }  \honemm ,         \label{eq_Maxwell_courant_time}
    \end{align}
    and depends basically on the grid spacings $ \Delta x $, $ \Delta y $, $
    \Delta z $ for the three-dimensional grid, which is taken to be equidistant
    in each dimension, and the Courant number $ S_{\mathrm{CFL,max}} $. The
    Courant number $ S_{\mathrm{CFL,max}} $ varies for different propagation
    methods. In case of FDTD in three dimensions, $ S_{\mathrm{CFL,max}} $ is
    equal to one \cite{Courant_1928,Courant_1967}. We have performed convergence tests of our
    Riemann-Silberstein time-evolution and have found that we can
    chose a Courant number slightly larger than one.\\
    To summarize this section, we have introduced a formulation of Maxwell's equations in Riemann-Silberstein
    form, which allows to use time-evolution techniques that have been developed for
    the time-stepping of the time-dependent Schr\"odinger equation. Since
    the same grid is used for the electric and the magnetic fields, this simplifies
    the implementation, in particular the coupling to matter, and allows for larger
    time steps compared to FDTD since higher order finite-difference discretizations can be employed.
    Choosing the
    Hamiltonian-like kernel $ \hat{\mathcal{H}} $ and the microscopic paramagnetic
    current $ J_{\mathrm{pmc}}^{k} $, diamagnetic current $
    J_{\mathrm{dmc}}^{k} $ and magnetization currents $ J_{\mathrm{mc}}^{k} $
    as inhomogeneity, allows to evolve the Maxwell fields
    coupled to matter. On the other hand, replacing $\hat{ \mathcal{H}} $ with an
    integral kernel for a linear medium $ \hat{\mathcal{H}}_{\mathrm{lm}} $,
    and using a free classical current density, allows to simulate with the same implementation
    macroscopic Maxwell fields. In principle it is also possible to combine
    both cases in the same real-time simulation. Such a combined
    microscopic-macroscopic propagation can be used for example to describe 
    a molecule inside an optical cavity geometry or 
    nanoplasmonic systems close to an interface.

  \subsection{Maxwell boundary conditions} \label{subsec_maxwell_boundary_conditions}

    For the Riemann-Silberstein implementation, it is very important to have flexible
    boundary conditions for the electromagnetic fields.  Since we are dealing
    with a finite simulation box for the radiation fields it is of course necessary
    to have absorbing boundaries, such that outgoing
    electromagnetic fields propagate without any reflections at the box
    boundaries. While this is a standard procedure in FDTD simulations, we
    emphasize here that such absorbing boundaries effectively allow to turn our
    coupled light-matter system into an open quantum system from first principles. 
    No artificial bath degrees of freedom have to be introduced as commonly done
    in the description of open quantum systems. Another class of
    boundary conditions, that arise from the different length scales of typical
    wavelengths for light and matter, are incident wave boundary conditions.
    Such boundaries allow to feed electromagnetic waves with much larger
    wavelength into the microscopic simulation box. Finally, the boundaries can
    also serve as electromagnetic detectors.  Provided the charge densities and
    currents of the matter system have decayed sufficiently and are effectively
    zero in the boundary region, the propagation of the electromagnetic fields
    from there on then only corresponds to a propagation in vacuum. In other words, whatever
    arrives in the boundary region would propagate to the far field and
    contributes to what can be measured in the far field by a detector. \\
    In the following, we discuss the different types of boundary conditions 
    relevant for our coupled light-matter simulations.

    \subsubsection{Boundary regions}   \label{subsubsec_boundary_regions}

\begin{figure}[h!]
        \center
        \begin{minipage}{0.46 \textwidth}
          \includegraphics[scale=0.6]{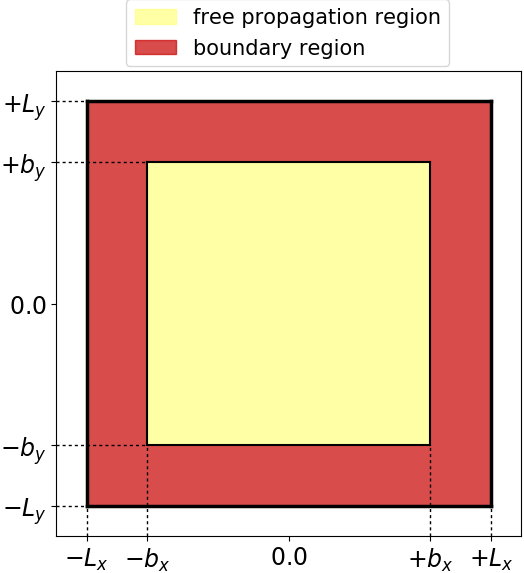}
          \caption{Schematic two-dimensional cut through the simulation box. The yellow area represents the inner free
                   Maxwell propagation area which is limited by the inner box limits 
                   $ b_x $ and $ b_y $ of the boundary region. 
                   The whole box is limited by the outer limits $ L_x $ and $ L_y $.}
          \label{fig_boundaries}
        \end{minipage}
      \end{figure}

    To properly define different boundary conditions, we separate the
    simulation box into two regions, the inner free Maxwell-propagation region
    and the outer boundary region with specified Maxwell's equation to fulfill
    the appropriate simulation condition. Such a region splitting is shown for
    a two-dimensional cut of the three-dimensional simulation box in Fig.
    \ref{fig_boundaries}.  The outer box limits are determined by $ L_u $ for
    direction $ u \in (x,y,z) $, whereas the boundary region is limited by $
    b_u $. We note that $ L_u $ and $ b_u $ are always positive and the box
    center is always located at the Cartesian origin. The total box dimension
    in each direction is $ - L_u $ to $ + L_u $, and the inner borders of the
    free propagation region are $ - b_u $ and $ + b_u $. Hence, the area
    between $ b_u < |u| <  L_u  $ describes the boundary region which contains
    different conditions to fulfill the corresponding simulation system which
    we consider in the following.
    
    \subsubsection{Absorbing boundaries by mask function}   \label{subsubsec_ab_mask_boundaries}
 
      A simple, robust and easily applicable method for absorbing boundaries
      can be reached by multiplying the Riemann-Silberstein vector after each time step
      by a mask
      function which decreases the electromagnetic fields at the boundaries. A
      proper mask function $ f_{\rm{mask}}^{1\mathrm{D}} (u) $ in direction $ u
      \in (x,y,z) $, which is already implemented in Octopus and satisfies the
      condition of damping the fields smoothly in the boundary
      region, is given by \cite{Octopus_2015}
      \begin{align}
          f_{\rm{mask}}^{1\mathrm{D}} (u) 
        = \left\{
            \begin{array}{l l l} 
             & 1                                                                     
             & \hspace{1mm} \mathrm{for} \hspace{2mm} |u| \le b_u                              \\
             & 1 - \mathrm{sin} \left( \frac{ \pi |u - b_u| }{ 2 |L_u - b_u| } \right)^2     
             & \hspace{1mm} \mathrm{for} \hspace{2mm} b_u < |u| < L_u \hspace{1mm} .           \
            \end{array}
          \right.        \label{eq_mask_function_1D}
      \end{align}
      For a three-dimensional simulation box, the corresponding mask function
      $ f_{\rm{mask}}^{3\mathrm{D}} (x, y, z) $ is factorized into three
      one-dimensional mask functions for each direction and therefore takes the
      form 
      \begin{align}
           f_{\rm{mask}}^{3\mathrm{D}} (x, y, z)
        &= f_{\rm{mask}}^{1\mathrm{D}} (x) \cdot f_{\rm{mask}}^{1\mathrm{D}} (y) \cdot f_{\rm{mask}}^{1\mathrm{D}} (z)   \hspace{1mm} .
      \end{align}
      The only parameter that improves the absorbing effect of the mask
      function is the boundary width $ L_u - b_u $ in the corresponding
      direction. In case of relatively large absorbing boundary regions and
      small outgoing electromagnetic field amplitudes compared to the inner box fields,
      absorbing boundaries in terms of mask functions are a simple and effective method
      for simulating open Maxwell systems. Whenever larger electromagnetic field
      amplitudes have to be absorbed, or when using larger padding regions for
      the mask function become computationally too costly, it becomes
      advantageous to use a perfectly matched layer, which we discuss next.

    \subsubsection{Absorbing boundaries by perfectly matched layer}   \label{subsubsec_ab_pml_boundaries}
 
      A more accurate method for open Maxwell systems is the
      perfectly matched layer (PML) absorbing boundary condition. We have
      implemented such a PML analogous to the Ber\`enger layer for the Yee
      finite difference time domain algorithm
      \cite{taflove2005computational,bérenger2007perfectly}, but now modified
      for the Riemann-Silberstein Maxwell propagation. The basic idea of the
      Ber\`enger layer is to complement Maxwell's equations with an artificial
      lossy layer, which is described by a non-physical electric conductivity $
      \sigma_{\mathrm{el}} $ and non-physical magnetic conductivity $
      \sigma_{\mathrm{mag}} $. These conductivities are defined such as to
      yield minimal reflections at the boundaries, but have no physical meaning otherwise. The loss due to the conductivity parameters is
      linear in the corresponding $ \vec{E} (\vec{r},t) $ and $
      \vec{B}(\vec{r},t) $, so that Faraday's and Amp\`ere's law without
      current density take the form 
      \begin{align}
        \partial_{t} E^{k}(\vec{r},t) = - \frac{1}{\epsilon_0 \mu_0} \epsilon^{klm} \partial_{l} B_{m}(\vec{r},t)
                                          - \sigma_{\mathrm{el}} E^{k}(\vec{r},t)    \honemm ,                                    \label{eq_Ampere_PML_lossy_layer}
      \end{align}
    
      \begin{align}
        \partial_{t} B^{k}(\vec{r},t) =   \epsilon^{klm} \partial_{l} E_{m}(\vec{r},t) 
                                          - \sigma_{\mathrm{mag}} B^{k}(\vec{r},t)   \honemm .                                    \label{eq_Faraday_PML_lossy_layer}
      \end{align}
      We note here, that the PML is in
      principle not restricted to vacuum conditions but also valid for other
      homogeneous dielectric conditions. In other words, the Riemann-Silberstein PML that we have 
      implemented also works in a linear medium, but with the
      constraint that $ \epsilon $ and $ \mu $, and consequently $ c $ are
      constant at the border and inside the boundaries. 
      Transforming Eq.~(\ref{eq_E_field_from_RS}-\ref{eq_B_field_from_RS}) 
      to frequency domain, where $ \tilde{F} $ denotes the
      Riemann-Silberstein vector in frequency space, leads to the underlying
      Riemann-Silberstein Maxwell's equation for the absorbing layer 
      \begin{equation}
        \begin{alignedat}{3}
          &
          && - \omega \tilde{F}_{\pm}^{k}(\vec{r},\omega)
          && = \mp c_0 \epsilon^{klm} \partial_{l} \tilde{F}_{\pm,m}(\vec{r},\omega)                                                         \\
          & 
          &&
          && \quad - \mi \frac{1}{2} \sigma_{\rm{e}}
             \Big( \tilde{F}_{+}^{k}(\vec{r},\omega) + \hspace{-0.5mm} \tilde{F}_{-}^{k}(\vec{r},\omega) \Big)                             \\
          &
          &&
          && \quad \mp \mi \frac{1}{2} \sigma_{\mathrm{mag}}
             \Big( 
               \tilde{F}_{+}^{k}(\vec{r},\omega) - \hspace{-0.5mm} \tilde{F}_{-}^{k}(\vec{r},\omega) 
             \Big)   \hspace{1mm} ,                                                                \label{eq_RS_Mx_equation_PML_lossy_layer}
        \end{alignedat}
      \end{equation}
      where the first term on the right-hand side describes the curl operation.
      The principle of a PML is to propagate the respective field components in
      the absorbing boundary region which are necessary for a correct
      propagation inside the free Maxwell region, and to damp the remaining
      components without causing strong reflections back into the free Maxwell
      region. For this purpose,
      Ber\`enger's method splits up Maxwell's equations for each direction in
      two equations which form the basis for the so-called split PML
      \cite{BERENGER1996363,bérenger2007perfectly,taflove2005computational}.
      The field component in $ k $-direction is split into one component
      for $ l $ and one for $ m $ with $ k \neq l \neq m $, so that the vector k component
      $ \tilde{F}_{\pm}^{k} $ is given by
      \begin{align}
          \tilde{F}_{\pm}^{k}(\vec{r},\omega)
        = \tilde{F}_{\pm}^{k,(l)}(\vec{r},\omega)
          + \tilde{F}_{\pm}^{k,(m)}(\vec{r},\omega)    \honemm .     \label{eq_RS_field_split}
      \end{align}
      The field component $ \tilde{F}_{\pm}^{k} $ is split such that the $ \tilde{F}_{\pm}^{k,(l)} $
      part is responsible for the field propagation parallel to direction $ l $ and accordingly
      $ \tilde{F}_{\pm}^{k,(m)} $ parallel to direction $ m $. In other words, there are two
      separate propagations which simulate only the free propagation along the corresponding
      direction. Thus, one propagation could be where the field enters the PML region while the
      other part is still in the free propagation box. The damping of the fields is applied by the 
      electric and magnetic conductivities $ \sigma_{\mathrm{el}} $, $
      \sigma_{\mathrm{mag}} $ which are artificially modified and depend now on the
      splitted direction, i.e., $l $ direction for $ \tilde{F}_{\pm}^{k,(l)} $, not the field direction.
      In addition, each equation only contains one part of the two curl terms. Applying all
      these considerations to the six components of the Maxwell's equations
      in Riemann-Silberstein form yields twelve relations for the PML. Explicitly, the two equations 
      for the $ x $ component in equation
      (\ref{eq_RS_Mx_equation_PML_lossy_layer}) are
      \begin{equation}
        \begin{alignedat}{2}
            - \omega \tilde{F}_{\pm}^{x,(y)}(\vec{r},\omega)
          & = \pm c_0 \partial_y 
              \Big(
                \hspace{-0.7mm} \tilde{F}_{\pm}^{z,(x)}(\vec{r},\omega) \hspace{-0.7mm} 
                + \hspace{-0.7mm} \tilde{F}_{\pm}^{z,(y)}(\vec{r},\omega) \hspace{-0.7mm}
              \Big)                                                                                \\
          & \quad - \mi \frac{1}{2} \sigma_{\mathrm{el}}^{(y)} 
              \Big(
                \hspace{-0.7mm} \tilde{F}_{+}^{x,(y)}(\vec{r},\omega) \hspace{-0.7mm}
                + \hspace{-0.7mm} \tilde{F}_{-}^{x,(y)}(\vec{r},\omega) \hspace{-0.7mm}
              \Big)                                                                                \\
          & \quad \mp \mi \frac{1}{2} \sigma_{\mathrm{mag}}^{(y)}
              \Big(
                \hspace{-0.7mm} \tilde{F}_{+}^{x,(y)}(\vec{r},\omega) \hspace{-0.7mm}
                - \hspace{-0.7mm} \tilde{F}_{-}^{x,(y)}(\vec{r},\omega) \hspace{-0.7mm}
              \Big)                                                                                \honemm ,
        \end{alignedat}    \label{eq_RS_split_x_y}
      \end{equation}
 
      \begin{equation}
        \begin{alignedat}{2}
            - \omega \tilde{F}_{\pm}^{x,(z)}(\vec{r},\omega) 
          & = \mp c_0 \partial_z
              \Big(
                \hspace{-0.7mm} \tilde{F}_{\pm}^{y,(x)}(\vec{r},\omega) \hspace{-0.7mm} 
                + \hspace{-0.7mm} \tilde{F}_{\pm}^{y,(z)}(\vec{r},\omega) \hspace{-0.7mm} 
              \Big)                                                                                \\
          & \quad - \mi \frac{1}{2} \sigma_{\mathrm{el}}^{(z)}
              \Big(
                \hspace{-0.7mm} \tilde{F}_{+}^{x,(z)}(\vec{r},\omega) \hspace{-0.7mm}
                + \hspace{-0.7mm} \tilde{F}_{-}^{x,(z)}(\vec{r},\omega) \hspace{-0.7mm}
              \Big)                                                                                \\
          & \quad \mp \mi \frac{1}{2} \sigma_{\mathrm{mag}}^{(z)}
              \Big(
                \hspace{-0.7mm} \tilde{F}_{+}^{x,(z)}(\vec{r},\omega) \hspace{-0.7mm}
                - \hspace{-0.7mm} \tilde{F}_{-}^{x,(z)}(\vec{r},\omega) \hspace{-0.7mm}
              \Big)                                                                                \honemm ,
        \end{alignedat}    \label{eq_RS_split_x_z}
      \end{equation}
      for the $ y $ component
      \begin{equation}
        \begin{alignedat}{2}
            - \omega \tilde{F}_{\pm}^{y,(z)}(\vec{r},\omega)
          & = \pm c_0 \partial_z
              \Big( \hspace{-0.7mm} \tilde{F}_{\pm}^{x,(y)}(\vec{r},\omega) \hspace{-0.7mm}
                + \hspace{-0.7mm} \tilde{F}_{\pm}^{x,(z)}(\vec{r},\omega) \hspace{-0.7mm}
              \Big)                                                                                \\
          & \quad - \mi \frac{1}{2} \sigma_{\mathrm{el}}^{(z)}
              \Big(
                \hspace{-0.7mm} \tilde{F}_{+}^{y,(z)}(\vec{r},\omega) \hspace{-0.7mm}
                + \hspace{-0.7mm} \tilde{F}_{-}^{y,(z)}(\vec{r},\omega) \hspace{-0.7mm}
              \Big)                                                                                \\
          & \quad \mp \mi \frac{1}{2} \sigma_{\mathrm{mag}}^{(z)}
              \Big(
                \hspace{-0.7mm} \tilde{F}_{+}^{y,(z)}(\vec{r},\omega) \hspace{-0.7mm}
                - \hspace{-0.7mm} \tilde{F}_{-}^{y,(z)}(\vec{r},\omega) \hspace{-0.7mm}
              \Big)                                                                                \honemm ,
        \end{alignedat}    \label{eq_RS_split_y_z}
      \end{equation}
 
      \begin{equation}
        \begin{alignedat}{2}
            - \omega \tilde{F}_{\pm}^{y,(x)}(\vec{r},\omega)
          & = \mp c_0 \partial_x
              \Big(
                \hspace{-0.7mm} \tilde{F}_{\pm}^{z,(x)}(\vec{r},\omega) \hspace{-0.7mm}
                + \hspace{-0.7mm} \tilde{F}_{\pm}^{z,(y)}(\vec{r},\omega) \hspace{-0.7mm}
              \Big)                                                                                \\
          & \quad - \mi \frac{1}{2} \sigma_{\mathrm{el}}^{(x)}
              \Big( \hspace{-0.7mm} \tilde{F}_{+}^{y,(x)}(\vec{r},\omega) \hspace{-0.7mm}
              + \hspace{-0.7mm} \tilde{F}_{-}^{y,(x)}(\vec{r},\omega) \hspace{-0.7mm}
            \Big)                                                                                  \\
          & \quad \mp \mi \frac{1}{2} \sigma_{\mathrm{mag}}^{(x)}
            \Big( 
              \hspace{-0.7mm} \tilde{F}_{+}^{y,(x)}(\vec{r},\omega) \hspace{-0.7mm}
              - \hspace{-0.7mm} \tilde{F}_{-}^{y,(x)}(\vec{r},\omega) \hspace{-0.7mm}
            \Big)                                                                                  \honemm ,
        \end{alignedat}    \label{eq_RS_split_y_x}
      \end{equation}
      and for the $ z $ component
      \begin{equation}
        \begin{alignedat}{2}
            - \omega \tilde{F}_{\pm}^{z,(x)}(\vec{r},\omega)
          & = \pm c_0 \partial_x 
              \Big(
                \hspace{-0.7mm} \tilde{F}_{\pm}^{y,(x)}(\vec{r},\omega) \hspace{-0.7mm} 
                + \hspace{-0.7mm} \tilde{F}_{\pm}^{y,(z)}(\vec{r},\omega) \hspace{-0.7mm}
              \Big)                                                                                \\
          & \quad - \mi \frac{1}{2} \sigma_{\mathrm{el}}^{(x)}
              \Big(
                \hspace{-0.7mm} \tilde{F}_{+}^{z,(x)}(\vec{r},\omega) \hspace{-0.7mm}
                + \hspace{-0.7mm} \tilde{F}_{-}^{z,(x)}(\vec{r},\omega) \hspace{-0.7mm}
              \Big)                                                                                \\
          & \quad \mp \mi \frac{1}{2} \sigma_{\mathrm{mag}}^{(x)}
              \Big( 
                \hspace{-0.7mm} \tilde{F}_{+}^{z,(x)}(\vec{r},\omega) \hspace{-0.7mm}
                - \hspace{-0.7mm} \tilde{F}_{-}^{z,(x)}(\vec{r},\omega) \hspace{-0.7mm}
              \Big)    \honemm ,
        \end{alignedat}    \label{eq_RS_split_z_x}
      \end{equation}
 
      \begin{equation}
        \begin{alignedat}{2}
            - \omega \tilde{F}_{\pm}^{z,(y)}(\vec{r},\omega) 
          & = \mp c_0 \partial_y 
              \Big(
                \hspace{-0.7mm} \tilde{F}_{\pm}^{x,(y)}(\vec{r},\omega) \hspace{-0.7mm}
                + \hspace{-0.7mm} \tilde{F}_{\pm}^{x,(z)}(\vec{r},\omega) \hspace{-0.7mm}
              \Big)                                                                                \\
          & \quad - \mi \frac{1}{2} \sigma_{\mathrm{el}}^{(y)}
              \Big(
                \hspace{-0.7mm} \tilde{F}_{+}^{z,(y)}(\vec{r},\omega) \hspace{-0.7mm}
                + \hspace{-0.7mm} \tilde{F}_{-}^{z,(y)}(\vec{r},\omega) \hspace{-0.7mm}
              \Big)                                                                                \\
          & \quad \mp \mi \frac{1}{2} \sigma_{\mathrm{mag}}^{(y)}
              \Big(
                \hspace{-0.7mm} \tilde{F}_{+}^{z,(y)}(\vec{r},\omega) \hspace{-0.7mm}
                - \hspace{-0.7mm} \tilde{F}_{-}^{z,(y)}(\vec{r},\omega) \hspace{-0.7mm}
              \Big)    \honemm .
        \end{alignedat}    \label{eq_RS_split_z_y}
      \end{equation} 
      Analogous to Ber\`enger's split field PML derivation for the
      Yee-Algorithm, we want to include also in our case the frequency $ \omega
      $ and the electric and magnetic conductivity $ \sigma_{\mathrm{el}} $ and
      $ \sigma_{\mathrm{mag}} $ in a factor multiplied by the corresponding
      split field
      \cite{BERENGER1996363,bérenger2007perfectly,taflove2005computational}
      before we recombine the two split field equations.  Using the two factors
      \begin{align}
          \tilde{\eta}^{(l)}(\omega)
        = - \frac{\mi \omega ( \sigma_{\mathrm{el}}^{(l)} + \sigma_{\mathrm{mag}}^{(l)} - 2 \mi \omega ) }
                 {2 ( \sigma_{\mathrm{el}}^{(l)} - \mi \omega )( \sigma_{\mathrm{mag}}^{(l)} - \mi \omega ) }   \honemm ,
      \end{align}
 
      \begin{align}
          \tilde{\xi}^{(l)}(\omega)
        = \frac{\mi \omega ( \sigma_{\mathrm{mag}}^{(l)} - \sigma_{\mathrm{el}}^{(l)} ) }
               {2 ( \sigma_{\mathrm{el}}^{(l)} -\mi \omega ) ( \sigma_{\mathrm{mag}}^{(l)} - \mi \omega ) }    \honemm ,
      \end{align}
      the system of the split equations in Eqs. (\ref{eq_RS_split_x_y} -
      \ref{eq_RS_split_z_y}) can be rearranged equivalently to 
      \begin{equation}
        \begin{alignedat}{2}
            - \omega \tilde{F}_{\pm}^{x,(y)}(\vec{r},\omega)
          & = \pm c_0 \tilde{\eta}^{(y)}(\omega) \partial_y 
              \Big(
                \hspace{-0.7mm} \tilde{F}_{\pm}^{z,(x)}(\vec{r},\omega) \hspace{-0.7mm}
                + \hspace{-0.7mm} \tilde{F}_{\pm}^{z,(y)}(\vec{r},\omega) \hspace{-0.7mm}
              \Big)                                                                                \\
          & \quad \pm c_0 \tilde{\xi}^{(y)}(\omega) \partial_y 
              \Big( 
                \hspace{-0.7mm} \tilde{F}_{\mp}^{z,(x)}(\vec{r},\omega) \hspace{-0.7mm}
                + \hspace{-0.7mm} \tilde{F}_{\mp}^{z,(y)}(\vec{r},\omega) \hspace{-0.7mm}
              \Big)   \honemm ,
        \end{alignedat}    \label{eq_RS_split_x_y_2}
      \end{equation}
 
      \begin{equation}
        \begin{alignedat}{2}
            - \omega \tilde{F}_{\pm}^{x,(z)}(\vec{r},\omega)
          & = \mp c_0 \tilde{\eta}^{(z)}(\omega) \partial_z 
              \Big( 
                \hspace{-0.7mm} \tilde{F}_{\pm}^{y,(x)}(\vec{r},\omega) \hspace{-0.7mm} 
                + \hspace{-0.7mm} \tilde{F}_{\pm}^{y,(z)}(\vec{r},\omega) \hspace{-0.7mm}
              \Big)                                                                                \\
          & \quad \mp c_0 \tilde{\xi}^{(z)}(\omega) \partial_z 
              \Big(
                \hspace{-0.7mm} \tilde{F}_{\pm}^{y,(x)}(\vec{r},\omega) \hspace{-0.7mm}
                + \hspace{-0.7mm} \tilde{F}_{\pm}^{y,(z)}(\vec{r},\omega) \hspace{-0.7mm}
              \Big)
        \end{alignedat}    \label{eq_RS_split_x_z_2}
      \end{equation}
      for the $ x $ component, and
      \begin{equation}
        \begin{alignedat}{2}
            - \omega \tilde{F}_{\pm}^{y,(z)}(\vec{r},\omega)
          & = \pm c_0 \tilde{\eta}^{(z)}(\omega) \partial_z 
              \Big(
                \hspace{-0.7mm} \tilde{F}_{\pm}^{x,(y)}(\vec{r},\omega) \hspace{-0.7mm}
                + \hspace{-0.7mm} \tilde{F}_{\pm}^{x,(z)}(\vec{r},\omega) \hspace{-0.7mm}
              \Big)                                                                                \\
          & \quad \pm c_0 \tilde{\xi}^{(z)}(\omega) \partial_z 
              \Big(
                \hspace{-0.7mm} \tilde{F}_{\mp}^{x,(y)}(\vec{r},\omega) \hspace{-0.7mm}
                + \hspace{-0.7mm} \tilde{F}_{\mp}^{x,(z)}(\vec{r},\omega) \hspace{-0.7mm}
              \Big)   \honemm ,
        \end{alignedat}    \label{eq_RS_split_y_z_2}
      \end{equation}
 
      \begin{equation}
        \begin{alignedat}{2}
            - \omega \tilde{F}_{\pm}^{y,(x)}(\vec{r},\omega) 
          & = \mp c_0 \tilde{\eta}^{(x)}(\omega) \partial_x 
              \Big(
                \hspace{-0.7mm} \tilde{F}_{\pm}^{z,(x)}(\vec{r},\omega) \hspace{-0.7mm}
                + \hspace{-0.7mm} \tilde{F}_{\pm}^{z,(y)}(\vec{r},\omega) \hspace{-0.7mm}
              \Big)                                                                                \\
          & \quad \mp c_0 \tilde{\xi}^{(x)}(\omega) \partial_x 
              \Big(
                \hspace{-0.7mm} \tilde{F}_{\pm}^{z,(x)}(\vec{r},\omega) \hspace{-0.7mm}
                + \hspace{-0.7mm} \tilde{F}_{\pm}^{z,(y)}(\vec{r},\omega) \hspace{-0.7mm}
              \Big)
        \end{alignedat}    \label{eq_RS_split_y_x_2}
      \end{equation}
      for the $ y $ component, and
      \begin{equation}
        \begin{alignedat}{2}
            - \omega \tilde{F}_{\pm}^{z,(x)}(\vec{r},\omega)
          & = \pm c_0 \tilde{\eta}^{(x)}(\omega) \partial_x 
              \Big(
                \hspace{-0.7mm} \tilde{F}_{\pm}^{y,(x)}(\vec{r},\omega) \hspace{-0.7mm}
                + \hspace{-0.7mm} \tilde{F}_{\pm}^{y,(z)}(\vec{r},\omega) \hspace{-0.7mm}
              \Big)                                                                                \\
          & \quad \pm c_0 \tilde{\xi}^{(x)}(\omega) \partial_x 
              \Big(
                \hspace{-0.7mm} \tilde{F}_{\mp}^{y,(x)}(\vec{r},\omega) \hspace{-0.7mm}
                + \hspace{-0.7mm} \tilde{F}_{\mp}^{y,(z)}(\vec{r},\omega) \hspace{-0.7mm}
              \Big)   \honemm ,
        \end{alignedat}    \label{eq_RS_split_z_x_2}
      \end{equation}
 
      \begin{equation}
        \begin{alignedat}{2}
            - \omega \tilde{F}_{\pm}^{z,(y)}(\vec{r},\omega) 
          & = \mp c_0 \tilde{\eta}^{(y)}(\omega) \partial_y
              \Big(
                \hspace{-0.7mm} \tilde{F}_{\pm}^{x,(y)}(\vec{r},\omega) \hspace{-0.7mm}
                + \hspace{-0.7mm} \tilde{F}_{\pm}^{x,(z)}(\vec{r},\omega) \hspace{-0.7mm}
              \Big)                                                                                \\
          & \quad \mp c_0 \tilde{\xi}^{(y)}(\omega) \partial_y
              \Big(
                \hspace{-0.7mm} \tilde{F}_{\pm}^{x,(y)}(\vec{r},\omega) \hspace{-0.7mm}
                + \hspace{-0.7mm} \tilde{F}_{\pm}^{x,(z)}(\vec{r},\omega) \hspace{-0.7mm}
              \Big)
        \end{alignedat}    \label{eq_RS_split_z_y_2}
      \end{equation}
      for the $ z $ component. Finally, adding each of the two
      Eqs.~(\ref{eq_RS_split_x_y_2}-\ref{eq_RS_split_x_z_2}),
      (\ref{eq_RS_split_y_x_2}-\ref{eq_RS_split_y_z_2}), and
      (\ref{eq_RS_split_z_x_2}-\ref{eq_RS_split_z_y_2}) using
      Eq.~(\ref{eq_RS_field_split}) yields the PML equations in
      frequency domain for the Riemann-Silberstein representation 
      \begin{equation}
        \begin{alignedat}{2}
            - \omega
            \begin{pmatrix}
              \tilde{F}_{+}^{k}(\vec{r},\omega)    \\[2mm]
              \tilde{F}_{-}^{k'}(\vec{r},\omega)   \\
            \end{pmatrix}
         &= c_0
            \begin{pmatrix}
              \sum \limits_{l,m} - \epsilon^{klm} \tilde{\eta}_{l}(\omega) \partial_{l} \tilde{F}_{+,m}(\vec{r},\omega)       \\[4mm]
              \sum \limits_{l',m'} \epsilon^{klm} \tilde{\eta}_{l'}(\omega) \partial_{l'} \tilde{F}_{-,m'}(\vec{r},\omega)
            \end{pmatrix}                                                                                                \\[2mm]
         &\quad + c_0
            \begin{pmatrix}
              \sum \limits_{l,m} - \epsilon^{klm} \tilde{\xi}_{l}(\omega) \partial_{l} \tilde{F}_{-,m}(\vec{r},\omega)         \\[4mm]
              \sum \limits_{l',m'} \epsilon^{k'l'm'} \tilde{\xi}_{l'}(\omega) \partial_{l'} \tilde{F}_{+,m'}(\vec{r},\omega)
            \end{pmatrix}    \honemm .
        \end{alignedat}     \label{eq_RS_PML_Hamiltonian_frequency_domain_convolution}
      \end{equation}
      In our PML implementation for simplicity, we do not introduce new
      correction terms in $ \tilde{\eta}^{(l)} $ or $ \tilde{\xi}^{(l)} $ to
      improve the PML and to reduce low-frequency reflections like it is
      commonly applied for the Yee algorithm
      \cite{BERENGER1996363,bérenger2007perfectly,taflove2005computational}.
      While such extensions are possible in future refinements of our
      implementation, we found the simple form without correction terms already to provide good
      absorbance at the boundaries.  By back transforming 
      Eq.~(\ref{eq_RS_PML_Hamiltonian_frequency_domain_convolution}) from frequency
      domain into time domain, we arrive at
      \begin{equation}
        \begin{alignedat}{2}
             \eta^{(l)}(t)
          &= \delta(t)
             - \frac{1}{2} 
             \left(
               \sigma_{\mathrm{el}}^{(l)} \me^{- \sigma_{\mathrm{el}}^{(l)} t}
               + \sigma_{\mathrm{mag}}^{(l)} \me^{- \sigma_{\mathrm{mag}}^{(l)} t}
             \right) \Theta(t)                                                           \\[2mm]
          &= \delta(t) + \zeta(t) \Theta(t)   \honemm ,
        \end{alignedat}
      \end{equation}
 
      \begin{equation}
        \begin{alignedat}{2}
             \xi^{(l)}(t)
          &= - \frac{1}{2}
             \left( 
               \sigma_{\mathrm{el}}^{(l)} \me^{- \sigma_{\mathrm{el}}^{(l)} t} 
               - \sigma_{\mathrm{mag}}^{(l)} \me^{- \sigma_{\mathrm{mag}}^{(l)} t}
             \right) \Theta(t)                                                           \\[2mm]
          &= \xi(t) \Theta(t)  \honemm .
        \end{alignedat}
      \end{equation}
      The electric conductivity $ \sigma_{\mathrm{el}} $ and the magnetic
      conductivity $ \sigma_{\mathrm{mag}} $ have to be chosen such that the
      reflection becomes minimal. As is well-known in FDTD
      \cite{BERENGER1996363,bérenger2007perfectly,taflove2005computational}, 
      the relation between the electric conductivity $
      \sigma_{\mathrm{el}} $ and the magnetic conductivity $
      \sigma_{\mathrm{mag}} $ to minimize the reflection coefficient has to
      obey
 
      \begin{align}
          \frac{\sigma_{\mathrm{el}}}{\epsilon_0} 
        = \frac{\sigma_{\mathrm{mag}}}{\mu_0}            \label{eq_ep_mu_PML_condition}
      \end{align}
      at the border between the free
      Maxwell simulation box and the absorbing boundaries. Using this relation
      between the two conductivities, it is convenient to use only one
      conductivity with $ \sigma = \sigma_{\mathrm{el}} $, and the updated
      forms of the expressions $\zeta^{(l)}(t)$, and $ \xi^{(l)}(t) $ are
      \begin{align}
          \zeta^{(l)}(t)
        = - \frac{1}{2} \sigma^{(l)} \me^{- \sigma^{(l)} t} 
          \left( 1 + \frac{\mu_0}{\epsilon_0} \me^{- (\mu_0/\epsilon_0-1) \sigma^{(l)} t} \right)   \honemm ,
      \end{align}

      \begin{align}
          \xi^{(l)}(t)
        = - \frac{1}{2} \sigma^{(l)} \me^{- \sigma^{(l)} t} 
          \left( 1 - \frac{\mu_0}{\epsilon_0} \me^{- (\mu_0/\epsilon_0-1) \sigma^{(l)} t} \right)   \honemm ,
      \end{align}
      As a result, the back transformation of
      Eq.~(\ref{eq_RS_PML_Hamiltonian_frequency_domain_convolution}) becomes
      \begin{equation}
        \begin{alignedat}{2}
            \mi c_0 \partial_0
            \begin{pmatrix}
              F_{+}^{k}(\vec{r},t)    \\[2mm]
              F_{-}^{k'}(\vec{r},t)   \\
            \end{pmatrix}
         &= c_0
            \begin{pmatrix}
              \sum \limits_{l,m} - \epsilon^{klm} \Big( \delta \ast \partial_{l} F_{+,m}(\vec{r}) \Big)(t)                \\[4mm]
              \sum \limits_{l',m'} \epsilon^{k'l'm'} \Big( \delta \ast \partial_{l'} F_{-,m'}(\vec{r}) \Big)(t)
            \end{pmatrix}                                                                                      \\[2mm]
         & \hspace{-4mm} + c_0
            \begin{pmatrix}
              \sum \limits_{l,m} - \epsilon^{klm} \Big( \zeta^{(l)} \ast \partial_{l} F_{+,m}(\vec{r}) \Big)(t)             \\[4mm]
              \sum \limits_{l',m'} \epsilon^{k'l'm'} \Big( \zeta^{(l')} \ast \partial_{l'} F_{-,m'}(\vec{r}) \Big)(t)
            \end{pmatrix}                                                                                      \\[2mm]
         & \hspace{-4mm} + c_0
            \begin{pmatrix}
              \sum \limits_{l,m} - \epsilon^{klm}    \Big( \xi^{(l)} \ast \partial_{l} F_{-,m}(\vec{r}) \Big)(t)            \\[4mm]
              \sum \limits_{l',m'} \epsilon^{k'l'm'} \Big( \xi^{(l')} \ast \partial_{l'} F_{+,m'}(\vec{r}) \Big)(t)
            \end{pmatrix}    \honemm .
        \end{alignedat}                                                                  \label{eq_RS_PML_Hamiltonian_time_domain_convolution_1}
      \end{equation}
      which contains several convolutions in time. Whereas the first convolution on the right-hand side in (\ref{eq_RS_PML_Hamiltonian_time_domain_convolution_1})
      is simply
      \begin{align}
        \Big( \delta \ast \partial_{l} F_{\pm,m}(\vec{r}) \Big)(t) = \partial_{l} F_{\pm,m}(\vec{r},t)    \honemm ,
      \end{align}
      the remaining convolutions are explicitly given by
      \begin{equation}
        \begin{alignedat}{2}
            \Big( \zeta^{(l)} \ast \partial_{l} F_{\pm,m}(\vec{r}) \Big)(t)
          = \int \limits_{0}^{t} \zeta^{(l)}(t-\tau) F_{\pm,m}(\vec{r},\tau) \md \tau   \honemm ,
        \end{alignedat}                                                                            \label{eq_RS_PML_convolution_zeta}
      \end{equation}

      \begin{equation}
        \begin{alignedat}{2}
            \Big( \xi^{(l)} \ast \partial_{l} F_{\pm,m}(\vec{r}) \Big)(t)
          = \int \limits_{0}^{t} \xi^{(l)}(t-\tau) F_{\pm,m}(\vec{r},\tau) \md \tau   \honemm .
        \end{alignedat}                                                                            \label{eq_RS_PML_convolution_xi}
      \end{equation}
      This completes the construction of the PML for our Riemann-Silberstein formulation.
      Since we have already illustrated how to include a linear medium in the Riemann-Silberstein
      time-evolution in the previous sections, it becomes now straight forward to combine the PML with our existing
      implementation. Adding the adequate PML expressions to the integral kernel for the Riemann-Silberstein
      Hamiltonian  $ \hat{\bar{\mathcal{H}}} $ in Eq.~(\ref{eq_RS_Hamiltonian}), we arrive at a propagation
      scheme with perfectly matched layer boundaries
      \begin{equation}
        \begin{alignedat}{2}
             \hat{\bar{\mathcal{H}}}_{\mathrm{PML}}(\vec{r},\vec{r} \hspace{0.5mm}' \hspace{-1mm}, t, t')
          &= \hat{\bar{\mathcal{H}}}_{(0)}(\vec{r},\vec{r} \hspace{0.5mm}' \hspace{-1mm}, t, t')
             + \hat{\bar{\mathcal{K}}}(\vec{r},\vec{r} \hspace{0.5mm}' \hspace{-1mm}, t, t')                  \\
          & \quad  + \delta(\vec{r} - \vec{r} \,') \delta(t-t')
            \bm{\hat{\mathcal{G}}}(\vec{r} \,', t) \bm{\mathcal{F}}(\vec{r} \,',t')  \honemm ,       \label{eq_RS_Hamiltonian_PML}
        \end{alignedat}
      \end{equation}
      with
      \begin{align}
          \bm{\hat{\mathcal{G}}}(\vec{r}, t) \bm{\mathcal{F}}(\vec{r},t)
        = \int \limits_{0}^{t} \md \tau \hat{\mathcal{G}}(\vec{r}, t, \tau) \mathcal{F}(\vec{r},\tau)  \honemm ,
      \end{align}
      where the 6x6 PML matrix $ \hat{\bar{\mathcal{G}}}(\vec{r}, t, \tau) $ is given by
      \begin{equation}
        \begin{alignedat}{2}
             \hspace{-5mm}
             \hat{\bar{\mathcal{G}}}(\vec{r},t,\tau)
          &= \begin{pmatrix}
               \begin{array}{rr}
                 1 & \honemm 0 \\[2mm] 0 & \honemm -1
               \end{array}
             \end{pmatrix}_{\hspace{-1mm} 2 \times 2}
             \hspace{-3mm} \otimes \hspace{-1mm}
             \left(
               - \mathrm{i} \hbar c_0 \hspace{-0.5mm}
               \left[ \sum_{k=1}^{3} \zeta^{(k)}(t-\tau) S^{k} \partial_{k} \right]  
             \right)_{\hspace{-1mm} 3 \times 3}                                                         \\
          &  \hspace{-5mm} +
             \begin{pmatrix}
               \begin{array}{rr}
                 0 & \honemm -1 \\[2mm] 1 & \honemm 0
               \end{array}
             \end{pmatrix}_{\hspace{-1mm} 2 \times 2}
             \hspace{-3mm} \otimes \hspace{-1mm}
             \left(
               - \mathrm{i} \hbar c_0 \hspace{-0.5mm}
               \left[ \sum_{k=1}^{3} \xi^{(k)}(t-\tau) S^{k} \partial_{k} \right] 
             \right)_{\hspace{-1mm} 3 \times 3}  \hspace{-2mm} .                                        \label{eq_RS_Hamiltonian_PML_convolution}
        \end{alignedat}
      \end{equation}
      The left factor of the second Kronecker product in
      Eq.~(\ref{eq_RS_Hamiltonian_PML_convolution}) has entries in the
      off-diagonal, and therefore the two Riemann-Silberstein vectors $
      \vec{F}_{\pm} $ always couple in the PML region.\\
      In principle, the PML terms in Eq.~(\ref{eq_RS_Hamiltonian_PML}) have to
      be calculated for each time step, which massively increases computational
      cost. However, taking a closer look at Eqs.~(\ref{eq_RS_PML_convolution_zeta})
      and (\ref{eq_RS_PML_convolution_xi}), we notice that the two functions $
      \zeta^{(k)}(t-\tau) $ and $ \xi^{(k)}(t-\tau) $ contain exponential factors.
      Therefore it is possible to obtain a rather accurate approximation of
      the terms by using a recursive-convolution method
      \cite{Luebbers_Hunsberger} with finite time steps $ \Delta t $.
      The recursive-convolution method allows to express integrals of the form
      \begin{align}
        g(t) = \int \limits_{0}^{t} \md \tau \me^{- \alpha (t-\tau)} f_1(t-\tau) f_2(\tau)
      \end{align}
      in terms of
      \begin{equation}
        \begin{alignedat}{2}
             g(m \Delta t)
          &= \int \limits_{0}^{m \Delta t} \md \tau \me^{- \alpha ((m+1) \Delta t -\tau)}
             f_1((m+1) \Delta t - \tau) f_2(\tau)                                                                                                    \\
          &\hspace{-10mm} = \me^{- \alpha \Delta t} 
             \underbrace{ \hspace{-2mm} \int \limits_{0}^{(m-1) \Delta t} \hspace{-1mm} \md \tau
             \me^{- \alpha ((m-1) \Delta t -\tau)} f_1( (m-1) \Delta t - \tau) f_2(\tau) }_{g((m-1) \Delta t)}                                       \\
          &  + \hspace{-4mm} \int \limits_{(m-1) \Delta t}^{m \Delta t} \hspace{-5mm} \md \tau \me^{- \alpha ((m-1) \Delta t -\tau)}
             f_1((m-1) \Delta t - \tau) f_2(\tau)                                                                                                    \\[4mm]
          &\hspace{-10mm} = \me^{-\alpha \Delta t} g((m-1) \Delta t)                                                                                       \\
          &  + \hspace{-4mm} \int \limits_{(m-1) \Delta t}^{m \Delta t} \hspace{-5mm} \md \tau \me^{- \alpha ((m+1) \Delta t -\tau)}
             f_1((m+1) \Delta t - \tau) f_2(\tau)   \honemm ,
        \end{alignedat}                                                                                      \label{eq_convolution_recursive}
      \end{equation}
      where we have taken $ t = m \Delta t $.
      For finite yet sufficiently small time steps $ \Delta t $, it can be
      assumed that the function $ f_2(\tau) $ in the last integral term on the
      right-hand side of equation (\ref{eq_convolution_recursive}) is constant.
      This allows to take $ f_2((m-1) \Delta t) $ outside of the integral. In
      the next step, we substitute $\alpha = \sigma$ and the functions $ f_1(t) $ and $f_2(t) $ with
      the corresponding ones in Eq.~(\ref{eq_RS_PML_convolution_zeta}) and Eq.~(\ref{eq_RS_PML_convolution_xi}).
      Therefore, we obtain $ f_{1,+,l}(t), f_{2,+,l}(t) $ for Eq.~(\ref{eq_RS_PML_convolution_zeta})
      and $ f_{1,-,l}(t), f_{2,-,l}(t) $ for Eq.~(\ref{eq_RS_PML_convolution_xi}) with
      \begin{align}
        f_{1,\pm,l}(t) &= - \frac{1}{2} \sigma^{(l)} \left( 1 \pm \frac{\mu_0}{\epsilon_0} \me^{- (\mu_0/\epsilon_0-1) \sigma^{(l)} t} \right)   \honemm ,
      \end{align}

      \begin{align}
        f_{2,\pm,l}(t) &= \partial_{l} F_{\pm}^{l}(\vrt)   \honemm .
      \end{align}
      However, the above
      recursive convolution applied to Eqs.~(\ref{eq_RS_Hamiltonian_PML}) and
      (\ref{eq_RS_Hamiltonian_PML_convolution}) does not allow to express the term $ \hat{\boldsymbol{\mathcal{G}}} \boldsymbol{\mathcal{F}} $ in
      Eq.~(\ref{eq_RS_Hamiltonian_PML}) as a matrix vector multiplication with an
      approximated matrix $ \hat{\mathcal{G}} $ and vector $ \mathcal{F} $. Nevertheless,
      it is possible to replace the whole $ \hat{\boldsymbol{\mathcal{G}}} \boldsymbol{\mathcal{F}} $ term by a 6x6
      dimensional matrix, denoted as $ \tilde{\mathcal{G}} $
      \begin{align}
          \tilde{\mathcal{G}}(m \Delta t)
        = - \mi \hbar c_0
          \begin{pmatrix}
            \hspace{1mm} \tilde{g}_{k,l} (m \Delta t) & \tilde{g}_{k,l'} (m \Delta t)   \\
            \hspace{1mm} \tilde{g}_{k',l}(m \Delta t) & \tilde{g}_{k',l'}(m \Delta t)
          \end{pmatrix}   \honemm .
      \end{align}
      The matrix $ \tilde{\mathcal{G}} $ contains four 3x3 dimensional
      matrices, which are defined recursively and depend on the current $ t = j
      \Delta t $ and the previous time $ t' = (j-1) \Delta t $. These recursive
      matrices are given by
      \begin{equation}
        \begin{alignedat}{2}
             \tilde{g}_{k,l}(m \Delta t)
          &= a_{k} \tilde{g}_{k,l}((m \hspace{-1mm} - \hspace{-1mm} 1) \Delta t) \delta_{kl}                 \\
          & \qquad - b_{+,k} \epsilon^{qlp} \partial_p F_{+}(\vec{r},m \Delta t) \delta_{kq}                 \honemm ,  \\[3mm]
             \tilde{g}_{k,l'}(m \Delta t)
          &= a_{k} \tilde{g}_{k,l'}((m \hspace{-1mm} - \hspace{-1mm} 1) \Delta t) \delta_{kl'}               \\
          & \qquad - b_{+,k} \epsilon^{ql'p} \partial_p F_{+}(\vec{r},m \Delta t) \delta_{kq}                \honemm ,  \\[3mm]
             \tilde{g}_{k',l}(m \Delta t) 
          &= -a_{k'} \tilde{g}_{k',l}((m \hspace{-1mm} + \hspace{-1mm} 1) \Delta t) \delta_{k'l}             \\
          & \qquad - b_{-,k'} \epsilon^{qlp} \partial_p F_{-}(\vec{r},m \Delta t) \delta_{k'q}               \honemm ,  \\[3mm]
             \tilde{g}_{k',l'}(m \Delta t)
          &= -a_{k'} \tilde{g}_{k',l'}((m \hspace{-1mm} + \hspace{-1mm} 1) \Delta t) \delta_{k'l'}           \\
          & \qquad - b_{-,k'} \epsilon^{qlp} \partial_p F_{-}(\vec{r},m \Delta t) \delta_{k'q}               \honemm .
        \end{alignedat}
      \end{equation}

      The auxiliary variables $ a_{k} $ and $ b_{\pm,k} $ result from the last
      line of Eq. (\ref{eq_convolution_recursive}).  Adapting them to the
      corresponding PML equation yields

      \begin{align}
        a_{k} = \me^{- \sigma^{(k)} \Delta t}   \honemm ,
      \end{align}

      \begin{align}
        b_{\pm,k} = \frac{1}{2} \me^{-2 \sigma \Delta t} \left( 1 - \me^{\sigma \Delta t} \right) 
                    \pm \frac{1}{2} \me^{-2 \tfrac{\mu_0}{\epsilon_0} \sigma \Delta t } \left( 1 - \me^{\tfrac{\mu_0}{\epsilon_0} \sigma \Delta t} \right)
      \end{align}

      Collecting all steps, we can express the Maxwell Riemann-Silberstein
      Hamiltonian $ \hat{\mathcal{H}}_{\mathrm{PML}} $ from Eq. 
      (\ref{eq_RS_Hamiltonian_PML}) in discretized form
      
      \begin{align}
          \hat{\bar{\mathcal{H}}}_{\mathrm{PML}}(\vec{r},\vec{r} \hspace{0.5mm}' \hspace{-1mm},m \Delta t, t')
        = \hat{\bar{\mathcal{H}}}_{(0)} + \hat{\mathcal{K}}(\vec{r},\vec{r} \hspace{0.5mm}' \hspace{-1mm}, m \Delta t, t')
          + \tilde{\mathcal{G}}(\vec{r}, m \Delta t)  \honemm .       \label{eq_RS_Hamiltonian_PML_numerical}
      \end{align}

      With this definition it is then straightforward to insert the above PML
      expression $ \hat{\bar{\mathcal{H}}}_{\mathrm{PML}} $ in the Maxwell propagator $
      \bm{\hat{\mathcal{U}}} $ of the numerical propagation equations in Eq.
      (\ref{eq_RS_time_evolution_equation_recursive_numerical}) or Eq.
      (\ref{eq_RS_time_evolution_equation_recursive_numerical_etrs}) to enable
      the simulation of open quantum systems via PML absorption. 

      In the last step, we have to determine the conductivity $ \sigma_{(u)} $
      adequately to get an optimal PML.  In FDTD, several useful profiles for
      the conductivity $ \sigma_{(u)} $ were found and we have chosen for our
      Riemann-Silberstein PML the FDTD polynomial grading profile which has the
      form \cite{taflove2005computational}

      \begin{align}
        \sigma_{(u)}(u) = \left( \frac{|u| - b_{u}}{L_{u} - b_{u}} \right)^{q} \sigma_{(u),\mathrm{max}}
      \end{align}

      with direction coordinate $ u \in (x,y,z) $ where $ b_u $ and $ L_u $
      denote the corresponding boundary dimensions in Fig.
      \ref{fig_boundaries}. The last variable $ \sigma_{\mathrm{max}} $ for
      the grading profile is determined by 

      \begin{align}
          \sigma_{(u),\mathrm{max}}
        = - \frac{\epsilon (q+1) \mathrm{ln}(R(0))}{2 \mu (L_{u} - b_{u})}   \honemm,
      \end{align}

      where the tolerated reflection error for normal angle incidence equal to
      zero can be set manually. The parameter $ q $ for the grading profile
      is usually set in a range between $ 2.0 \le q \le 4.0 $ to get numerically sufficient
      absorbance.

    \subsubsection{Incident waves boundaries}   \label{subsubsec_incident_waves_boundaries}

\begin{figure}[h!]
        \center
        \begin{minipage}{0.46 \textwidth}
          \includegraphics[scale=0.6]{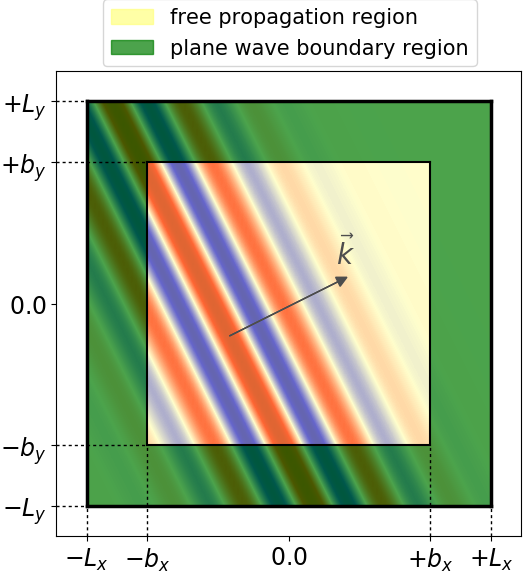}
          \caption{The figure shows an incident wave simulation where Dirichlet boundary conditions
                   are imposed in the green boundary area. In the shown example, an electromagnetic plane wave is approaching
                   the simulation box. The analytically known values for electric and magnetic fields 
                   of the plane wave are updated in the boundary region in each time step and in the
                   interior region of the simulation box we solve Maxwell's equations in Riemann-Silberstein
                   form numerically.}
          \label{fig_plane_wave_boundaries}
        \end{minipage}
      \end{figure}

      A very frequently found experimental situation corresponds to cases where
      the matter system is a molecule or nanoparticle on the scale
      of a few tens or hundreds of {\AA}ngtr{\"o}ms, while the incident laser
      fields are typically in the infrared, optical, or ultraviolet range. This
      corresponds to a spatial extension of the optical waves which is a few
      orders of magnitude larger than the matter system. Simulating both, radiation
      and matter, in the same simulation box, would therefore require boxes
      with an unfeasible size.  In addition, the used light pulse can frequently
      be approximated with a
      mathematically closed description such that the time-evolution is known 
      analytically. For such situations, it is not necessary to chose a large
      Maxwell simulation box such that the external light signal is completely
      inside the box. A convenient method to simulate such
      electromagnetic waves in our Riemann-Silberstein simulation box can be
      obtained by using the boundary region to match with the outside free evolution of,
      e.g., a laser pulse, by updating the values of the grid points
      according to the Dirichlet boundary conditions of the free evolution. In Fig.
      \ref{fig_plane_wave_boundaries}, we illustrate such a scenario. We show in a 2D cut the
      analytically calculated outer frame around the simulation box. In the interior region a
      Gaussian shaped pulse envelope propagates parallel to its wavevector $
      \vec{k} $, which is performed numerically. The transition between analytical boundary
      and numerical electromagnetic wave in the interior is seamless.
 
      In general, an arbitrary
      shaped analytical wave can be obtained by a superposition of
      different linearly independent waves with 
 
      \begin{align}
          \mathcal{F}_{\mathrm{pw}}(\vrt)
        = \sum \limits_{i} \mathcal{F}_{\mathrm{pw},(i)}
          \mathrm{exp} \left( \mi ( - k_{(i)}^{l} r_{l} - \omega_{(i)} t ) \right)   \honemm ,
      \end{align}

      where the $ i^{th} $ wave is represented by its wavevector $ \vec{k}_{(i)} $ and frequency
      $ \omega_{(i)} $ and by a Riemann-Silberstein vector $
      \mathcal{F}_{\mathrm{pw},(i)} $ as initial vector.  We select the width of the boundary region where
      this incident analytical wave is prescribed as Dirichlet boundary condition
      according to the number of the grid points that are used for the stencil
      of the finite-difference discretization. In this way artefacts at the
      interface of boundary region and interior region can be avoided.
 
      We emphasize here that the incident plane-wave boundary condition
      amounts to  simulating an open quantum system since energy enters the system
      through the analytically prescribed boundaries.  Time-reversal
      symmetry does not hold for open systems, especially in the presence of
      magnetic fields, and consequently the ETRS
      propagator in 
      Eq.~(\ref{eq_RS_time_evolution_equation_recursive_numerical_etrs}) does not
      hold. However, we assume in the present work that the full coupled
      Hamiltonian stays approximately time-reversal since the main breaking of the symmetry
      arises if we consider the magnetic field propagation without any
      back reaction of the matter. 
      
      The incident waves boundaries give us also the opportunity to simulate
      pump-probe experiments if we propagate two different signals with
      arbitrary angles and time delay that hit the molecule and calculate the
      resulting electromagnetic field.

    \subsubsection{Incident waves boundaries plus absorbing boundaries}   \label{subsubsec_incident_waves_plus_ab_boundaries}

\begin{figure}[h]
        \center
        \begin{minipage}{0.46 \textwidth}
          \includegraphics[scale=0.6]{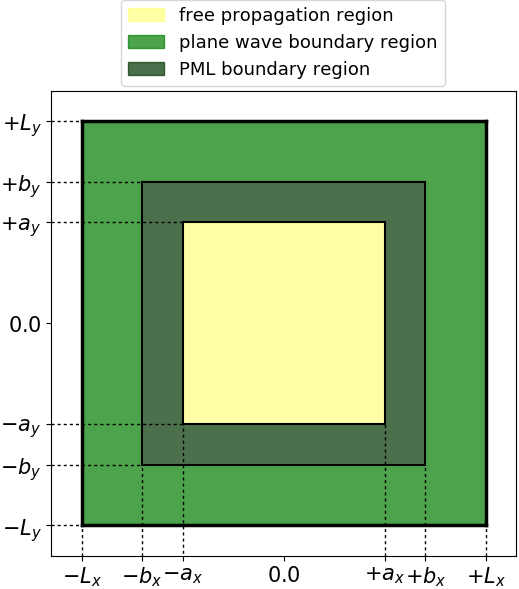}
          \caption{In the 2-dimensional cut of the simulation box with total dimensions $ L_x $, and $ L_y $, the boundary region is split
                   into two subareas with analytical waves and PML boundary conditions. The outer boundaries with the limits $ L_x $, $ b_x $, and
                   $ L_y $, $ b_y $ are the analytical waves boundaries, whereas the inner boundaries determined by $ a_x $, $ b_x $, and $ a_y $, $ b_y $
                   build the PML region.} 
          \label{fig_plane_wave_and_pml_boundaries}
        \end{minipage}
      \end{figure}

      The incident waves boundaries that we just introduced allow to simulate
      signals that enter the simulation box. On the other hand, the PML dampens all outgoing electromagnetic fields. However, for
      most purposes it is desirable to combine both boundary conditions.  Due
      to the analytical behavior of the incident waves, we know the incoming
      fields for all times. In addition, the outgoing electromagnetic signals
      should not cause any reflection at the boundaries. In the previous
      section, we have shown how the PML looks like only for the absorbing
      boundary condition. Since the Maxwell's equations for the incoming fields
      is just linear we can add it to the matter-coupled (internal) field that
      is absorbed at the boundaries. At the same time, if we have given
      the total electromagnetic field (internal and incoming fields) we can
      easily subtract the incoming field in the whole simulation box to determine
      the internal fields. This allows us to apply the PML only to the internal
      fields of the coupled light-matter system. To combine
      incident waves with absorbing boundaries, we therefore split the boundary
      region into two regions: an outer region for the incident waves,
      and an inner one for the PML as it is shown in Fig.
      (\ref{fig_plane_wave_and_pml_boundaries}).
 
      A time step is then performed such that first the freely
      propagating wave that arises from the incident wave is subtracted from
      the numerically propagated Riemann-Silberstein vector. Next, the PML is
      applied to the remaining field.  In the last step, the freely propagating wave
      is added again to the field. We note, that there are two options to compute
      the values of the incident field. As first option, one uses simply the
      analytical wave values for the current time step. The second option
      requires a second auxiliary propagation, where the incident wave is
      propagated on the numerical grid, however, without any coupling to the
      matter. In other words, in this case, the free wave propagation is
      calculated by the discretized Maxwell time-evolution operator. This
      method avoids numerical reflection artefacts at the boundary due to the
      fact that there are always small numerical discrepancies between the analytical
      wave and a numerically calculated one.

%
 
      \begin{figure}[h]
        \center
        \begin{minipage}{0.46 \textwidth}
          \includegraphics[scale=0.6]{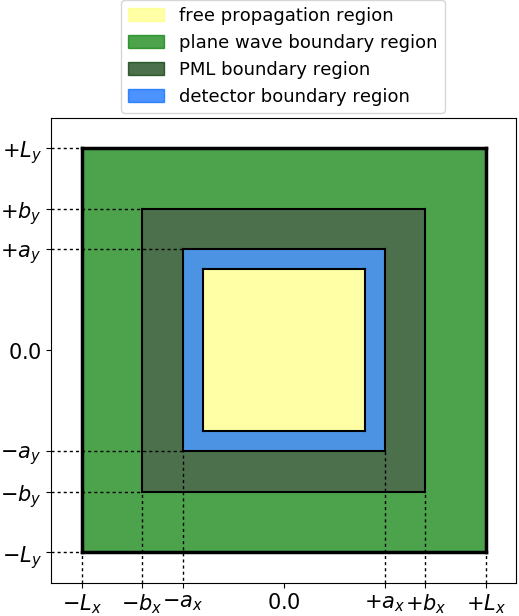}
          \caption{A small blue closed area at boundaries of the inner free Maxwell propagation area illustrates the detector region, where the included
                   grid point values are used to analyze the Maxwell far-field.}
          \label{fig_plane_wave_and_pml_boundaries_and_detector}
        \end{minipage}
      \end{figure}

  \subsection{Time-evolution operator for Kohn-Sham orbitals} \label{subsec_time_evolution_Kohn_Sham_orbitals}

    Since various time-stepping schemes for the time-dependent Kohn-Sham
    equations are already well established and have been extensively discussed
    in the literature \cite{Castro2004,Pueyo2018}, we here only
    briefly summarize the basic equations and introduce the corresponding
    notation.\\ According to Eq.~(\ref{eq_RN26_a}), the one-particle
    MPKS Hamiltonian takes the
    form
    \begin{equation}
      \begin{alignedat}{2}
           \hat{h}_{\mathrm{MPKS}}^{(n)}
        &= - \frac{1}{2 M_{(n)}} \tilde{\mathpzc{P}}_{(n)}^k \tilde{\mathpzc{P}}_{(n),k}
          + q_{(n)} \left( a^{0} + a_{\mathrm{Mxc}}^{0} \right)                                                        \\
        & \quad - \frac{q_{(n)} \hbar}{2 M_{(n)} c} 
          S^{(n)}_{k} \left[\epsilon^{klp} \partial_{l}
          \left( a_{p} + a^{\mathrm{Mxc}}_{p} \right) \right]  \honemm ,
      \end{alignedat}                                                                                                  \label{eq_RN26_b}
    \end{equation}
    where the canonical momentum is given by
    \begin{align*}
        \tilde{\mathpzc{P}}_{(n)}^k
      = - \mathrm{i} \hbar \partial^{k} + \frac{q_{(n)}}{c}
        \left( a^{k} + a^{k}_{\mathrm{Mxc}} \right)   \honemm .
    \end{align*}
    Summing these one-particle non-interacting Hamiltonians $
    \hat{h}_{\mathrm{MPKS}}^{(n)}(\vec{r}_i) $, gives the non-interacting
    many-particle MPKS Hamiltonian for the species $n$, i.e.,
    \begin{equation}
      \hat{H}_{\mathrm{MPKS}}^{(n)} = \sum \limits_i \hat{h}_{\mathrm{MPKS}}^{(n)}(\vec{r}_i)    \honemm .                        \label{eq_MPKS_hamiltonian}
    \end{equation}
    The time evolution of the Kohn-Sham orbitals $ \phi_{(n,i)} $ from
    starting time $ t_0 $ to time $ t $ is given by
    \begin{align}
      \phi_{(n,i)}(\vec{r}, t) = \hat{u}_{\mathrm{MPKS}}^{(n)}(t,t_0) \phi_{(n,i)}(\vec{r}, t_0),      \label{eq_matter_time_evolution_equation}
    \end{align}
    where $\hat{u}_{\mathrm{MPKS}}^{(n)}(t,t_0)$ denotes the corresponding time-ordered MPKS
    time-evolution operator for the species $n$
    \begin{align}
      \hat{u}_{\mathrm{MPKS}}^{(n)}(\vec{r},t,t_0) = \hat{\mathcal{T}}\mathrm{exp}
      \left[ - \mi \int \limits_{t_0}^{t} \md \tau \hat{h}_{\mathrm{MPKS}}^{(n)}(\vec{r}, \tau) \right].    \label{eq_matter_time_evolution_operator}
    \end{align}
    Based on this, the Kohn-Sham wave function evolves in time according to
    \begin{align}
    \Phi^{(n)}(t) = {\hat U}^{(n)}(t,t_0) \Phi^{(n)}(t_0),
    \end{align}
    where
    \begin{align}
    {\hat U}^{(n)}(t,t_0)= \otimes_{q=1}^{N^{(n)}} \hat{u}_{\mathrm{MPKS}}^{(n)}(\vec{r}_q,t,t_0),
    \end{align}
    and $N^{(n)}$ denotes the number of occupied Kohn-Sham orbitals for species ${(n)}$.
    Note that the evolution operators do not need to be (anti-)symmetrized, since the 
    symmetry of the initial state is kept.


    For our coupled Maxwell-Kohn-Sham system, we want to use for both, matter
    and radiation, consistent time-evolution techniques. We therefore select,
    as in the case of the Riemann-Silberstein time-stepping, the ETRS
    time-evolution operator to propagate the Kohn-Sham orbitals.  The numerical
    ETRS time-evolution equation for the $ (m + 1) \Delta t $ time step of the
    Kohn-Sham orbital $ \phi_{(n,i)}(\vec{r}, (m+1) \Delta t) $ is given by
    \begin{equation}
      \begin{alignedat}{2}
          \phi_{(n,i)} \hspace{-0.5mm} (\vec{r}, \hspace{-0.5mm} (m \hspace{-0.5mm} + \hspace{-0.5mm} 1) \Delta t) &  \\
       & \hspace{-10mm} = \hat{u}_{\mathrm{MPKS}}^{\mathrm{ETRS},(n)} \hspace{-0.5mm}
          ( \hspace{-0.5mm} (m \hspace{-0.5mm} + \hspace{-0.5mm} 1) \Delta t,
            \hspace{-0.5mm} m \Delta t) \phi_{(n,i)}(\vec{r}, \hspace{-0.5mm} m \Delta t)            \label{eq_matter_time_evolution_equation_recursive_etrs}
      \end{alignedat}
    \end{equation}
    where we have introduced the ETRS time-evolution operator\cite{Castro2004} 
    \begin{equation}
      \begin{alignedat}{2}
          \hat{u}_{\mathrm{MPKS}}^{\mathrm{ETRS},(n)}((m \hspace{-0.8mm} + \hspace{-0.8mm} 1) \Delta t, m \Delta t) =                                \\
        & \hspace{-45mm} \mathrm{exp} \hspace{-0.5mm}
          \bigg[ \hspace{-1.0mm} - \mi \frac{\Delta t}{2} \hat{h}_{\mathrm{MPKS}}^{(n)}( (m \hspace{-0.8mm} + \hspace{-0.8mm} 1) \Delta t) \bigg]
          \mathrm{exp} \hspace{-0.5mm} 
          \bigg[ \hspace{-1.0mm} - \mi \frac{\Delta t}{2} \hat{h}_{\mathrm{MPKS}}^{(n)}(m \Delta t) \bigg]   \honemm .
      \end{alignedat}                                                                              \label{eq_matter_time_evolution_operator_recursive_etrs}
    \end{equation}

    Similar to the Maxwell time-evolution, the time step parameter $ \Delta t $
    has to be selected to yield a stable and accurate propagation.
    However, in contrast to the Maxwell system, there is no CFL criterion for
    the Kohn-Sham evolution since the speed of matter waves in our
    non-relativistic setting is not capped by the speed of light. Nevertheless,
    since we restrict our considerations to low-energy problems where the
    non-relativistic approximation is justified, by construction the Kohn-Sham
    orbitals evolve much slower than the speed of light.  Hence, the maximum
    $ \Delta t $ is in most cases much larger than the one for the Maxwell fields 
    (for more details see the different case studies presented below).

  \subsection{Full minimal coupling and multipole expansion}  \label{subsec_multipole_expansion}

    In this section we discuss the different levels of theory that can
    be used to couple light and matter. \\
    Full minimal coupling 
    extends the Dirac equation to include the coupling to an
    electromagnetic field, taking both, Lorentz- and gauge-invariance into account. 
    Although we partially violate the Lorentz-invariance property in our
    non-relativistic limit that leads to the MPKS approach based on the
    Hamiltonian in Eq.~(\ref{eq_RN26_b}), it still represents
    an accurate approximation provided the kinetic-energies of the particles are well
    below relativistic values. In that case higher-order relativistic corrections to
    the Pauli equation become important. Such semi-relativistic version of the
    Pauli-Fierz Hamiltonian exist
    and can then be used instead. But for most
    applications we envision, the MPKS approach should be sufficiently accurate.
    Let us in the following see how we get from this full minimal-coupling
    description to simplified couplings that are often employed and that we want
    to use later to investigate the influence of these approximations on physical
    processes and observables.\\
    As first step, we separate the Hamiltonian of Eq.~(\ref{eq_MPKS_hamiltonian}) into a kinetic
    Hamiltonian $ \hat{H}_{\mathrm{kin}} $ and an interacting Hamiltonian $
    \hat{H}_{\mathrm{int}} $ which includes Maxwell and matter variables  

    \begin{align}
        \hat{H}_{\mathrm{MPKS}}^{(n)}
      = \hat{H}_{\mathrm{MPKS}}^{\mathrm{kin},(n)} 
      + \hat{H}_{\mathrm{MPKS}}^{\mathrm{int},(n)}
      = \sum \limits_{i} \hat{h}_{\mathrm{MPKS}}^{\mathrm{kin},(n)}
        + \sum \limits_{i} \hat{h}_{\mathrm{MPKS}}^{\mathrm{int},(n)}   \honemm ,        \label{eq_MPKS_hamiltonian}
    \end{align}

    where the kinetic piece is given by
    \begin{equation}
      \begin{alignedat}{2}
          \hat{h}_{\mathrm{MPKS}}^{\mathrm{kin},(n)}
        = \frac{\hbar^2}{2 M_{(n)}} \partial^{k} \partial_{k}   \honemm ,
      \end{alignedat}                                                                           \label{eq_MPKS_kinetic_hamiltonian}
    \end{equation}

    and the light-matter coupling is contained in 
    \begin{equation}
      \begin{alignedat}{2}
           \hat{h}_{\mathrm{MPKS}}^{\mathrm{int},(n)}
        &= \frac{- \mi \hbar q_{(n)}} {M_{(n)}} a_{\mathrm{KS}}^{k} \partial_{k}
           + \frac{q_{(n)}^{2}}{2 M_{(n)}} a_{\mathrm{KS}}^{k} a_{\mathrm{KS},k}
           + q_{(n)} a_{\mathrm{KS}}^{0}                                                        \\
        & \quad - \frac{q_{(n)} \hbar}{2 M_{(n)} c_0} 
           \sigma_{k} \epsilon^{klm} \partial_{l} a_{\mathrm{KS},m}    \honemm .
      \end{alignedat}                                                                            \label{eq_MPKS_interaction_hamiltonian}
    \end{equation}
    
    The total vector potential  $ a_{\mathrm{KS}}^{\mu} = a^{\mu} + A^{\mu} + a^{\mu}_{\rm{xc}} $ in
    Eq.~(\ref{eq_MPKS_interaction_hamiltonian}) is determined by the
    Riemann-Silberstein vector via
    Eq.~(\ref{eq_vector_potential_via_spatial_integral}) or
    (\ref{eq_vector_potential_via_time_integral}).  We note here, that the
    total vector potential, especially the scalar potential component $ a^{0}_{\mathrm{KS}} $,
    in principle includes all electronic and nuclear potentials.\\ 
    
    In many applications, the full minimal coupling is expensive to calculate
    or not needed since the length scales of matter and radiation differ
    vastly. Therefore, the minimal coupling is often approximated by a
    multipole expansion using the electric and magnetic fields variables. As is
    well-known, the ubiquitous electric dipole approximation is equivalent to
    the lowest order term of this multipole expansion.  Since this type of
    multipole approximations are ubiquitous in quantum physics it is very
    interesting to investigate how observables change order-by-order.  In the
    following, we briefly summarize the derivation of the multipole expansion
    based on the Power-Zienau-Woolley transformation
    (cf. chapter 5.2 of Ref.~\cite{loudon1988}) and adapt it 
    to the present MPKS case. As a first step, we introduce the
    polarization 
%
%
%
    \begin{align}
      \vec{P}^{(n)}(\vec{r}) =  \frac{q_{(n)}}{c_0} \sum \limits_{i} \vec{r}_{i} \int \limits_{0}^{1} \delta (\vec{r} - \alpha \vec{r}_{i}) \md \alpha   \honemm .
    \end{align}

    \noindent
    In Coulomb gauge with $ \vec{\nabla} \cdot \vec{a}_{\mathrm{KS}} = 0 $, the vector
    potential is always transversal and hence the  unitary Power-Zienau-Woolley
    transformation $ \hat{U}_{\mathrm{PZW}} $ is defined by

    \begin{equation}
    \label{eq:PZW}
      \begin{alignedat}{2}
           \hat{U}_{\mathrm{PZW}} 
        &= \mathrm{exp} 
           \left[ 
             \frac{\mi}{\hbar} \int \md^3r\vec{P}_{\perp}^{(n)}(\vec{r}) \cdot \vec{a}_{\rm{KS}}(\vec{r}, t)
           \right]            \\
        &= \mathrm{exp}
           \left[
              \frac{\mi q_{(n)}}{\hbar c_0} 
             \sum \limits_{i} \int \limits_{0}^{1}
             \vec{r}_{i} \cdot \vec{a}_{\rm{KS}}(\alpha \vec{r_i}, t) \md \alpha
           \right]    \honemm .
      \end{alignedat}
    \end{equation}

    \noindent
    By then using the MPKS equation, however, transformed with the unitary operator $\hat{U}_{\rm{PZW}}$ for the wave function $\Phi'^{(n)} = \hat{U}_{\mathrm{PZW}}^{-1} \Phi^{(n)}$, leads together with 
    \begin{equation}
      \begin{alignedat}{2}
          \hat{U}_{\mathrm{PZW}}^{-1} \vec{\nabla}_i \hat{U}_{\mathrm{PZW}} = \vec{\nabla}_i + \frac{\mi q_{(n)}}{\hbar c_0} \int \limits_{0}^{1}
          \left(\vec{\nabla}_i \vec{r}_{i} \cdot \vec{a}_{\rm{KS}}(\alpha \vec{r_i}, t) \md \alpha\right)\honemm ,
      \end{alignedat}
    \end{equation}
    as well as an extra term due to $\partial_t \hat{U}_{\mathrm{PZW}}^{-1} (t)$ to
    \begin{equation}
      \begin{alignedat}{2}
          \hspace{-2mm} \mi \hbar \partial_t \phi'_{(n,i)}(\vec{r}s_{(n)},t) &    \\
        & \hspace{-25mm} =\left\{ \frac{1}{2M_{(n)}}
          \left( - \mi \hbar \vec{\nabla} + \frac{q_{(n)}}{c_0}
          \int_0^1 \alpha \vec{r}\times\vec{b}_{\rm{KS}}(\alpha \vec{r},t) \rm{d} \alpha \right)^2     \right. 
          \\
        & \hspace{-25mm} +q_{(n)}a^0_{\rm{KS}}(\vec{r},t) + q_{(n)}\int_{0}^{1} \vec{r}\cdot\vec{E}_{\perp, \rm{KS}}(\alpha \vec{r},t) \rm{d} \alpha  
          \\
        & \hspace{-25mm} \left.- \frac{q_{(n)} \hbar}{2 M_{(n)}}\vec{S}^{(n)}\cdot \vec{b}_{\rm{KS}}(\vec{r}s_{(n)},t) \right\} \phi'_{(n,i)}(\vec{r},t).
      \end{alignedat}
      \label{eq:PWZMPKS}
    \end{equation}

    \noindent
    In this form it becomes easy to perform a multipole expansion for the electric dipole term
%
%
%
    \begin{align}
        \hat{h}_{\mathrm{MPKS}}^{\mathrm{ED},(n)}(\vec{r},\vec{r}_{0})
      &= q_{(n)} \hspace{0.5mm} \vec{r} \cdot \vec{E}_{\perp, \rm{KS}}(\vec{r}_{0},t) 
    \end{align}
 
    \noindent
    the magnetic dipole term 
 
    \begin{align}
           \hat{h}_{\mathrm{MPKS}}^{\mathrm{MD},(n)}(\vec{r},\vec{r}_{0})
        &= 
              \mathrm{i} \frac{q_{(n)} \hbar}{M_{n} c_0} \vec{b}_{\rm{KS}}(\vec{r}_{0},t)
             \cdot \left( \vec{r}_{i} \times \vec{\nabla} \right)
       \label{eq_magnetic_dipole_hamiltonian}
    \end{align}
 
    \noindent
    and the electric quadrupole term
 
    \begin{align}
         \hat{h}_{\mathrm{MPKS}}^{\mathrm{EQ},(n)}(\vec{r},\vec{r}_{0})
      &= \frac{1}{2} q_{(n)}
         \Big( \vec{r} \cdot \vec{\nabla} \Big) 
         \vec{r} \cdot \left( \vec{E}_{\perp,\rm{KS}}(\vec{r},t) \right) \bigg|_{\vec{r}=\vec{r}_{0}}    \\
       \label{eq_electric_quadrupole_hamiltonian}
    \end{align}
    all expanded around a point $\vec{r}_{0}$, which can be chosen
    in good approximation either as center of mass or center of charge of the
    matter system. We note that we here used a multipole expansion based on
    classical (Kohn-Sham) fields. Without time-dependence in the
    classical-field case no electric transversal field can appear. If we
    instead performed the Power-Zienau-Woolley transformation on the quantized
    level (so the field in Eq.~\eqref{eq:PZW} is a time-independent operator)
    two major differences appear: First, instead of with electric fields we
    work with displacement fields and a novel term, the dipole self-energy term
    appears~\cite{loudon1988,rokaj2018light}. This term is physically and
    mathematically necessary. Secondly, since the fields are now (unbounded)
    operators, the Taylor expansion in its usual form is not applicable. Thus
    in the case of a Power-Zienau-Woolley transformation with quantized fields
    extra care needs to be taken~\cite{rokaj2018light,spohn2004}.

    \noindent
    For our present MPKS implementation, we have neglected higher order
    multipole terms that contain non-linear orders of Maxwell-field variables
    and the Stern-Gerlach term. Whereas the magnetic dipole term
    depends on the total magnetic field, the electric dipole and quadrupole
    terms depend only on the transverse component of the electric field.
    Consequently, we have to decompose the Riemann-Silberstein vector into its
    transverse and longitudinal components. In general, the
    Helmholtz-decomposition for the
    Riemann-Silberstein vector is

    \begin{equation}
      \begin{alignedat}{2}
           \vec{F}^{\perp}(\vrt)
        &= \vec{\nabla} \times \int \limits_{V}
           \frac{ \vec{\nabla}_{\vec{r}'} \times \vec{F}(\vec{r}\hspace{0.5mm}',t) }
                { 4 \pi | \vec{r} - \vec{r}\hspace{0.5mm}' | }
           \md \vec{r}'                                                                                      \\
        &\quad - \frac{1}{4 \pi} \oint \limits_{S} \vec{\hat{n}} \times 
           \frac{ \vec{F}(\vec{r}\hspace{0.5mm}',t) }{ 4 \pi | \vec{r} - \vec{r}\hspace{0.5mm}' | }
           \md S'
      \end{alignedat}                                                                                        \label{eq_helmholtz_decomposition}
    \end{equation}

    \noindent
    The first term on the right-hand side of Eq.
    (\ref{eq_helmholtz_decomposition}) can be computed efficiently by a Poisson
    solver since it is the solution of the Poisson equation. The second
    integral in Eq. (\ref{eq_helmholtz_decomposition}) is a surface integral
    which is necessary, if the Riemann-Silberstein vector does not vanish at
    the simulation box boundaries, such as for, e.g., periodic systems. Since the Riemann-Silberstein vector in the multipole Hamiltonian does only depend on the expansion point $\vec{r}_0 $ of
    the multipole expansion 
    and its corresponding Riemann-Silberstein vector inside the
    box, it is sufficient to calculate the values of the surface integral only
    for the stencil points that correspond to the expansion center $ \vec{r}_0
    $ of the multipole expansion. This reduces the computational cost for the
    boundary term significantly.



  \subsection{Multi-scale implementation} \label{subsec_multi_scale_implementation}
    
      In the previous sections we have already seen that the time-evolution 
      for matter and electromagnetic fields can be expressed mathematically in a very similar way.  
      While both share a similar Schr{\"o}dinger-type form, the physical
      parameters for many applications of the theory imply rather different
      length scales for light and matter.
      For example, when molecular systems interact with infrared, optical, or
      ultraviolet laser pulses, the matter wave functions are mainly
      localized in a relatively small volume with possibly rather strong
      fluctuations. In contrast, the Maxwell fields that result from typical
      experimental resonators are localized on the scale of the pulse envelope
      and are often smoother than the matter wave functions. For optical lasers,
      the length scales of radiation and matter waves differ by about two or three
      orders of magnitude.
      To handle such situations we use a finer grid
      for the matter wave functions and coarser grids for the Riemann-Silberstein vector.
      Our MPKS implementation in Octopus is however not restricted to such laser-molecule
      interaction applications. In the following, we introduce the possible combinations
      of matter and Maxwell grids that can be selected for MPKS simulations with Octopus.

    \subsubsection{Multi-grid types}

      \begin{figure}[h!]
        \begin{minipage}{0.46 \textwidth}
          \includegraphics[scale=0.60]{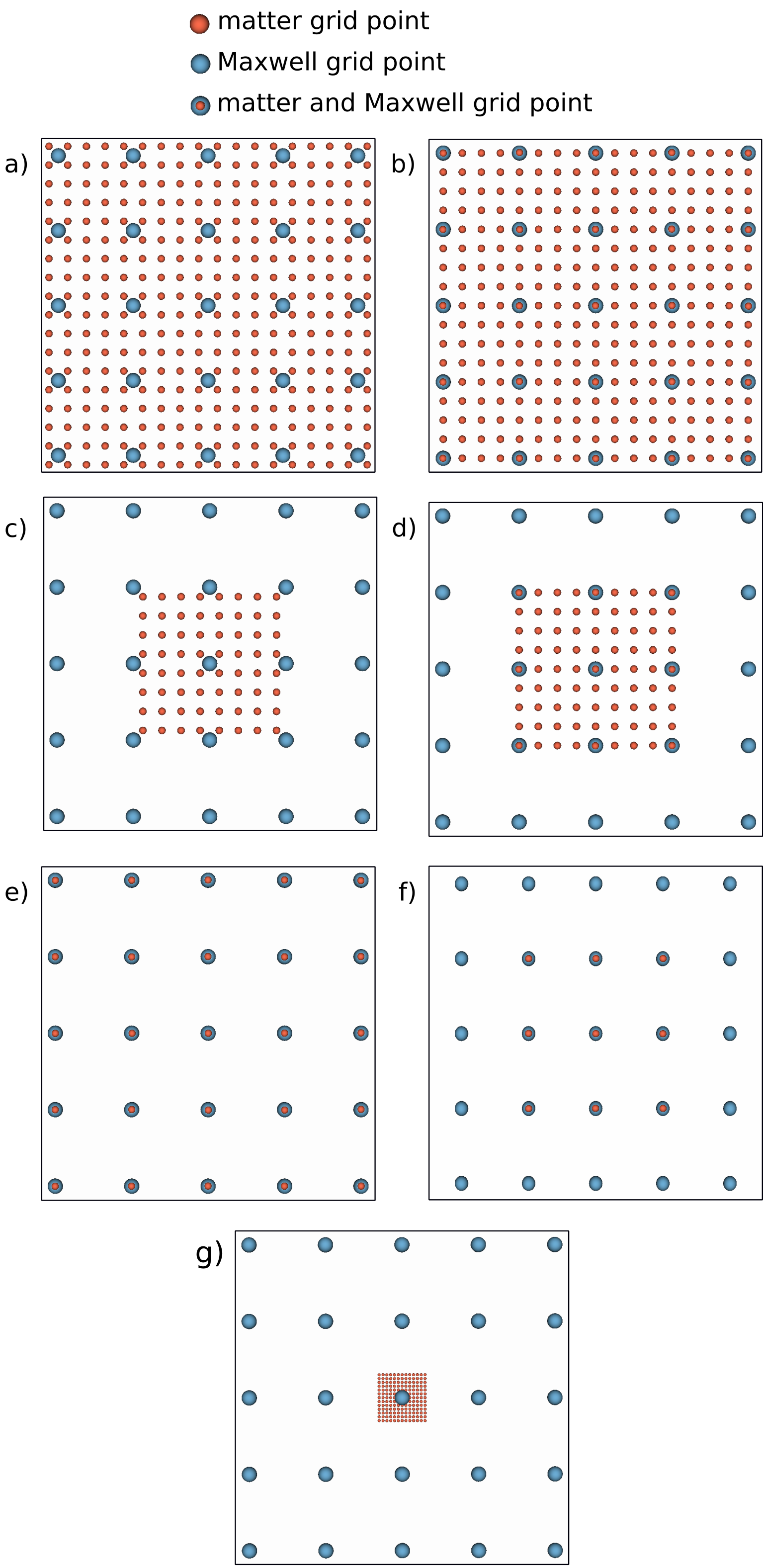}
          \caption{An overview of several possible multiscale grid types. The red dots represent grid points
                   for the Kohn-Sham variables and the blue dots are used for grid points of the Maxwell field variables. 
                   In most relevant applications, the Kohn-Sham grid is finer than the Maxwell grid, as in cases $ ( a, b, c, d ) $
                   and $ ( g ) $. The grids $ ( e ) $ and $ ( f ) $ describe the special case where both grids lie on top of
                   each other, but they do not have to be equally sized which is illustrated in $ ( f ) $. 
                   The multigrid designs where the Maxwell grid is larger than the Kohn-Sham grid $ ( c, d, f, g ) $
                   are suitable grids for bound non-periodic systems like molecules or nanoparticles, whereas only the remaining schemes
                   $( a, b, e ) $ can be used to simulate periodic systems. }
          \label{fig_multiscale_grids}
        \end{minipage}
      \end{figure}

      In Fig.~\ref{fig_multiscale_grids}, we illustrate the different
      possibilities to combine matter and Maxwell grids. We limit the graphical
      illustration here to the case when the matter grid spacing is smaller or equal
      to the Maxwell grid spacing. Our implementation is however not limited to
      these cases and can also be used for the reverse situation which is
      suitable to simulate, e.g., hard x-rays.\\
      The grids a) and b) in Fig. \ref{fig_multiscale_grids} correspond
      to simulation boxes with identical dimensions for both systems. In
      general, like in a), the Maxwell grid points do not necessarily have to
      lie on top of a matter grid point. For both grid types a) or b), it
      is not possible to use the field values at given grid points directly in
      the respective coupling terms of the propagation equations for light and
      matter.  Instead, as is used routinely for multigrid methods in
      literature, prolongation and interpolation schemes have to be used.  For
      example, the finer matter grid points have to be grouped in clusters
      which map to the next nearest Maxwell grid point and different weighted
      mean schemes can be applied to yield the coupling value for that Maxwell
      point. Vice versa, a Taylor series expansion of the Maxwell values can be
      used to determine the coupling values for the matter points. These grid
      types are well suited to simulate periodic systems.\\
      The two schemes in Figure Fig. \ref{fig_multiscale_grids} c) and d) show
      matter grid types, in the first case without common grid points, and in
      the second case with common grid points, but compared to the previous
      grids in a) and b) with smaller dimensions for the matter grid than for
      the Maxwell one.  As before, the values for the corresponding coupling
      terms have to be calculated by weighted mean and interpolation. Such grid
      combinations can describe efficiently bound molecules and nanoparticles,
      especially when focussing on electromagnetic far-fields. \\
      If near field effects are of interest, where the electromagnetic field
      fluctuations correlate strongly with the matter wave functions, it is a
      good choice to select the same grid spacings for both grids and to place
      matter and Maxwell grid points on top of each other as shown in
      Fig.  \ref{fig_multiscale_grids} e) and f).  In this case, the values for
      both respective coupling terms can be obtained directly from the field
      values at the respective grid point.  Besides near-field simulations,
      the grid type f) with larger Maxwell grid dimensions is
      suited to study the onset of the electromagnetic far-field and allows
      to define electromagnetic detectors at the box boundaries. \\
      Finally, a further grid combination is illustrated in Fig.
      \ref{fig_multiscale_grids} g), where the matter grid is chosen much finer
      than the Maxwell grid.  Only one Maxwell grid point lies in the middle of
      the matter grid. Here, it is assumed that the Maxwell field is
      approximately constant for all matter grid points. Vice versa, the
      coupling value for the Maxwell grid is obtained by the mean value of all
      matter points.

    \subsubsection{Multi-scale in time}

      So far we have focussed on different length scales and the related
      choices for spatial grid spacings and box dimensions. In this section we
      turn our attention to the different time scales and the related choices
      for numerical time steps for the matter and Maxwell propagation.\\ While
      it is in principle possible to chose identical time steps for the matter
      and Maxwell propagation, given the constraint that both propagators have
      to run stable, this is for most physical cases not the most efficient
      choice.  The reason for this can be seen directly in the Maxwell
      Hamiltonian-like form in
      Eq.~(\ref{eq_single_particle_photon_Hamiltonian}).  The gradient in this
      expression is multiplied with the speed of light $c_0$ (roughly 137 in
      atomic units). This factor imposes the speed for the electromagnetic
      waves on the grid.  On the other hand, the matter Hamiltonian in our
      non-relativistic Pauli limit is lacking the factor of $c_0$, yielding a
      much smaller spectral range of the maximum and minimum eigenvalue of the
      Hamiltonian for a given grid and stencil, and hence a much slower
      wave motion.  As consequence of the "fast" photon motion compared to the
      "slow" motion of matter, in the Maxwell case a much smaller maximum time
      step $ \Delta t_{\mathrm{Mx,max}} $ has to be selected compared to the
      maximum time step of the matter $ \Delta t_{\mathrm{KS,max}} $. \\ Such a
      situation of different physical time scales arises already in
      electron-nuclear dynamics, where the large nuclear mass leads to a rather
      slow motion of the nuclei compared to the faster motion of the lighter
      electrons. Coupling the propagation of the electromagnetic fields also
      with the electron-nuclear dynamics is adding a third time-scale to the
      problem.  In our numerical time-stepping scheme, we exploit the different
      time-scales explicitly to increase the computational efficiency.  Our
      simulations have shown, that the coupled propagation of nuclei,
      electrons, and electromagnetic fields keeps relatively accurate, stable
      and converged, if we perform several Maxwell propagation steps $ \Delta
      t_{\mathrm{Mx}} $ in-between the Kohn-Sham propagation steps for the
      electrons and Ehrenfest steps for the nuclei \cite{andrade_2009}. For convenience, we select
      the Kohn-Sham time step $ \Delta t_{\mathrm{KS}} < \Delta
      t_{\mathrm{KS,max}} $ as the basic time step parameter for the coupled
      MPKS system, and the number $ N_{\mathrm{Mx-steps}} $ of intermediate
      Maxwell steps is automatically chosen such that
      \begin{align}
        \Delta t_{\mathrm{KS}} \le N_{\mathrm{Mx-steps}} \Delta t_{\mathrm{Mx,CFL}}   \honemm ,
      \end{align}
      where $ \Delta t_{\mathrm{Mx,CFL}} $ denotes the Courant time step for
      the Maxwell grid given in equation (\ref{eq_Maxwell_courant_time}). Performing
      these intermediate Maxwell steps,
      assumes that the intermediate Maxwell propagation between $ m \Delta
      t_{\mathrm{KS}} $ and $ (m+1) \Delta t_{\mathrm{KS}} $ does not affect 
      the matter propagation significantly, which means that the current density for the
      corresponding $ i^{th} $ intermediate time step is well approximated by the linear
      expansion
      \begin{equation}
        \begin{alignedat}{2}
            \mathcal{J}(\vec{r}, m \Delta t_{\mathrm{KS}} + i \Delta t_{Mx}) =                                                                       \\
          & \hspace{-38mm} \mathcal{J}(\vec{r}, m \Delta t_{\mathrm{KS}})
            + \left[ \frac{\mathcal{J}(\vec{r}, (m + 1)\Delta t_{\mathrm{KS}}) 
            - \mathcal{J}(\vec{r}, m \Delta t_{\mathrm{KS}} )}{N_{\mathrm{Mx-steps}}} \right] i   \honemm .
        \end{alignedat}
      \end{equation}
      The ETRS time-evolution equation for the $ i^{th} $ intermediate step then
      takes according to Eq.~(\ref{eq_RS_time_evolution_inhomogeneous}) the form
      \begin{equation}
        \begin{alignedat}{2}
            \mathcal{F}(\vec{r}, t_{m,i+1}) 
          & \approx \bm{\hat{\mathcal{U}}}_{(0)}(t_{m,i+1},t_{m,i})
            \bm{\mathcal{F}}(\vec{r}, t_{m,i})                                 \\
          & \quad - \frac{\Delta t_{\mathrm{Mx}}}{4} 
            \hat{\mathcal{U}}_{(0)}(t_{m,i+1},t_{m,i})
            \mathcal{J}(\vec{r} , t_{m,i})                                \\
          & \quad - \frac{\Delta t_{\mathrm{Mx}}}{4} 
            \hat{\mathcal{U}}_{(0)}(t_{m,i+1}, t_{m,i+1/2})
            \mathcal{J}(\vec{r}, t_{m,i+1/2})                             \\
          & \quad - \frac{\Delta t_{\mathrm{Mx}}}{4} 
            \hat{\mathcal{U}}_{(0)}(t_{m,i},t_{m,i+1/2})
            \mathcal{J}(\vec{r}, t_{m,i+1/2})                             \\
          & \quad - \frac{\Delta t_{\mathrm{Mx}}}{4} 
            \mathcal{J}(\vec{r}, t_{m,i})
        \end{alignedat}                                                        \label{eq_RS_time_evolution_equation_recursive_numerical_approx_etrs}
      \end{equation}

      \noindent
      with

      \begin{equation}
        \begin{alignedat}{2}
          t_{m,i}     &= m \Delta t_{\mathrm{KS}} + i \Delta t_{\mathrm{Mx}}         \honemm ,     \\
          t_{m,i+1}   &= m \Delta t_{\mathrm{KS}} + (i+1) \Delta t_{\mathrm{Mx}}     \honemm ,     \\
          t_{m,i+1/2} &= m \Delta t_{\mathrm{KS}} + (i+1/2) \Delta t_{\mathrm{Mx}}   \honemm .
        \end{alignedat}
      \end{equation}
      
      To reduce the computational cost even further, we can assume in most
      cases that the inhomogeneity terms in equation
      (\ref{eq_RS_time_evolution_inhomogeneous}) are approximately constant for
      all intermediate time steps during the time interval $ \Delta
      t_{\mathrm{KS}} $. Thus, we use in this case for the current density $
      \mathcal{J} $ in equation
      (\ref{eq_RS_time_evolution_equation_recursive_numerical_approx_etrs}) the
      arithmetic mean of $ \mathcal{J}(\vec{r}, m \Delta t) $, and  $
      \mathcal{J}(\vec{r}, (m+1) \Delta t) $ which reduces the amount of
      necessary computational expensive $ \bm{\hat{\mathcal{U}}} $ and $ \hat{\mathcal{U}}_{(0)} $
      operations. Clearly, these simplifications introduce approximations to
      the full time-evolution and convergence has always to be checked for a given application.

     \begin{figure*}[ht!]
        \begin{minipage}{0.46 \textwidth}
          {\bf Only forward coupling}
          \includegraphics[scale=0.45]{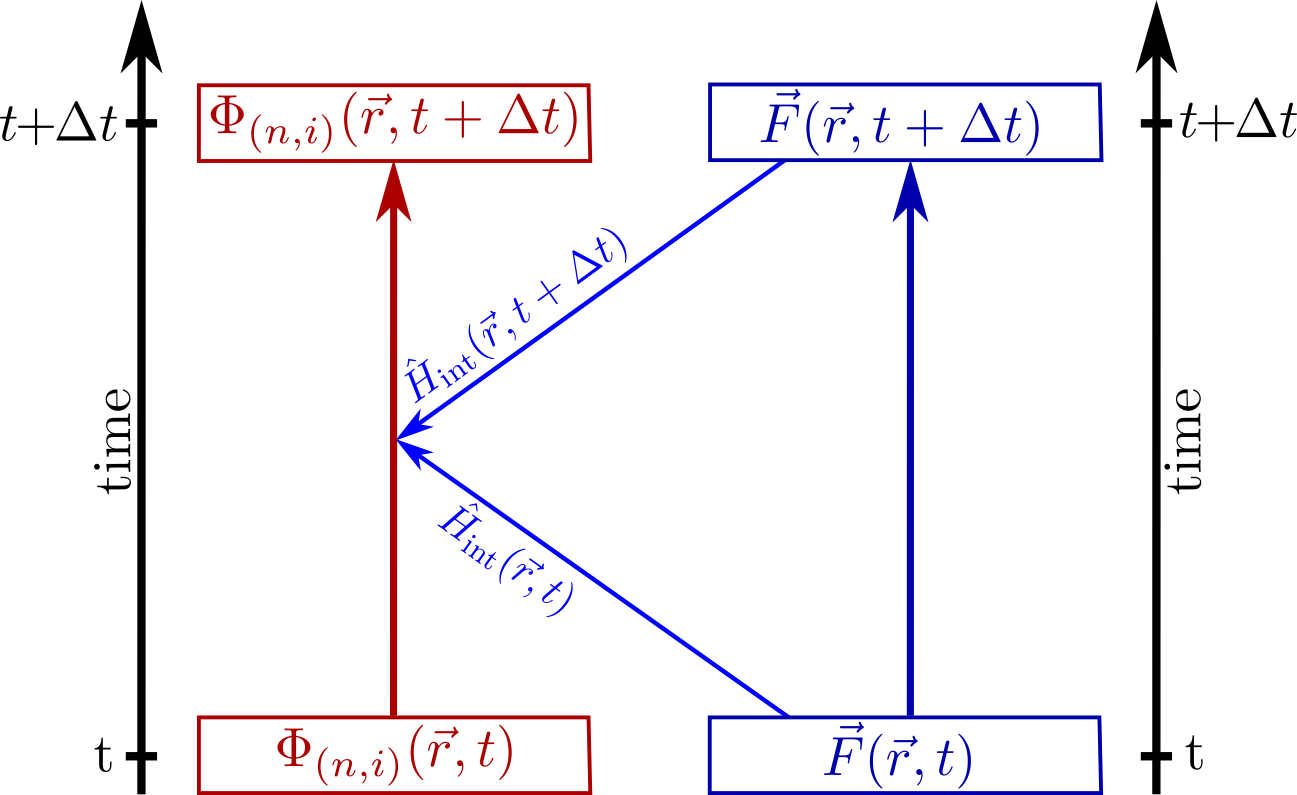}

        \end{minipage}
        \qquad
        \begin{minipage}{0.46 \textwidth}
          {\bf Self-consistent forward and backward coupling}
          \includegraphics[scale=0.45]{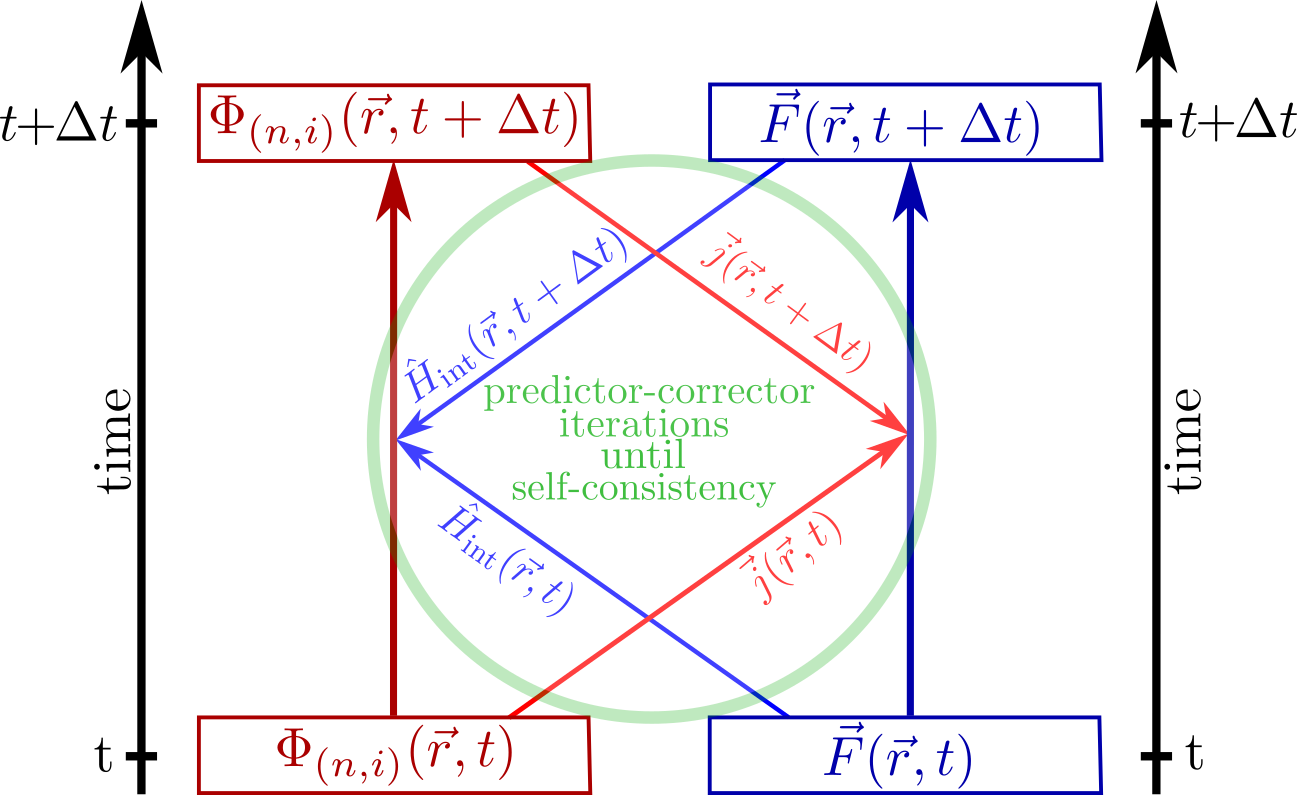}
        \end{minipage}
        \caption{
                 The figure on the left-hand side illustrates the most common
                 coupling situation for light-matter interactions in quantum
                 mechanical many-body simulations.  The electromagnetic fields
                 (in blue) propagate freely and only influence the propagation
                 of the matter (in red).  The back reaction of the matter
                 currents on the electromagnetic fields is neglected. Since the
                 electromagnetic fields are not influenced by matter currents,
                 there is not really a numerical propagation needed in this
                 case and the fields are typically taken from analytical
                 solutions of Maxwell's equations or the paraxial wave
                 equation.\\ In the figure on the right-hand side we illustrate
                 a fully self-consistent predictor-corrector scheme for a
                 coupled Maxwell-Pauli-Kohn-Sham time-stepping. As before, the
                 electromagnetic fields influence the propagation of the
                 matter (forward coupling). However, in addition, here the
                 currents from the matter propagation also influence the
                 propagation of the electromagnetic fields (backward coupling).
                 A given time-step for the matter wavefunctions and the
                 electromagnetic fields is repeated until self-consistency is
                 found (self-consistent forward-backward coupling). Only then
                 the simulation continues to perform the next time step.
                }
          \label{fig_predictor_corrector_method}
                 
      \end{figure*}

    \subsubsection{Parallelization strategies}
  
      For a speedup of the time-evolution of the Riemann-Silberstein vector, we employ a domain 
      parallelization for the real-space grid. Since we have implemented our propagation scheme 
      in Octopus by using the existing gradient operations that have been optimized for matter
      degrees of freedom, we immediately inherit the existing
      domain parallelization of the code \cite{doi:10.1002/pssb.200642067} also for the Maxwell case. A parallelization in
      "states" or "orbitals/k-points" as is used for matter wavefunctions, is not so effective
      for the Maxwell case, as there are always only six "orbitals" to propagate, namely the six
      vector components of the Riemann-Silberstein vector. We have therefore restricted ourselves for the
      calculations in the present work solely to the domain parallelization of the Maxwell grid. 

  \subsection{Predictor-corrector scheme} \label{subsec_predictor_corrector_scheme}

     So far we have described separately the propagation schemes for matter and
     electromagnetic fields as well as the coupling Hamiltonian. In the present
     section, we discuss the predictor-corrector method which we employ to
     enforce a self-consistent propagation of the subsystems.

    \subsubsection{Forward coupling}

      In most studies in the literature light-matter coupling is restricted to
      forward Maxwell-matter coupling: the electromagnetic fields influence the
      matter degrees of freedom, but the matter currents are not allowed to
      influence the propagation of the electromagnetic fields. This situation
      is illustrated on the left-hand side of
      Fig.~\ref{fig_predictor_corrector_method}.
      In the forward coupling case, the external Maxwell field propagates
      without any perturbation by the matter and is calculated  separately,
      either analytically or numerically. Looking at the time-evolution operator
      in equation (\ref{eq_matter_time_evolution_operator_recursive_etrs}), we note that the
      operator $ \hat{U}_{\mathrm{MPKS}}^{\mathrm{ETRS},(n)}((m+1) \Delta
      t_{\mathrm{KS}}, m \Delta t_{\mathrm{KS}}) $ depends on the Hamiltonian
      operator $ \hat{H}_{\mathrm{MPKS}}^{(n)}( (m+1) \Delta t_{\mathrm{KS}}) $
      at the future time $ t = (m+1) \Delta t_{\mathrm{KS}} $. This future
      Hamiltonian is not only determined by the external Maxwell fields but
      also by the motions of the ions and electrons and their interactions.
      Therefore, it is necessary to apply a predictor corrector cycle for the
      matter propagation. In a first step, the future Hamiltonian is estimated
      by an extrapolation \cite{marques2003} and the calculated time
      propagation returns an estimated Kohn-Sham potential which is again used
      for an updated extrapolation of the Hamiltonian. These steps are repeated
      until the absolute value of the variance of two subsequently Kohn-Sham
      potentials falls below a small threshold value. For our calculations, we
      set a threshold value of $ 1e^{-6} $ in atomic units for the potential
      variance and adjust the time step $ \Delta t_{\mathrm{KS}} $ for the
      propagation so that the matter system is converged in at least two
      iterations if the system is only disturbed very weakly by the external
      field.  During the full run and stronger perturbations, we notice that
      the number of iterations is barely larger than five.

    \subsubsection{Forward and backward coupling}

      The back-reaction of matter on the Maxwell fields appears due to the
      current density in equation (\ref{eq_RN25}) (in the MPKS reformulation due to its expectation value), which is caused by
      the motion of matter. The three electronic current types, paramagnetic, diamagnetic
      and magnetization current, as well as the ionic current influence the Maxwell propagation equation
      (\ref{eq_RS_time_evolution_inhomogeneous}). The influence of the
      paramagnetic current, the magnetization current and the external diamagnetic current as well as optional external
      currents result directly in the inhomogeneity $ \mathcal{J} $ term in
      equation (\ref{eq_RS_current}). The internal diamagnetic current implicitly
      influences the time-evolution due to the modified Maxwell time-evolution
      operator $ \bm{\hat{\mathcal{U}}} $ for this case given in
      (\ref{eq_RS_time_evolution_homogen_diamagnetic_current}).  The full forward and
      backward coupling scheme is shown on the right-hand side of Fig.
      \ref{fig_predictor_corrector_method}.

      \noindent
      In a fully self-consistent scheme, both systems and accordingly their
      time-evolution propagation equations in
      (\ref{eq_RS_time_evolution_inhomogeneous}) and
      (\ref{eq_matter_time_evolution_equation}) couple to each other.  First we
      apply the extrapolation of the future matter Hamiltonian to get a
      prediction for the Kohn-Sham orbitals. These orbitals and the initial
      ones give us the necessary current density which couples to the Maxwell
      fields.  Using the first predicted updated current density at time $
      (m+1) \Delta t_\mathrm{KS} $ leads to an updated Riemann-Silberstein
      vector.  At this point, the predictor-corrector loop restarts by updating
      the Kohn-Sham orbitals but now with a corrected matter Hamiltonian which
      includes the updated Riemann-Silberstein vector. As a consequence, the
      previously predicted variables get a correction closer to the values
      which make the coupled system self-consistent. We additionally check the consistency of
      the Maxwell fields by comparing the Maxwell energy inside the simulation
      box for two successive updated Riemann-Silberstein vectors. For consistency we
      use the same threshold value $1e^{-6}$ as for the matter case.
      Additionally, we chose the system propagation time $ \Delta t $ such that the
      predictor-corrector step iterates at least two times until the
      self-consistency thresholds are fulfilled for weak perturbations.  Again, from our experience
      the number of iterations for strong perturbation periods should not be
      larger than five steps. Otherwise the time steps have to be reduced.

    \begin{figure}[h!]
      \center
      \begin{minipage}{0.46 \textwidth}
        \includegraphics[scale=0.57]{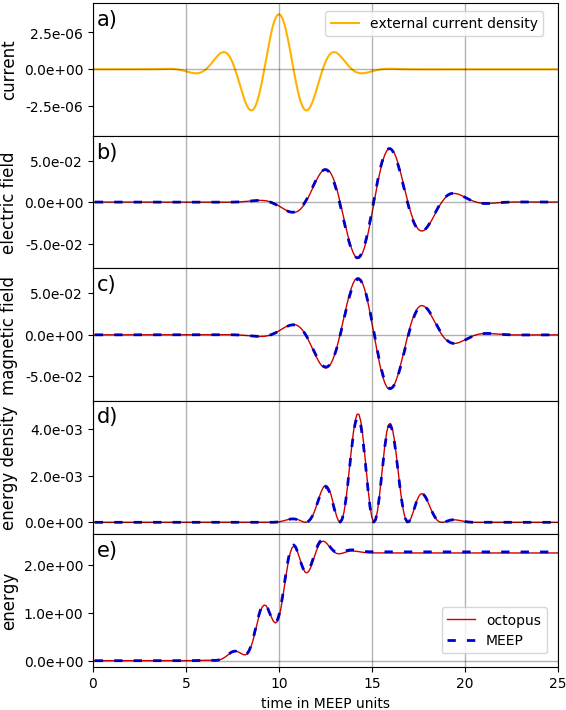}
        \caption{The figure shows a FDTD propagation performed with the electromagnetic simulation package MEEP
                 and compares with our Riemann-Silberstein implementation for Maxwell's equations in the Octopus code.
                 In both cases identical grid spacings and the same finite-difference order was chosen.
                 In panel $ a) $ we plot the external current value in x-direction at point $(5,0,0) $.
                 For the same point, the two following panels $ b) $, $ c) $ show the
                 electric field in z-direction and magnetic field in y-direction, and panel $ d) $ the corresponding
                 Maxwell energy density. The last plot in panel $ e) $ shows the integrated total energy density, i.e.
                 the Maxwell energy inside the simulation box without the PML boundary region. A very good agreement between
                 the two different propagation techniques is found.}
        \label{fig_numerical_dispersion}
      \end{minipage}
    \end{figure}

  \subsection{Validation and comparison to FDTD} \label{subsec_validation_and_comparison_to_FDTD}

    To validate our Maxwell implementation, we compare in this section our
    results directly with the MIT Electromagnetic Equation Propagation
    (MEEP) \cite{Meep_2010} package, which is a common simulation software for
    electromagnetic field propagation. The implemented Maxwell field
    propagation in MEEP is based on the FDTD
    method and the Yee-algorithm \cite{Yee_1966}.  The underlying
    electromagnetic simulation grid is split into two grids shifted by half of
    the grid spacing of the corresponding direction. As a consequence, the
    requested spatial and time derivative points for the propagation equation
    are in the middle of two adjacent grid points. Therefore, the center finite-difference method leads to the first-order derivative equation (here for simplicity
    discussed in one dimension)

    \begin{align}
        f'(x_{0})
      = \frac{f(x_{0} \hspace{-0.8mm} + \hspace{-0.8mm} \frac{\Delta x}{2}) \hspace{-0.8mm} - \hspace{-0.8mm}
              f(x_{0} \hspace{-0.8mm} - \hspace{-0.8mm} \frac{\Delta x}{2}) } {\Delta x}
        \hspace{-0.8mm} - \hspace{-0.9mm} \frac{1}{3!} f''' \hspace{-0.8mm} (x_0) \hspace{-1mm} 
        \left( \hspace{-1mm} \frac{\Delta x}{2} \hspace{-1mm} \right)^2 + \cdots   \honemm .
    \end{align}

    \noindent
    In case of the Yee grid, there are no grid points at the derivative points
    $ x_0 $ but next to it at $ x_0 - \Delta x/2 $ and $ x_0 + \Delta x/2 $. In
    our finite-difference discretization in Octopus, we also use the center
    finite-difference formula to get first-order derivatives, but the derivative
    points lie always on top of a grid point. Thus, the derivative equation
    takes the form 

    \begin{align}
        f'(x_{0})
      = \frac{f(x_{0} \hspace{-0.8mm} + \hspace{-0.8mm} \Delta x) \hspace{-0.8mm} - \hspace{-0.8mm}
              f(x_{0} \hspace{-0.8mm} - \hspace{-0.8mm} \Delta x) } { 2 \Delta x}
        \hspace{-0.8mm} - \hspace{-0.9mm} \frac{1}{3!} f''' \hspace{-0.8mm} (x_0) \hspace{-1mm} 
        \left( \Delta x \right)^2 + \cdots    \honemm ,
    \end{align}

    \noindent
    which means that the error of $ f'(x_0) $ is smaller for the Yee-algorithm
    if we consider the same order terms of the finite difference method.
    However, the MEEP finite difference stencil operation is always of order
    two whereas the Octopus stencil order can be set to higher orders to obtain
    better accuracy for the derivative operators.  
    

    As a test scenario, we simulate a spatial and temporal shaped external
    current density inside the simulation box, which is prescribed
    analytically, and plot several relevant physical variables.
    For this reference calculation, all physical units in Octopus are set equal
    to MEEP internal units. The spatial and
    time dependent current density that we impose externally takes the form

    \begin{align}
        \vec{j}(\vec{r},t)
      = \vec{e}_{z} 
        \mathrm{exp} \hspace{-1mm} \left( \frac{-x^2 \hspace{-1mm} - \hspace{-0.8mm} y^2 - \hspace{-0.8mm} z^2}{2} \right) \hspace{-0.8mm}
        \mathrm{exp} \hspace{-1mm} \left( \frac{-(t \hspace{-1mm} - \hspace{-0.8mm} t_0)^2}{2} \right) \hspace{-0.8mm}
        \mathrm{cos} \hspace{-0.8mm} \left( \omega (t \hspace{-1mm} - \hspace{-0.8mm} t_0) \right)   \honemm .    \label{eq_external_current_distribution}
    \end{align}

    \noindent
    We select a cubic simulation box of size $ 20.0 $ 
    and a PML boundary region width of $ 4.0 $ in
    each direction.  Referring to our sketch of the simulation box in Fig.
    \ref{fig_plane_wave_and_pml_boundaries_and_detector}, the corresponding
    parameters are $ L_x = L_y = L_z = 44.0 $ and $ a_x = a_y = a_z = 40.0 $.
    The grid spacing in each dimension is $ \Delta x = \Delta y = \Delta z = 0.2 $,
    which leads to a mutual time step in MEEP and Octopus of $ \Delta t = 0.1 $.
    We chose a time delay $ t_0 = 10.0 $ and a frequency $ \omega = 2.0 $. \\
    To compare our Octopus implementation to MEEP, we
    evaluate the electric field, the magnetic field, the electromagnetic energy
    density at the box point $ (5,0,0) $, and the energy inside the
    simulation box.

    Figure \ref{fig_numerical_dispersion} shows the comparison of MEEP and
    our Maxwell propagation implementation in Octopus. In the first panel $ a)
    $, we plot the initial current density $ \vec{j}(\vrt) $ in z-direction at
    point $ (5,0,0) $. It can be seen that the maximum is according to the time
    delay at time $ t = t_0 $.  Due to the spatial shape of the current
    density, the maximum value of $ \vec{j}(\vrt) $ is damped by the factor $
    e^{-5^2/2} $. The next two panels $ b) $ and $ c) $ in Fig.
    \ref{fig_numerical_dispersion} show the electric field in z-direction
    respectively the magnetic field in y-direction also both at the point $
    (5,0,0) $. Both field signals are shifted by $ \Delta t = 5.0 $ to the
    future compared to the maximum current density signal. This follows from the
    fact that the current maximum which causes the electromagnetic field
    reaction arises along the z-axis for $ x = y = 0 $ at time $ t = t_0 $.
    Hence, the
    center of the electromagnetic reaction signal arrives at $ t = t_0 + \Delta
    t = 15.0 $. We see that both simulations lead to the same electric and
    magnetic field behavior in time at the selected grid point. Furthermore, the
    Maxwell energy densities at this point calculated by Octopus and MEEP also
    match, and are plotted in Fig. \ref{fig_numerical_dispersion} $ d)
    $. The last two graphs in panel $ e) $ show the total Maxwell
    energy inside the box with $ -40.0 \le x \le 40.0 $, $ -40.0 \le y \le 40.0
    $, and $ -40.0 \le z \le 40.0 $. Also here we observe very good agreement
    with MEEP, which together with the previous cases shows the correctness of 
    our Riemann-Silberstein implementation of Maxwell's equations.

\section{Applications} \label{sec_applications}

   In this section, we demonstrate the significance of taking the fully
   self-consistent coupling of the time-dependent Kohn-Sham equations for the
   electrons, Ehrenfest equations for the nuclei, and Maxwell's equations for
   the electromagnetic fields into account.  We use our EMPKS implementation in
   the Octopus code that we introduced in the previous sections to study
   several different coupling scenarios. These range from conventional forward
   light-matter coupling in dipole approximation with clamped nuclei to
   a theory level with
   forward-backward self-consistent light-matter coupling including electric
   quadrupole and magnetic dipole terms where we also include the motion of the
   ions and classical Lorentz forces on the ions. An overview of the various
   EMPKS theory levels that we use in the present work can be found in
   Tab.~\ref{tab_Na_297_dimer_acronyms}. The possibility to switch on and off
   different degrees of freedom and their mutual coupling has the benefit that
   we can better isolate the impact and significance of physical mechanisms in
   various applications.

    \begin{table}[t!]
      \begin{tabular}{ l | l }
        \toprule
        Acronym & \hphantom{} Description \\
        \hline
        \hline
        \midrule
        \setulcolor{cblue}
        \ul{\mbox{\scriptsize{ F@ED }}} & 
          \begin{tabular}{l}
            \scriptsize{ \textbf{F}orward coupling with \textbf{E}lectric \textbf{D}ipole term }
          \end{tabular}                                                                                           \\ \hline
        \setulcolor{cgreen}
        \ul{\mbox{\scriptsize{ FB@ED }}} & 
          \begin{tabular}{l}
            \scriptsize{ \textbf{F}orward and \textbf{B}ackward coupling with \textbf{E}lectric } \\
            \scriptsize{ \textbf{D}ipole term }
          \end{tabular}                                                                                           \\ \hline
        \setulcolor{corange}
        \ul{\mbox{\scriptsize{ F@(ED+MD+EQ) }}} &
          \begin{tabular}{l}
            \scriptsize{ \textbf{F}orward coupling with \textbf{E}lectric \textbf{D}ipole, \textbf{M}ag- } \\
            \scriptsize{ netic \textbf{D}ipole and \textbf{E}lectric \textbf{Q}uadrupole term }
          \end{tabular}                                                                                           \\ \hline
        \setulcolor{cred}
        \ul{\mbox{\scriptsize{ FB@(ED+MD+EQ) }}} &
          \begin{tabular}{l}
            \scriptsize{ \textbf{F}orward and \textbf{B}ackward coupling with \textbf{E}lectric } \\
            \scriptsize{ \textbf{D}ipole, \textbf{M}agnetic \textbf{D}ipole and \textbf{E}lectric} \\
            \scriptsize{ \textbf{Q}uadrupole term }
          \end{tabular}                                                                                           \\ \hline
      \end{tabular}
      \caption{Table of acronyms that are used in the present work to indicate the level of EMPKS theory.}
      \label{tab_Na_297_dimer_acronyms}
    \end{table}

   \subsection{System and simulation parameters}

      \begin{figure}[h]
        \center
        \begin{minipage}{0.46 \textwidth}
          \center
          \includegraphics[scale=0.4]{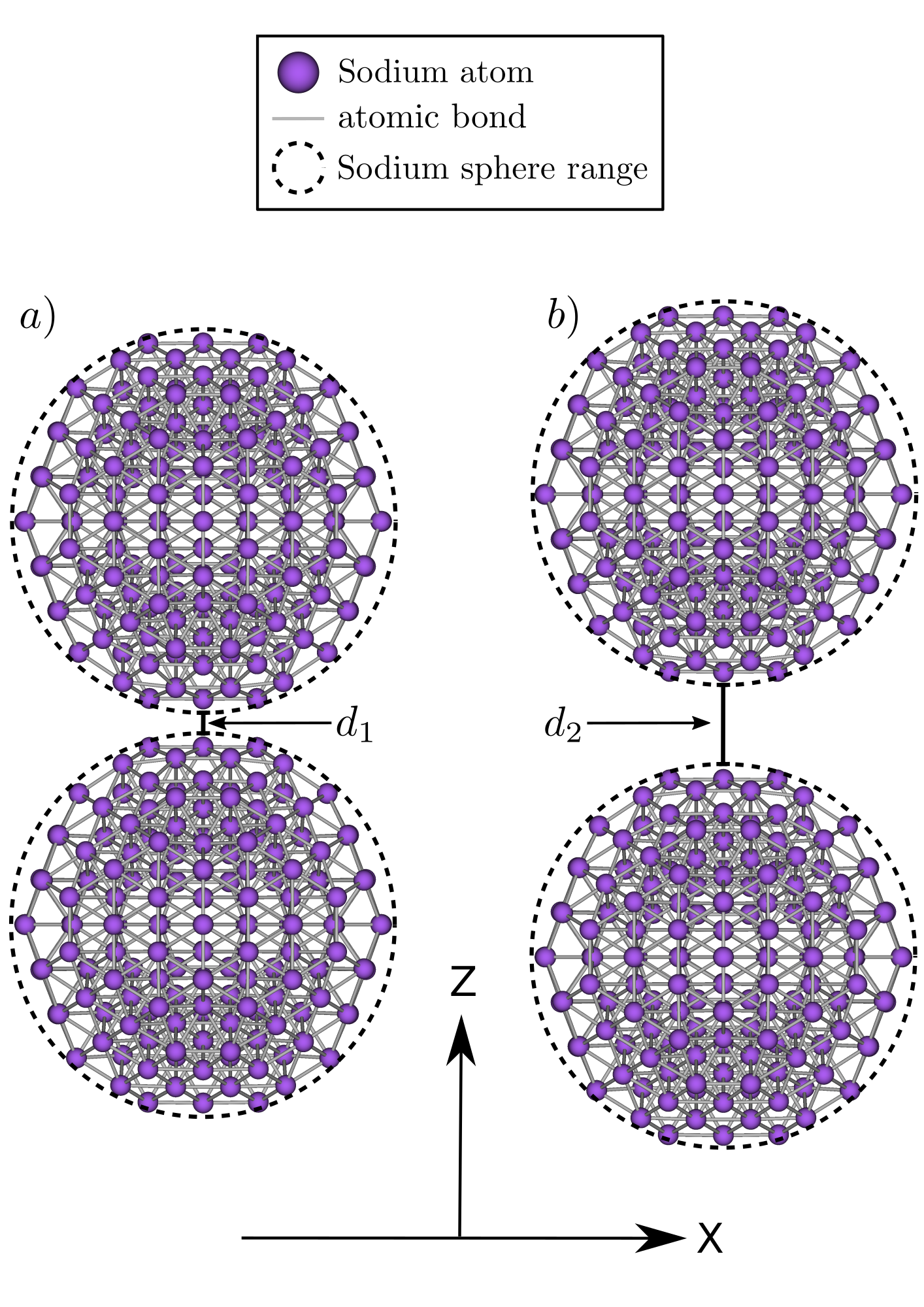}
          \caption{Geometry of the Na$_{297}$ dimer in E2E configuration with
          different distances $ d_1=0.1 $nm (left) and $ d_2=0.5 $nm (right) between the two
          effective spheres of the nanoparticles which are illustrated by the
          black dashed circle.}
          \label{fig_Na_297_dimer_geometry}
        \end{minipage}
      \end{figure}

    \subsubsection{Na$_{\mathrm{297}}$-dimer geometry}

    As a first test system for our EMPKS implementation, we have selected a
    nanoplasmonic system which was already investigated previously in a work by
    Varas et. al. \cite{doi:10.1021/acs.jpclett.5b00573}.
    This system consists of two almost spherical sodium nanoparticles with 297
    sodium atoms each, which are arranged in a dimer configuration. In
    Fig.~\ref{fig_Na_297_dimer_geometry}, we illustrate the geometry of the
    sodium dimer for two different distances between the sodium nanoparticles. We
    use the default Troullier-Martins pseudo-potentials in Octopus, which treat
    one valence electron per sodium atom.  Consequently, the total system
    comprises 594 sodium atoms and 594 valence electrons \cite{Noya2007,CCD}. \\
    We have performed standard geometry optimizations for the ground state of the
    system.  The icosahedral polyhedron is the most stable geometry for a
    single sodium nanoparticle. The quite large polyhedron is approximately a
    sphere with an effective diameter of $ 2R $. From the optimization we find
    an effective radius of $ R \approx 2.61$nm.  The key parameters for the
    composition of the dimer are the distance between the two nanoparticles and their
    relative orientation.  We use the distance $ b $ of the two centered sodium
    atoms of each icosahedron so that $ d $ is defined as $ d = b - 2 R $ and
    does not depend on the relative orientation of the two nanoparticles to
    each other.  The two icosahedrons can be oriented in several
    constellations. We have used a relative orientation such that the 3-atom
    edges of the hexagons on the surface of the nanoparticles are lying face to
    face. This so called E2E configuration
    \cite{doi:10.1021/acs.jpclett.5b00573} is illustrated in
    Fig.~\ref{fig_Na_297_dimer_geometry}.  The dimer axis is oriented parallel
    to the z-axis, therefore the dimer is symmetric with respect to x- and y-axis.

    To investigate the effect of the total internal dipole of the system on the
    coupled time-evolution, we consider in the following two different
    distances of the nanoparticles with $ d_1 = 0.1$ nm and $ d_2 = 0.5$ nm
    which result in different dipoles.  By computing the optical absorption
    cross section of the dimer, it was shown in Ref.
    \cite{doi:10.1021/acs.jpclett.5b00573} that the system
    with distance $d_1$, shows a prominent quadrupole (Q mode) localized
    surface-plasmon resonance, whereas the second system with distance $d_2$
    has a strong dipole (D mode) resonance.  In the following examples, we
    excite the nanoplasmonic dimer with an incoming laser pulse, where we
    select the corresponding resonance frequencies of the Q and D mode as
    carrier frequency of the laser.

      \begin{table}[t!]
        \begin{minipage}{0.46 \textwidth}
          \begin{tabular}{| c | c | c | c | c |}
            \hline
                                     &       \multicolumn{2}{c|}{distance $d_1$}   &         \multicolumn{2}{c|}{distance $d_2$}        \\  \cline{2-5}
            variable                 &              conv. units &           [a.u.] &             conv. units  &                  [a.u.] \\ \hline 
            $\omega$                 &                  3.05 eV &            0.112 &                  2.83 eV &                   0.104 \\
            $k_{x}$                  &  1.55 e$^{-11}$ m$^{-1}$ &    8.17 e$^{-4}$ &  1.43 e$^{-11}$ m$^{-1}$ &           7.59 e$^{-4}$ \\
            $\lambda$                &                 406.5 nm &          7681.84 &                 438.1 nm &                 8279.02 \\
            $E_{0,z}$                &        5.142 e$^{7}$ V/m &     1.0 e$^{-4}$ &        5.142 e$^{7}$ V/m &            1.0 e$^{-4}$ \\  
            Intensity                &        3.51 e$^{12}$ W/m$^2$ &    5.45 e$^{-4}$ &        3.51 e$^{12}$ W/m$^2$ &           5.45 e$^{-4}$ \\  
            $\xi $                   &               2034.08 nm &          38438.5 &               2034.08 nm &                 41395.1 \\
            $x_0$                    &               4068.16 nm &          76877.0 &               4381.07 nm &                 82790.2 \\
            $L_{\mathrm{KS},x}$      &                 1.993 nm &           37.658 &                 1.993 nm &                  37.658 \\
            $L_{\mathrm{KS},y}$      &                 1.993 nm &           37.658 &                 1.993 nm &                  37.658 \\
            $L_{\mathrm{KS},z}$      &                 3.347 nm &           63.258 &                 3.547 nm &                  67.037 \\
            $L_{\mathrm{Mx},x}$      &                 2.646 nm &           50.000 &                 2.646 nm &                  50.000 \\
            $L_{\mathrm{Mx},y}$      &                 2.646 nm &           50.000 &                 2.646 nm &                  50.000 \\
            $L_{\mathrm{Mx},z}$      &                 4.498 nm &           85.000 &                 4.498 nm &                  85.000 \\
            $a_{\mathrm{Mx},x}$      &                 2.170 nm &           41.000 &                 2.170 nm &                  41.000 \\
            $a_{\mathrm{Mx},y}$      &                 2.170 nm &           41.000 &                 2.170 nm &                  41.000 \\
            $a_{\mathrm{Mx},z}$      &                 4.022 nm &           76.000 &                 4.022 nm &                  76.000 \\
            $\Delta x_{\mathrm{KS}}$ &                 0.053 nm &            1.000 &                 0.053 nm &                   1.000 \\
            $\Delta x_{\mathrm{Mx}}$ &                 0.053 nm &            1.000 &                 0.053 nm &                   1.000 \\
            $\Delta t_{\mathrm{KS}}$ &        5.096 e$^{-3}$ fs &            0.211 &        5.096 e$^{-3}$ fs &                   0.211 \\
            $\Delta t_{\mathrm{Mx}}$ &        1.019 e$^{-4}$ fs &    4.21 e$^{-3}$ &        1.019 e$^{-4}$ fs &           4.21 e$^{-3}$ \\
            \hline
          \end{tabular}
        \end{minipage}
        \caption{Simulation parameters for the sodium dimer for distances $ d_1=0.1$ nm and $ d_2=0.5$ nm.}
        \label{tab_dimer}
      \end{table}

    \subsubsection{Simulation boxes and grid alignment}

      For the real-time simulations of the nanoplasmonic dimer, we use in the
      following a Maxwell-Kohn-Sham simulation box which corresponds to grid
      type f) in Fig.  \ref{fig_multiscale_grids}. The matter Kohn-Sham
      grid is smaller than the Maxwell grid, but both grids have the same grid
      spacings in each spatial direction and all Kohn-Sham grid points lie on
      top of Maxwell grid points.  The Kohn-Sham grid geometry is based on the
      so called minimum box construction \cite{doi:10.1002/pssb.200642067}.  The
      minimum box of a molecule consists of the union of all Cartesian grid
      points which lie inside a fixed radius around each ion of the system. For
      all simulations, we select a radius of $ R_{\mathrm{min}} = 0.794 $ nm
      ($15$ a.u.). Taking the corresponding geometries for the dimer distances
      $ d_1 $ and $ d_2 $ into account, we obtain maximal extensions $
      L_{\mathrm{KS},x} $, $ L_{\mathrm{KS},y} $, $ L_{\mathrm{KS},z} $ in each
      direction given in Tab.~\ref{tab_dimer}.  The matter grid is surrounded
      by a significantly larger parallelepiped shaped box for the Maxwell grid
      points with the extensions $ L_{\mathrm{Mx},x} $, $ L_{\mathrm{Mx},y} $,
      $ L_{\mathrm{Mx},z} $ in negative and positive direction which is
      illustrated in Fig.
      \ref{fig_plane_wave_and_pml_boundaries_and_detector}.  As grid spacing
      for both grids we select $ 0.053 $ nm ($1.0 $ a.u.) and a stencil order
      of 4 for the finite-difference discretization. \\ For the Kohn-Sham
      orbitals we use a zero Dirichlet boundary condition, whereas for the
      Maxwell grid we employ the combined incident plane-wave plus absorbing
      boundaries via PML as introduced in section
      \ref{subsubsec_incident_waves_plus_ab_boundaries}.  Hence, the Maxwell
      simulation grid is separated into two areas, one outer for the incident
      plane-wave boundaries and one inner for the PML. The incident plane-wave
      boundary width depends on the derivative order for the finite-difference
      stencil times the grid spacing. In the present case, the width of the plane-wave
      boundary region is $0.212$ nm ($4.0$ a.u.), and we use in addition
      $0.265$ nm ($5.0$ a.u.) as PML region. The total inner simulation box for
      the Maxwell propagation is therefore limited by $ a_{\mathrm{Mx},x} $, $
      a_{\mathrm{Mx},y} $, and $ a_{\mathrm{Mx},z} $ also given in Tab.
      \ref{tab_dimer}.  \\

      \subsubsection{Measurement regions}
      To distinguish near-field and
      far-field effects, we define different measurement regions in the
      simulation box where we record the electromagnetic fields, and the
      current density as function of time during the laser pulse propagation.
      Besides the mid point (mp) $ \vec{r}_{\mathrm{mp}} = (0,0,0) $ that was
      already considered in the work of Varas et.~al., we also consider two far-field points. The point
      (ffpx) $ \vec{r}_{\mathrm{ffpx}} = (1.957~$nm$,0,0) = (37.0~$a.u.$,0,0) $ is shifted along the laser
      propagation axis, and (ffpy) $ \vec{r}_{\mathrm{ffpy}} = (0,1.957~$nm$,0) = (0,37.0~$a.u.$,0) $ along
      the y-axis.
      Furthermore, we define a detector surface given by
      the parametrization %
      \begin{align}
        \vec{r}_{\mathrm{sfx}}(\alpha, \beta ) = \vec{r}_{\mathrm{ffpx}} + \alpha \vec{e}_{y} + \beta \vec{e}_{z}, 
        \hspace{5mm} 
        \left\{ 
          \begin{array}{c}
            - 37.0 \le \alpha \le 37.0 \\
              81.0 \le \beta  \le 71.0
          \end{array}
        \right.  \honemm .    \label{eq_Na_297_dimer_detector_surface}
      \end{align}
      The detector surface includes the far-field point $
      \vec{r}_{\mathrm{ffpx}} $ and the extension is determined by the box
      limits. We chose the limits such that all points have sufficient distance
      to the absorbing PML region.  For ease of comparison, we compute the
      average of the electric field over the detector surface.

    \subsubsection{Laser pulse shape}

      For the incident laser pulse that approaches our nanoplasmonic dimer, we use
      the following analytical expression for the external electric field
      \begin{equation}
        \begin{alignedat}{2}
            \vec{E}_{\mathrm{pw}}(\vrt) &=                                                         \\
          & \hspace{-10mm} \vec{e}_{z} \hspace{0.5mm} E_{0,z} \hspace{0.5mm}
            \mathrm{cos} (k_x (x \hspace{-0.5mm} - \hspace{-0.5mm} x_0) \hspace{-0.5mm} - \hspace{-0.5mm} \omega t) \hspace{0.5mm}
            \mathrm{cos} \hspace{-1mm} 
            \left( \hspace{-1mm}
              \frac{ \pi (x \hspace{-0.5mm} - \hspace{-0.5mm} 2 \xi \hspace{-0.5mm} - \hspace{-0.5mm} x_0 \hspace{-0.5mm} - \hspace{-0.5mm} c_0 t) }
                   {2 \xi} \hspace{-0.5mm} + \pi \hspace{-1mm}
            \right)                    \\
          & \hspace{-10mm} \cdot \theta \left( \xi \hspace{-0.5mm} - \hspace{-0.5mm} \frac{|k_x( x \hspace{-0.5mm} - \hspace{-0.5mm} x_0) - \omega t|}{|k_x|} \right),
        \end{alignedat}
      \end{equation}
      where $ \theta $  denotes the usual Heavyside step function.  The laser
      pulse propagates with wavevector $ \vec{k} = (k_x,0,0) $ along the
      x-axis.  The electric-field polarization is oriented along the z-axis and
      consequently the magnetic field oscillates parallel to the y-axis.  The
      corresponding magnetic field $ \vec{B}_{\mathrm{pw}} $ and
      Riemann-Silberstein vectors $ \vec{F}_{\pm,\mathrm{pw}} $ are with $
      \vec{k} \cdot \vec{E}_{\mathrm{pw}} = \vec{k} \cdot \vec{B}_{\mathrm{pw}}
      = 0 $ and Eq.~(\ref{eq_RS_vector_pm_2}) given by

      \begin{align}
          \vec{B}_{\mathrm{pw}}(\vrt) 
        = - \vec{e}_y \frac{1}{c_0} \vec{E}_{\mathrm{pw}}(\vrt)   \honemm ,
      \end{align}

      \begin{align}
          \vec{F}_{\pm,\mathrm{pw}}(\vrt)
        = \sqrt{\frac{\epsilon_0}{2}} \vec{E}_{\mathrm{pw}}(\vrt)
          \pm \mi \sqrt{\frac{1}{2 \mu_0}} \vec{B}_{\mathrm{pw}}(\vrt)   \honemm .
      \end{align}

      As discussed in section \ref{subsubsec_incident_waves_boundaries}, this
      analytical form of the Riemann-Silberstein vector $
      \vec{F}_{\pm,\mathrm{pw}} $ is used to update the incident plane wave
      boundaries for each propagation step. While the incoming laser pulse is
      prescribed analytically in the boundary region, we propagate the
      electromagnetic fields fully numerically in the interior of the
      simulation box.

    \subsubsection{Propagators}

    Inside the Maxwell propagation region with ($ -l_{\mathrm{Mx},x} \le x \le
    l_{\mathrm{Mx},x} $), ($ -l_{\mathrm{Mx},y} \le y \le l_{\mathrm{Mx},y} $),
    ($ -l_{\mathrm{Mx},z} \le z \le l_{\mathrm{Mx},z} $), we propagate the
    Kohn-Sham system with the matter ETRS propagator from
    Eq.~(\ref{eq_matter_time_evolution_operator_recursive_etrs}) using the
    Power-Zienau-Woolley transformed MPKS Hamiltonian from 
    Eq.~(\ref{eq:PWZMPKS}) with multipole expansion.  The Maxwell
    system is evolved in time by the Maxwell ETRS propagator from
    Eq.~(\ref{eq_RS_time_evolution_equation_recursive_numerical_etrs}) with
    corresponding Hamiltonian kernel $ \hat{\mathcal{H}} $ from
    Eq.~(\ref{eq_RS_Hamiltonian}), where we use only $\hat{\mathcal{H}}$ and
    switch off the internal diamagnetic current kernel $ \hat{\mathcal{K}} $, and the
    current density term $ \mathcal{J} $ given in Eq.~(\ref{eq_RS_current}) only contains the paramagnetic current contribution. We propagate the Riemann-Silberstein vector corresponding to the total vector potential $A^{k}_{\rm{tot}} = a^k + A^k$ of external and internal fields. The nuclei are treated classically
    and are propagated with Ehrenfest equations of motion \cite{andrade_2009}. Since we have the electromagnetic
    fields available in the simulation box, we also include
    the classical Lorentz force that acts on the ions. For the transversal Kohn-Sham field we use the mean-field approximation $a^{k}_{\rm{KS}} \approx a^{k} + A^{k}$ as well as the physical mass of the particles to take into account the bare vacuum fluctuations of the photon field. For the longitudinal Kohn-Sham field we use $a^{0}_{\rm{KS}} \approx a^{0} + A^{0} + a^{\rm{LDA}}_{\rm{xc}}$, where $a^{\rm{LDA}}_{\rm{xc}}$ is the adiabatic LDA exchange-correlation approximation.\\
    In addition to the fully coupled EMPKS simulation, we propagate in addition
    the unperturbed Maxwell system inside the inner simulation box to get the
    required values for the incident plane wave plus PML boundaries according
    to section \ref{subsubsec_incident_waves_plus_ab_boundaries}.  Hence, the
    Maxwell Hamiltonian has to be updated inside the PML boundaries by the
    additional PML kernel $ \hat{\mathcal{G}} $ in
    Eq.~(\ref{eq_RS_Hamiltonian_PML_convolution}).

      \begin{figure*}[ht!]
        \begin{minipage}{1.0 \textwidth}
          \includegraphics[scale=0.4]{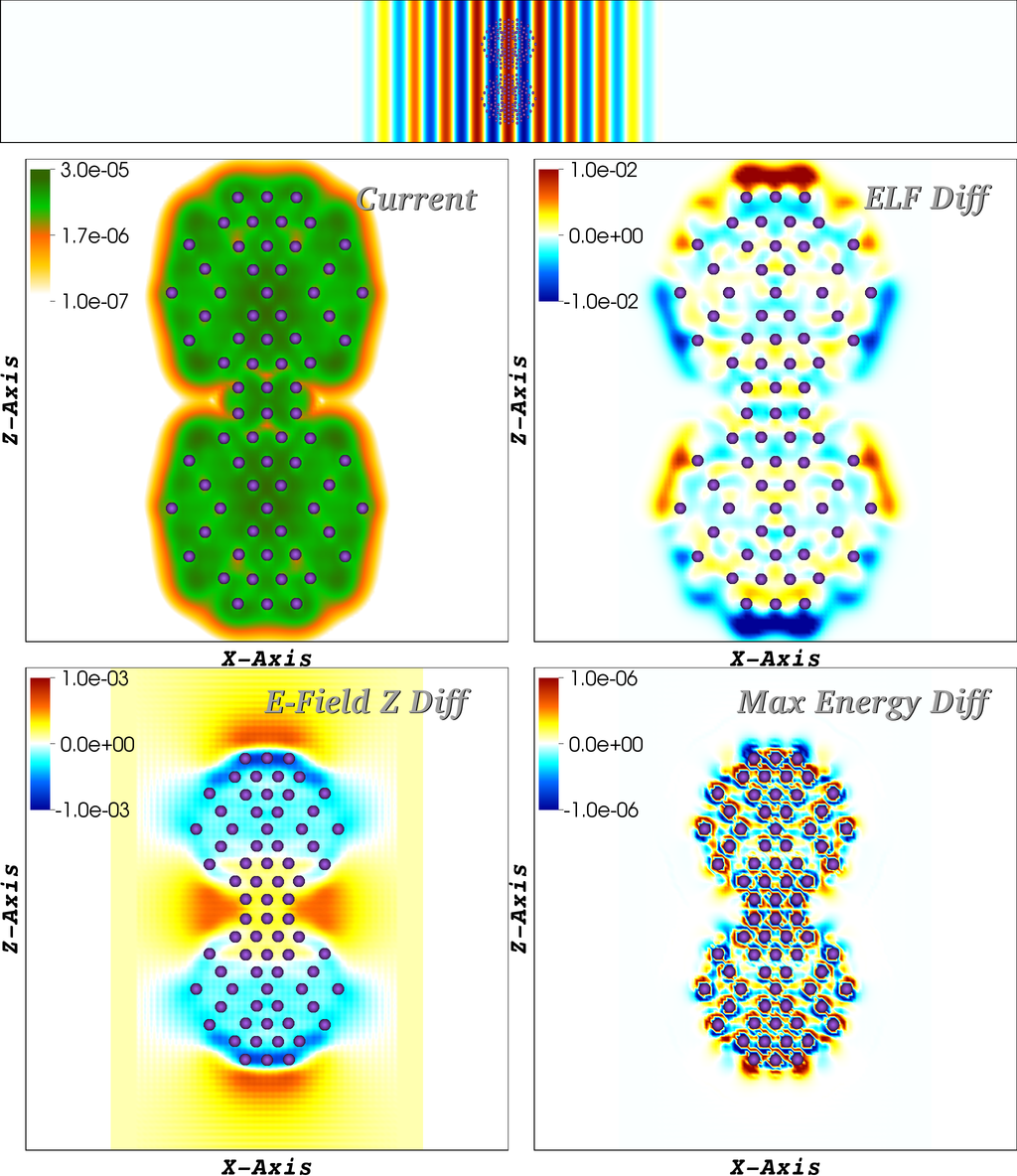}
          \caption{In Ref. \cite{zenodo.1489303}, we provide a movie that shows the real-time dynamics of the nanoplasmonic dimer
                   with distance $ d_1=0.1 $ nm. The time-evolution in the movie corresponds to the runs that we discuss in section \ref{sec_applications}.
                   In the figure, we show a frame of the movie at time 6.89 fs. The upper two panels show contour plots of matter variables, the absolute value of the
                   current density and the electron localization function (ELF). The most relevant Maxwell field variables, the
                   electric field along the laser polarization direction z and the total Maxwell
                   energy are presented in the lower panels. In the top of the figure, we show the incident laser pulse and at the
                   center the geometry of the nanoplasmonic dimer.}
          \label{fig_movie_frame_1}
        \end{minipage}
      \end{figure*}
       \begin{figure*}[ht!]
        \begin{minipage}{1.0 \textwidth}
          \includegraphics[scale=0.4]{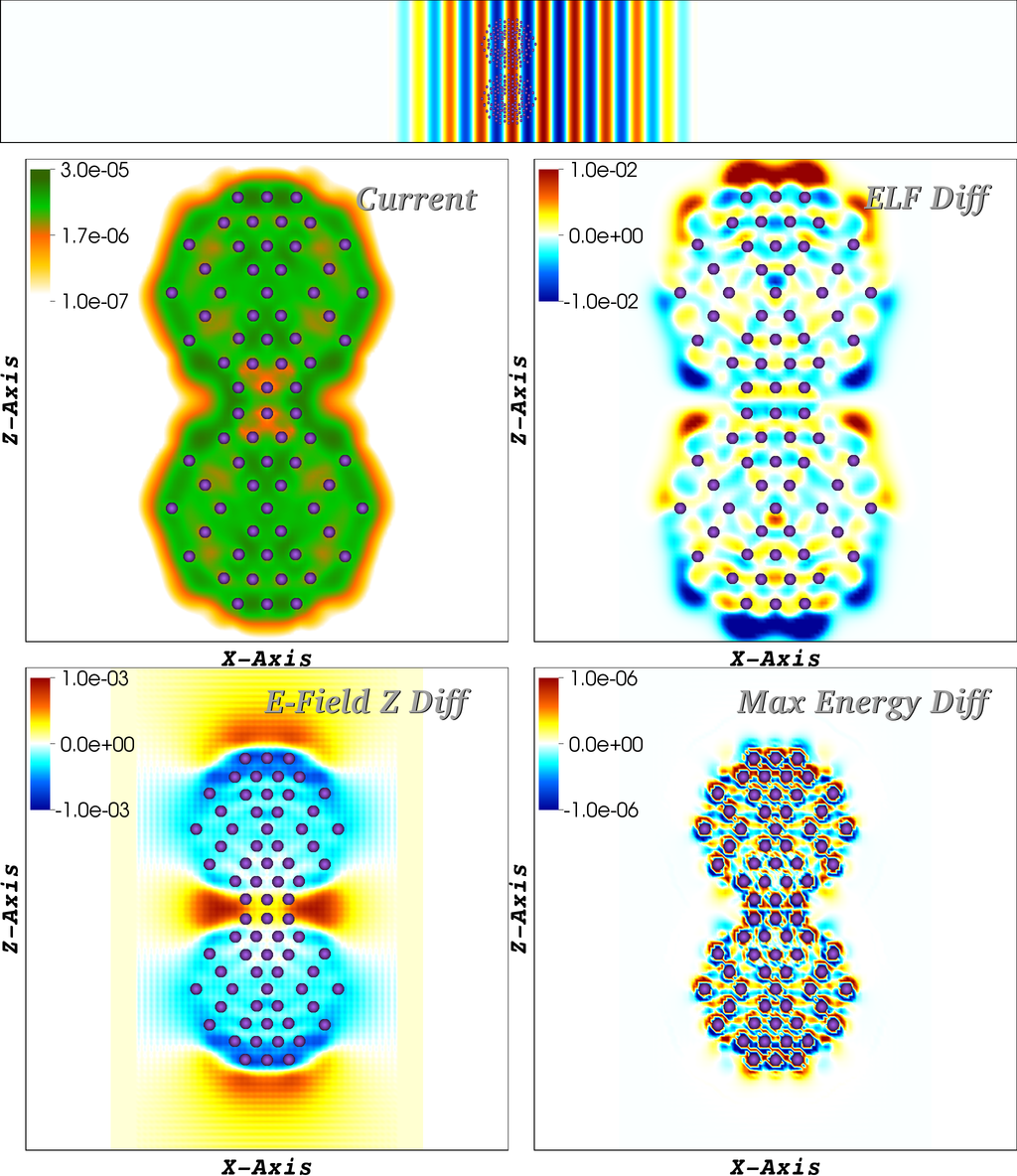}
          \caption{Similar movie frame as Fig.~(\ref{fig_movie_frame_1}), but with the frame at time 8.33 fs.}.
          \label{fig_movie_frame_2}
        \end{minipage}
      \end{figure*}

   \subsection{Results from EMPKS simulations}

    In the previous section, we have introduced the nanoplasmonic dimer, the
    incident laser pulse, and the parameters for the EMPKS time-stepping. In
    the following, we now focus on the actual EMPKS simulations for the dimer.
    In all cases, we compare our results from the different theory levels of
    self-consistently coupled light-matter propagations with the conventional
    forward-coupling only case with dipole approximation and highlight the
    differences. 

      \begin{figure*}[ht!]
        \begin{minipage}{1.0 \textwidth}
          \includegraphics[scale=0.5]{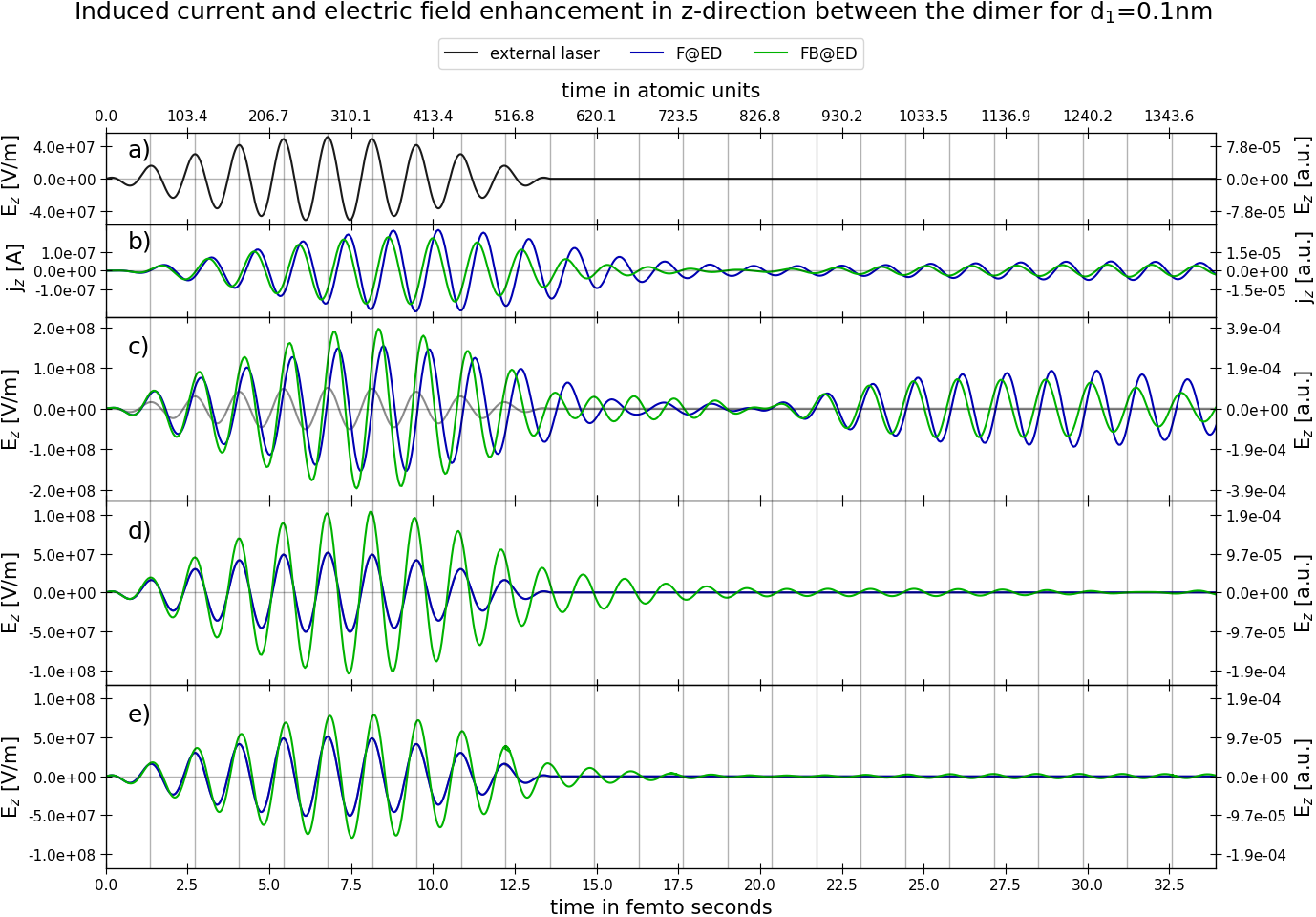}
          \caption{Electric field values and current density in z-direction in the mid point between the two sodium nanoparticles for $ d_1 = 0.1 $ nm.
                   The first panel $a)$ illustrates the incident cosinusoidal
                   laser pulse with frequency $ \omega_1 = 3.05 $ eV ($ 0.112 $
                   a.u) $ \lambda_1 = 406.5$ nm ($7681.84$ a.u.) and amplitude
                   of $ E_z^{0} = 5.142*10^{7} $ V/m ($10^{-4}$ a.u.) which
                   drives the system.  The second panel $b)$ displays the
                   current density at the mid point $ \vec{r}_{\mathrm{mp}}$
                   between the two nanoparticles. In panels $c)-e)$, we show
                   the electric field enhancement at the midpoint $
                   \vec{r}_{\mathrm{mp}}$, the electric field enhancement at
                   the far-field point $ \vec{r}_{\mathrm{ffpx}}$ and the
                   average of the electric field over the detector surface in
                   the far field, respectively. In all panels, F@ED refers to the
                   dipole-coupled case with only forward light-matter coupling,
                   whereas FB@ED denotes self-consistent forward-backward
                   light-matter coupling.  The curve in bright gray in panel
                   $c)$ has been added to simplify the comparison and is
                   identical to the laser pulse in panel $a)$. The period $T_1
                   = 1.36$ fs corresponding to the laser frequency $ \omega_1 $
                   is indicated with grey vertical lines. 
                   }
          \label{fig_Na_297_dimer_0_1_nm_electric_field_fixed_ions}
        \end{minipage}
      \end{figure*}

    To give an overview of the dynamics of the nanoplasmonic dimer, we
    provide in Ref.~\cite{zenodo.1489303} and Ref.~\cite{zenodo.1489305} movies 
    which show the interaction of the dimer with the incoming laser pulse.
    In Fig.~\ref{fig_movie_frame_1} and \ref{fig_movie_frame_2}, we show two
    representative snapshots at different time steps for the dimer with distance $d_1=0.1$ nm. 
    The movie frames are divided into four
    panels.  Each panel contains a contour plot which shows the
    x-y plane of the 3D data.  The upper left panel shows the absolute value of
    the induced paramagentic current. In the upper right panel, we show the electron
    localization function (ELF) to visualize the electron motion. In the lower
    left panel, we plot the electric field enhancement. We define this
    enhancement by subtracting the longitudinal electric field that corresponds
    to the ground state Kohn-Sham potential from the full induced electric
    field at a given time step.  The same difference is used in the lower right
    panel, which shows the Maxwell energy difference of the induced field and
    the ground state field energy.  Above these four panels, we also show a
    contour plot of the laser pulse propagation which includes at the center of
    the image also the geometry of the nanoplasmonic dimer.

    \subsubsection{Electric field enhancement}   \label{subsec_electric_field_enhancement}

      \begin{figure*}[ht!]
        \begin{minipage}{0.95 \textwidth}
          \includegraphics[scale=0.5]{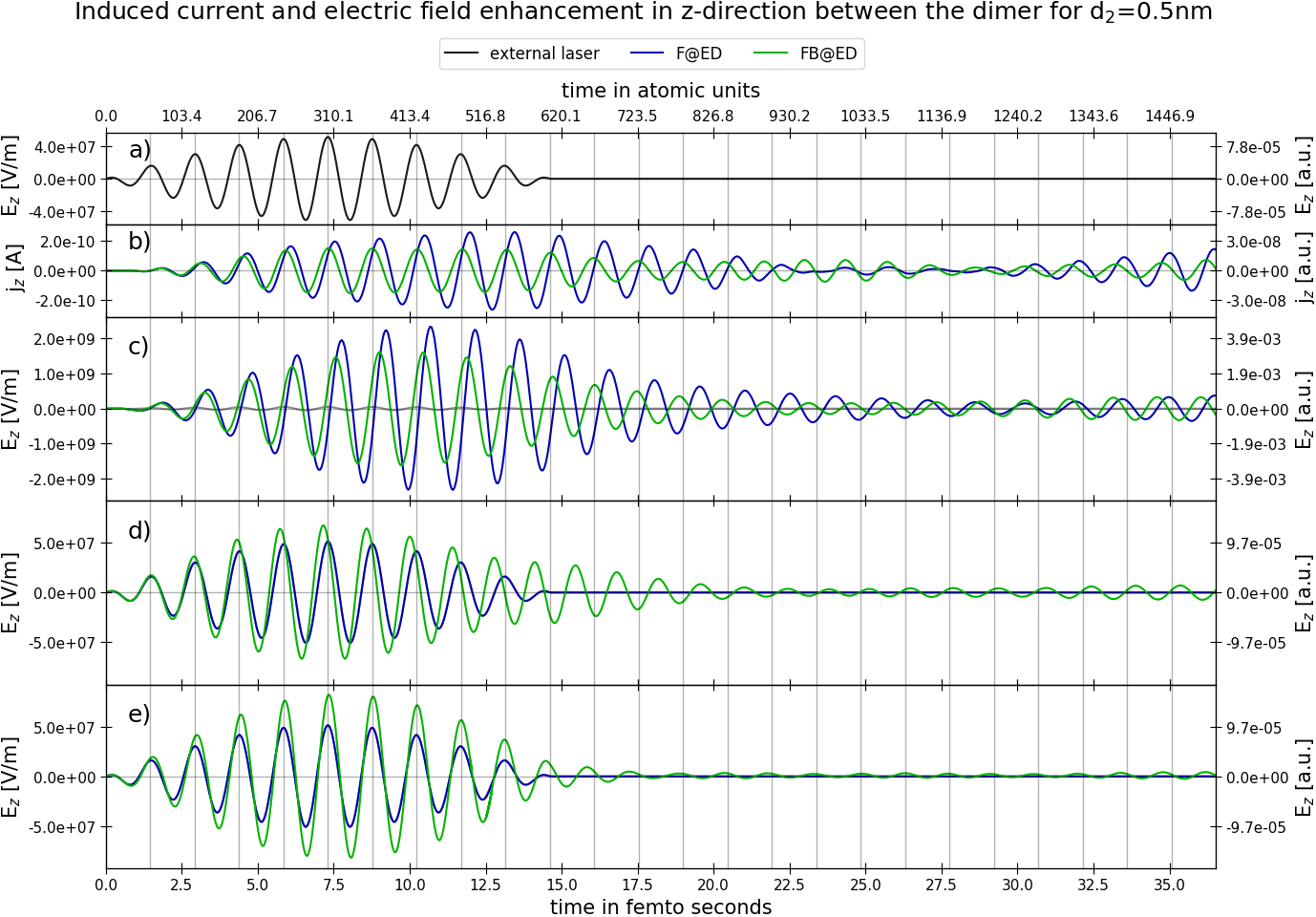}
          \caption{Similar to figure Fig. \ref{fig_Na_297_dimer_0_1_nm_electric_field_fixed_ions}, we show here the electric field enhancements and current densities for the
                   nanoplasmonic dimer with $ d_2 = 0.5 $nm. } 
          \label{fig_Na_297_dimer_0_5_nm_electric_field_fixed_ions}
        \end{minipage}
      \end{figure*}

      \begin{figure*}[ht]
        \begin{minipage}{1.0 \textwidth}
          \includegraphics[scale=0.5]{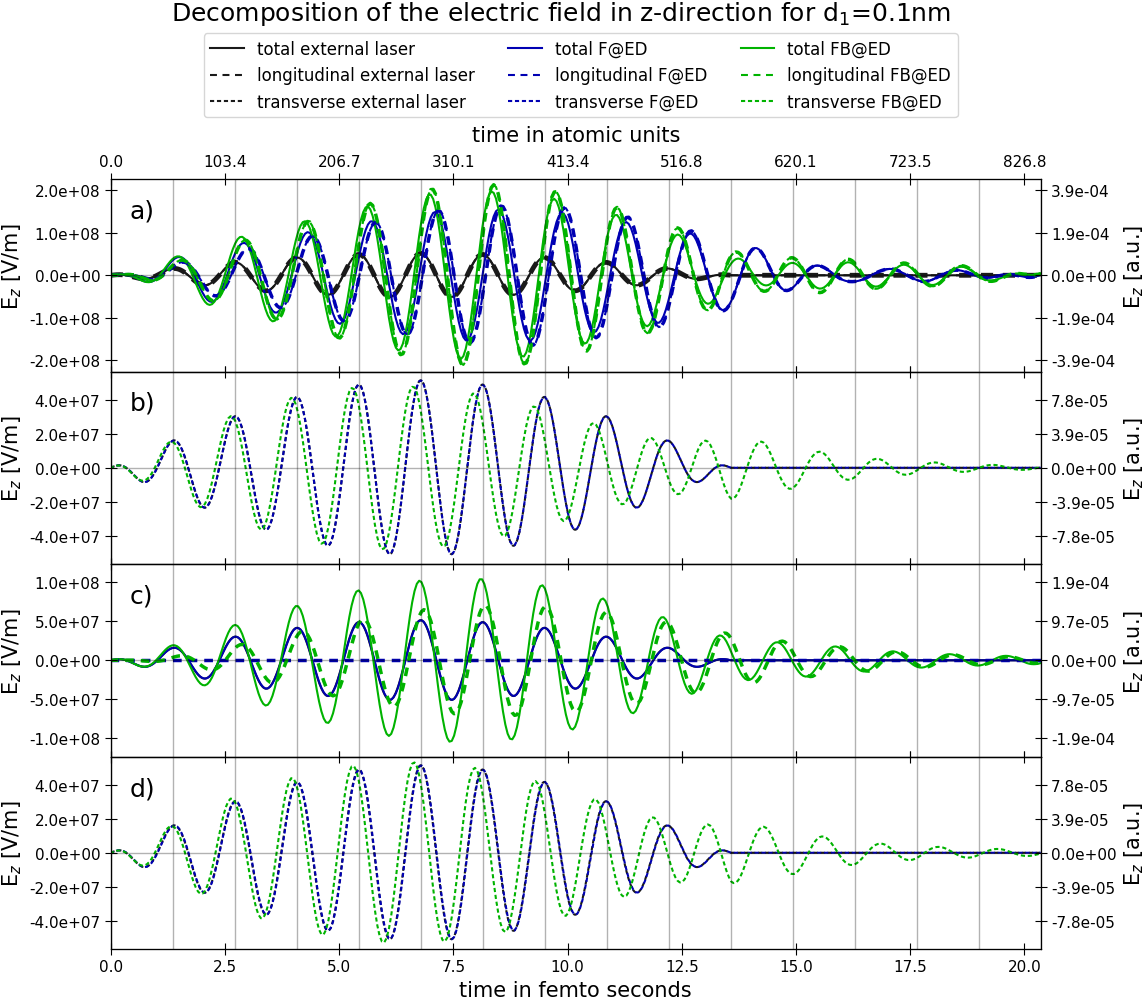}
        \end{minipage}
        \caption{Decomposition of the total electric field into transverse and longitudinal components.
                 Panel a) shows the total (solid lines) and the longitudinal (dashed lines) electric field in z-direction at $ \vec{r}_{\mathrm{mp}} $.
                 The corresponding transverse field (dotted lines) is plotted in b). The same field decomposition at the surface point
                 $ \vec{r}_{\mathrm{ffpx}} $ is plotted in c) and d).
                }
          \label{fig_Na_297_dimer_0_1_nm_electric_field_fixed_ions_decomposition}
      \end{figure*}

      \begin{figure*}[ht]
        \begin{minipage}{1.0 \textwidth}
          \includegraphics[scale=0.5]{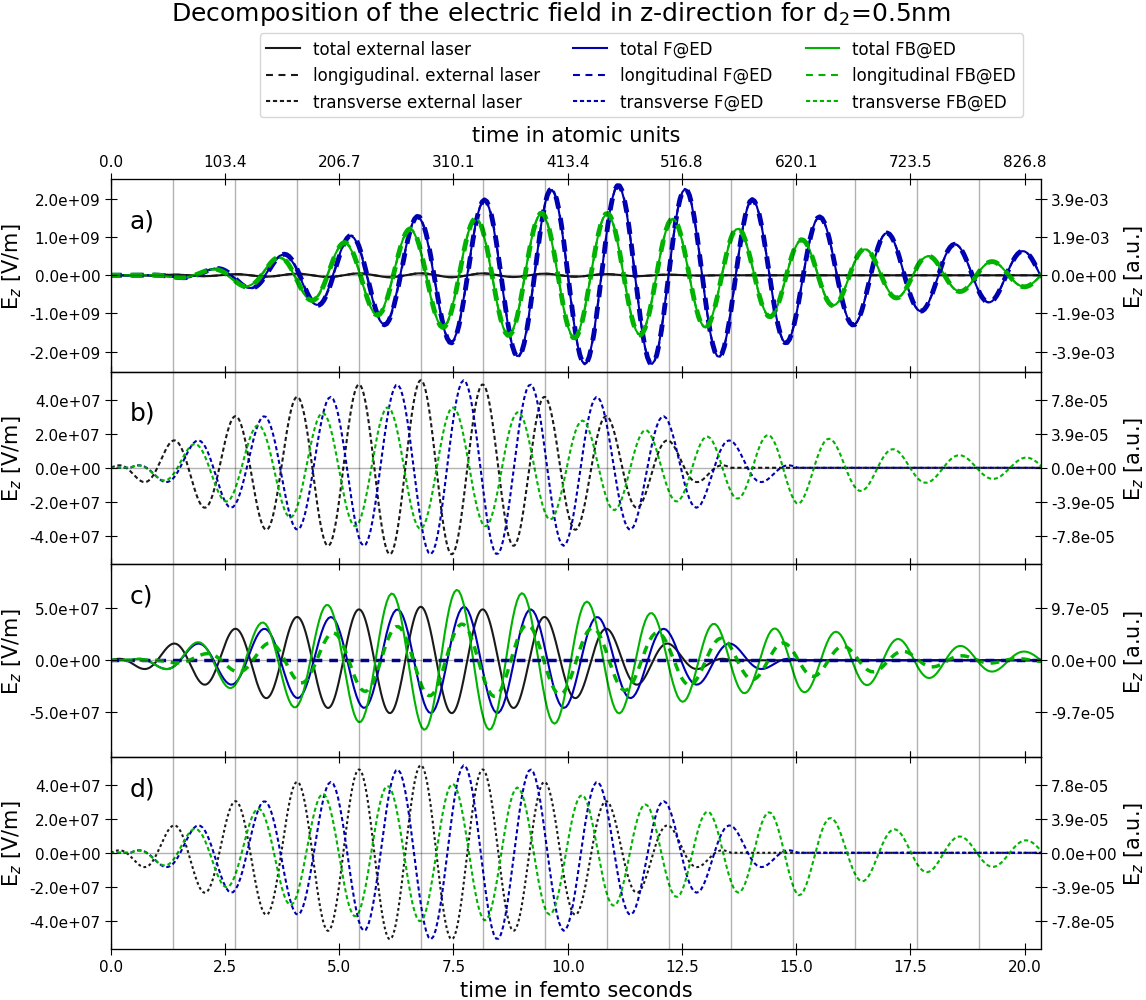}
        \end{minipage}
        \caption{ Decomposition of the electric field as in figure Fig.~\ref{fig_Na_297_dimer_0_1_nm_electric_field_fixed_ions_decomposition} for the
                  sodium dimer with $ d_2 = 0.5 $ nm.
                }
          \label{fig_Na_297_dimer_0_5_nm_electric_field_fixed_ions_decomposition}
      \end{figure*}

      In the work of Varas et.~al.~a large field enhancement right at the mid
      point between the two sodium nanoparticles was found. This enhancement
      originates solely from longitudinal photon degrees of freedom since in
      their work only a conventional time-dependent Kohn-Sham calculation
      without coupling to Maxwell's equations was performed. The longitudinal
      component of the electric field can be obtained in this case from the
      scalar Kohn-Sham potential  $ a_{\mathrm{KS}}^{0} = a^{0} + a^{0}_{\rm{xc}} + A^{0} $ according to
      \begin{align}
          \vec{E}_{\parallel}(\vrt)
        = - \vec{\nabla}  \left(a^{0}(\vrt) + A^{0}(\vrt)\right)  \honemm ,           \label{eq_transverse_electric_field_matter}
      \end{align}
      which is just the external and the Hartree potential. On the other hand, we can expect that also field enhancements occur for
      the transverse components of the electromagnetic field.  To investigate
      this, we use our EMPKS approach in the following to study the field
      enhancements in the fully coupled case.\\ In
      Fig.~\ref{fig_Na_297_dimer_0_1_nm_electric_field_fixed_ions}, we show
      in panel a) the electric field amplitude of the incident laser, in panel
      b) the current density at the mid point $ \vec{r}_{\mathrm{mp}}$, and in
      panels c)-e) the electric field at the mid point $
      \vec{r}_{\mathrm{mp}}$, the electric field at the far field point $
      \vec{r}_{\mathrm{ffpx}}$, and the average of the electric field over the detector
      surface, respectively.\\
      All blue curves in Fig.~\ref{fig_Na_297_dimer_0_1_nm_electric_field_fixed_ions} refer to the F@ED theory level.
      These results are fully identical with the previous results of Varas et.~al.,
      but have here been obtained from our EMPKS implementation by switching off 
      the matter to Maxwell back-reaction. The exact overlap with the data of Varas et. al. provides a further consistency-check
      of our implementation.\\
      The black curve in panel a) and the light gray curve in panel
      c) display the same initial cosinusoidal shaped free laser pulse that
      passes through the simulation box without any matter interaction. We
      include the pulse from panel a) again in panel c) to facilitate the comparison of the
      incident field amplitude (gray) in comparison to the actual electric
      field values when light-matter coupling is taking place (blue and green).  As was noted by
      Varas et al. for the F@ED theory level, a field enhancement of
      approximately a factor of three compared to the incident laser amplitude
      can be found at the mid point as can be seen from the blue curve in panel
      c). Varas et. al. also noticed a delay of the total induced electric field 
      maximum compared to the maximum field amplitude of the driving laser. This
      shift of maxima can also be seen in panel c) by comparing the maxima of the
      gray and blue curves.

      We now switch on the back-reaction of the matter on the electromagnetic
      fields. We denote this with theory level FB@ED (cf. Tab.
      \ref{tab_Na_297_dimer_acronyms}) and use green curves in the figure.
      Including the back-reaction, we find in panel c) a similar enhancement for very short
      times, a slightly increased enhancement for intermediate times and smaller beating
      for longer times. Overall, adding the
      back-reaction is increasing the field enhancement at the mid point while the external laser is active.
      Similar to the F@ED case of Varas et. al., a similar delay of the field
      maximum compared to the driving laser also appears in the
      forward-backward coupling case.  The situation is quite different in the
      far field which is illustrated in panel d). Here the fully
      self-consistent light matter coupling (green curve) shows a field
      enhancement which is almost twice as large as in the forward-only
      coupling case. While the selected far field point is lying on the laser
      propagation axis and could be considered special, this picture for the
      field enhancement in the far field is confirmed if we compare the field
      enhancements averaged over the detector surface as shown in panel e).  It
      is also worth noting, that the electric field is mostly in phase with the
      incident laser in the far field, for both F@ED and FB@ED. On the other hand a
      small phase and frequency shift occurs in the fully coupled case in the near-field in panel c).

      {\setlength{\extrarowheight}{160pt}
      \begin{figure*}[h]
        \begin{tabular}{cc}
          \hspace{1cm}
          \begin{minipage}{0.49 \textwidth}
            \begin{center}
              \hspace{-1.5cm} F@(ED+MD+EQ), t= 6.80 fs\\
              \includegraphics[trim={4cm 3cm 3cm 0},clip,width=\textwidth]{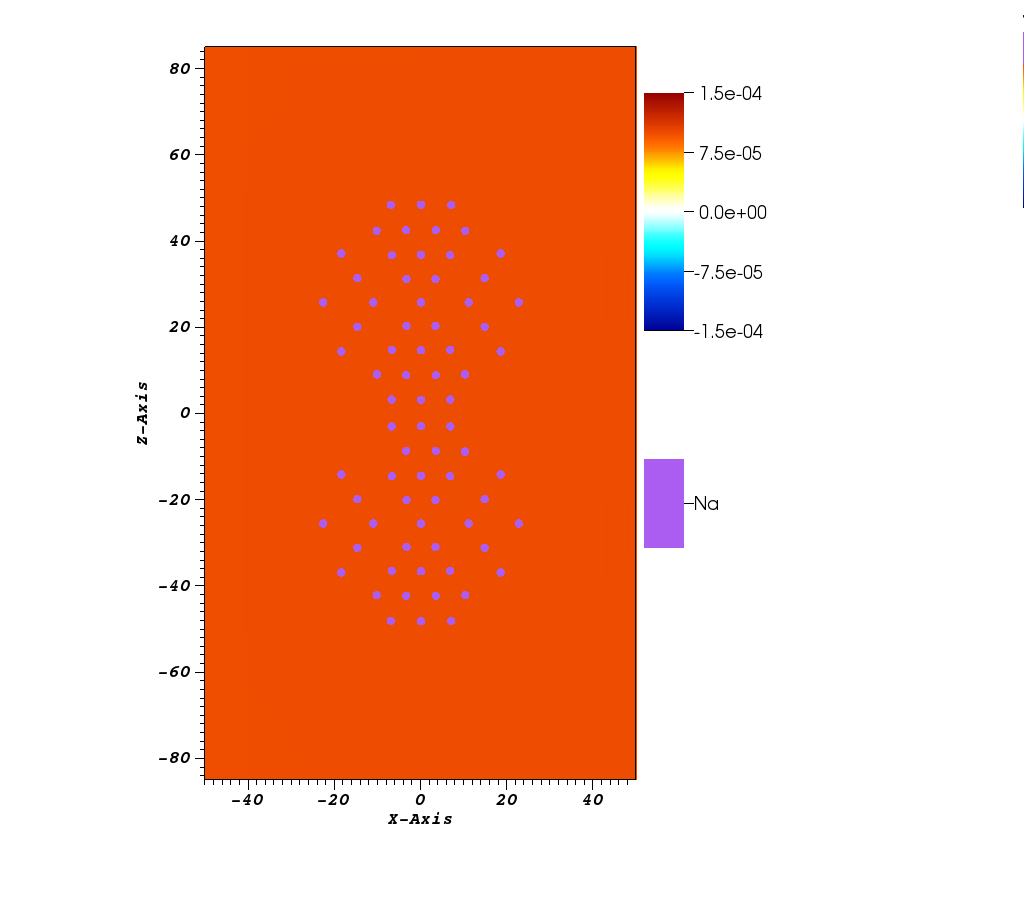}
            \end{center}
          \end{minipage}
            & 
          \begin{minipage}{0.49 \textwidth}
            \begin{center}
              \hspace{-1.5cm} FB@(ED+MD+EQ), t= 6.80 fs\\
              \includegraphics[trim={4cm 3cm 3cm 0},clip,width=\textwidth]{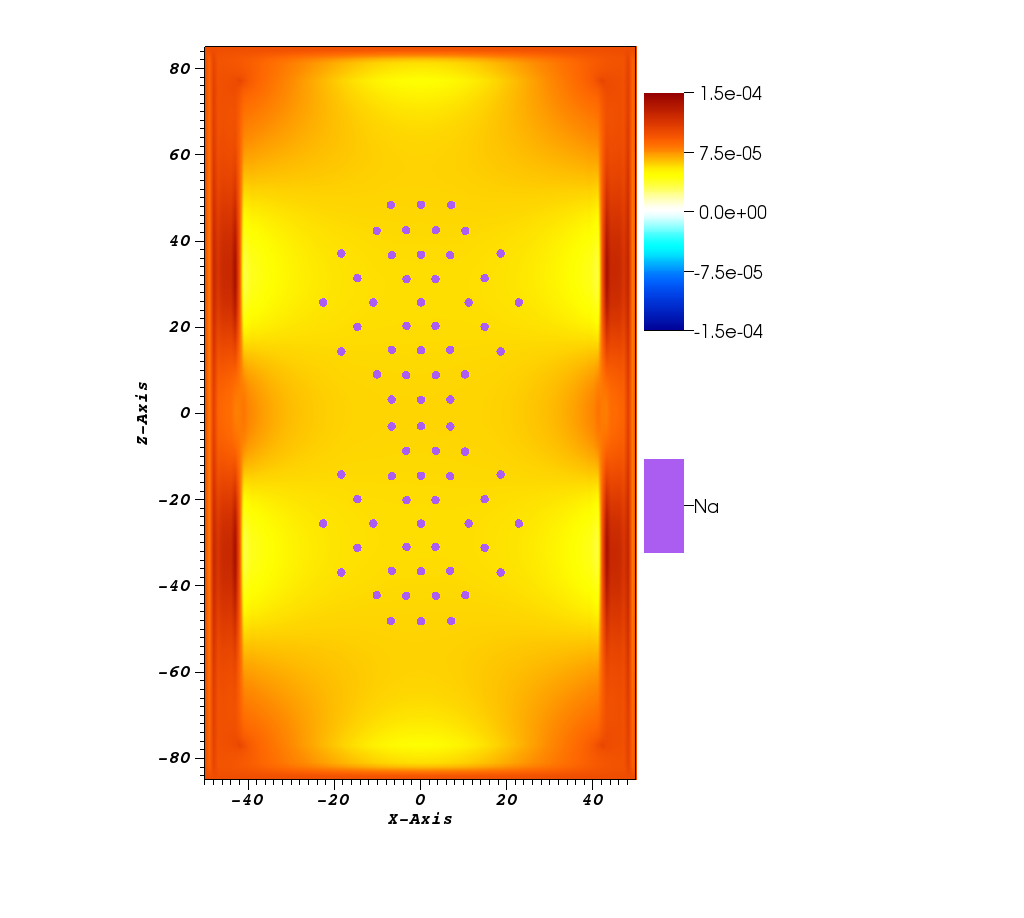}
            \end{center}
          \end{minipage}
            \\
          \hspace{1cm}
          \begin{minipage}{0.49 \textwidth}
            \begin{center}
              \hspace{-1.5cm} F@(ED+MD+EQ), t= 8.33 fs\\
              \includegraphics[trim={4cm 3cm 3cm 0},clip,width=\textwidth]{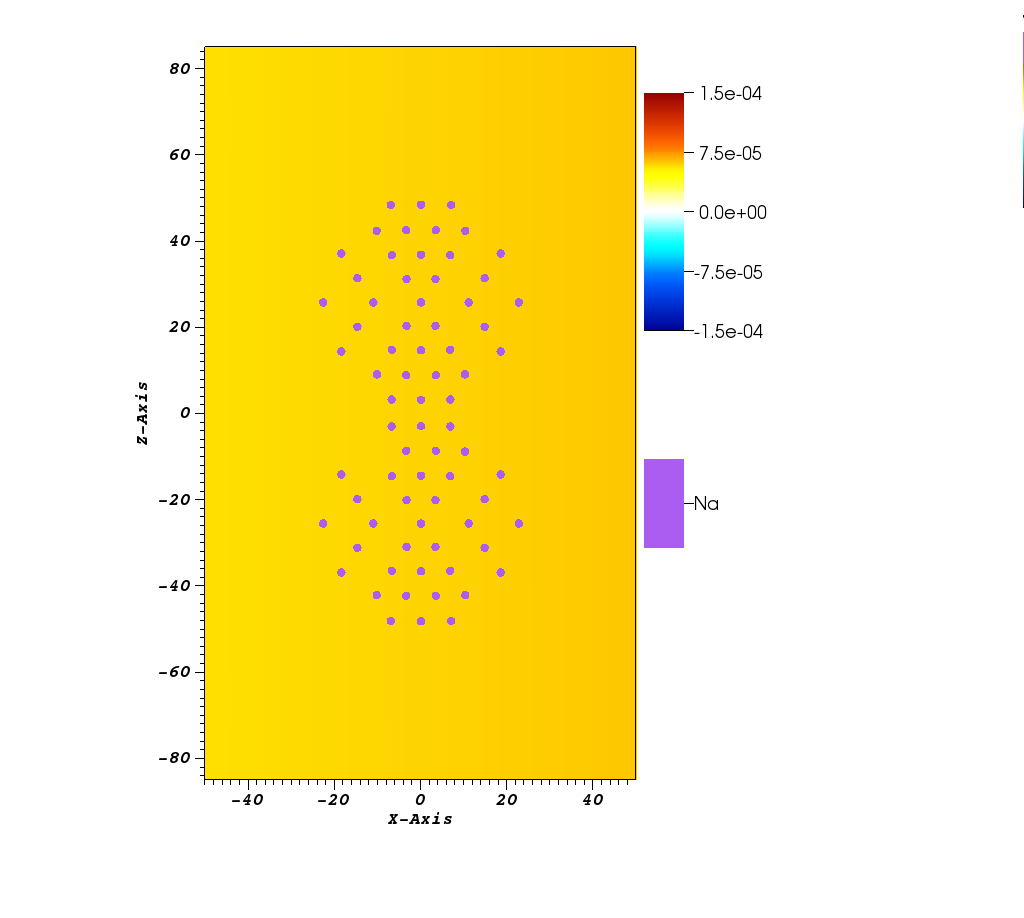}
            \end{center}
          \end{minipage}
            & 
          \begin{minipage}{0.49 \textwidth}
            \begin{center}
              \hspace{-1.5cm} FB@(ED+MD+EQ), t= 8.33 fs\\
              \includegraphics[trim={4cm 3cm 3cm 0},clip,width=\textwidth]{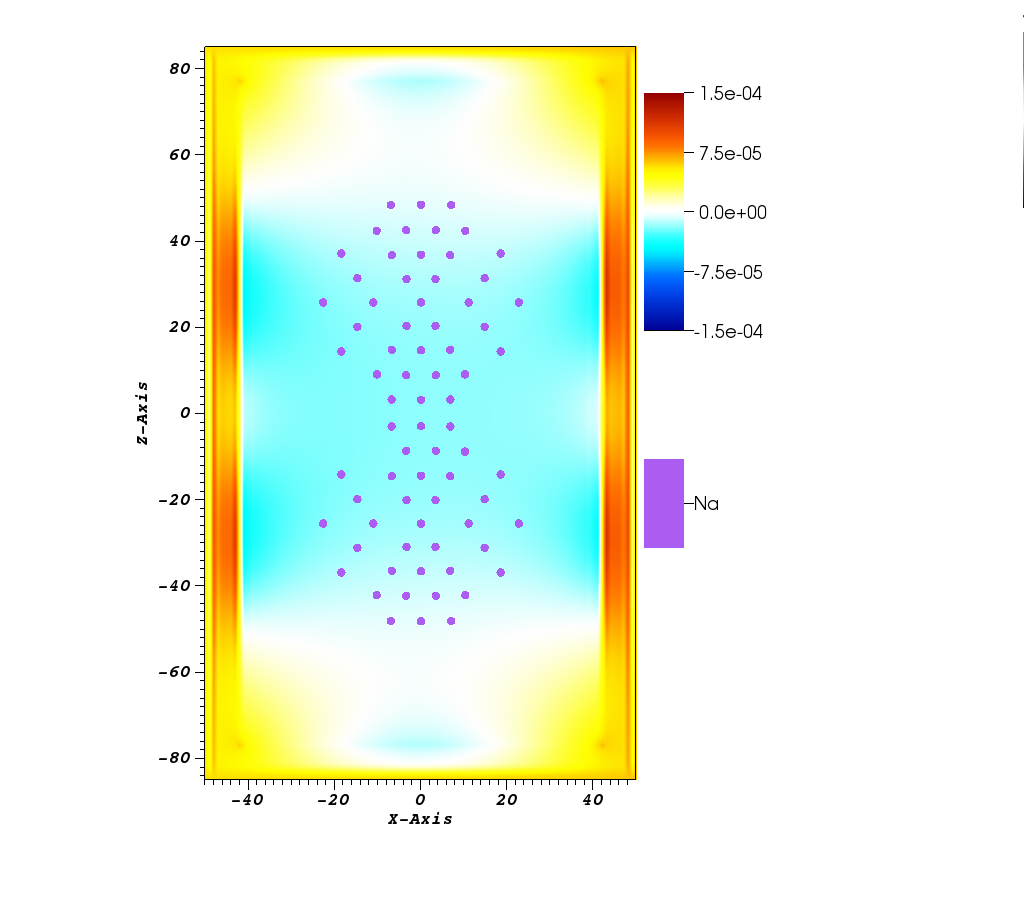}
            \end{center}
          \end{minipage}
        \end{tabular}
        \caption{Contour plots for the electric field enhancement in the x-z plane of the nanoplasmonic dimer. The upper two panels
                 show the field enhancement when the external laser reaches its maximum and in the lower two panels show the electric field 
                 when the field enhancement itself reaches its maximum. The two panels on the left (top and bottom) correspond to the
                 F@(ED+MD+EQ) theory level, whereas the two panels on the right (top and bottom) correspond to FB@(ED+MD+EQ) coupling. 
                 The electric field in the forward coupled case in the two plots on the left is rather uniform due to the optical wavelength 
                 which is large on the scale of the dimer. On the other hand for the forward-backward coupled case in the figures on the right 
                 hand side local field effects are spatially resolved and in particular in the bottom right panel the transversal
                 field contribution counter acts the longitudinal contribution since it has turned to a negative sign in most
                 regions of space.
                 }
        \label{fig_electric_field_enhancement_contour}
      \end{figure*}
      }

      \begin{figure*}[ht!]
        \begin{minipage}{1.0 \textwidth}
          \includegraphics[scale=0.5]{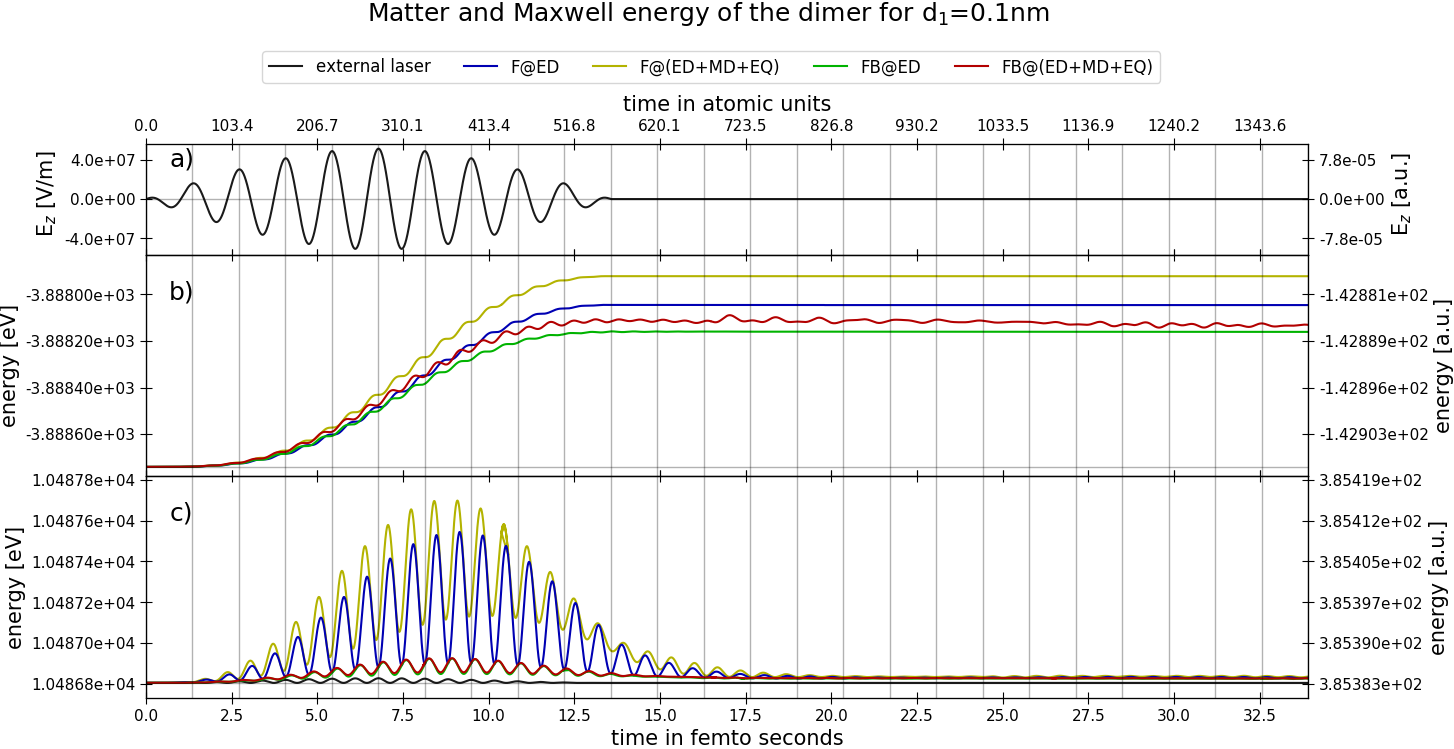}
          \caption{
          Matter and Maxwell energies for the $ d_1 = 0.1 $ nm separation of
          the sodium nanoparticles. In panel $a)$ we show as before the
          electric field amplitude of the incoming laser. Panel $b)$ displays
          the electronic energy for runs with different theory levels and in
          panel $c)$ we show the corresponding Maxwell energy density
          integrated over the simulation box.}
          \label{fig_Na_297_dimer_0_1_nm_energy}
        \end{minipage}
      \end{figure*}

      By construction, we know that the total electric field for F@ED coupling
      and consequently all related field enhancements in this case are only
      longitudinal.  To analyze further the nature of the field enhancement
      when the coupling to Maxwell's equations is enabled, we have performed a
      Helmholtz-decomposition of the electric field in the fully-coupled FB@ED
      case.  We find that also for FB@ED coupling the main contribution to the
      field enhancement arises from the longitudinal component. This implies
      that the additional back-reaction is causing the longitudinal components
      to leak much farther into the far-field.  There is also a small
      enhancement of the transverse field, but but this enhancement is an order
      of magnitude smaller compared to the longitudinal fields.  Since
      longitudinal fields are always bound to charges in space and since we do
      not consider electronic loss, the longitudinal field enhancement that we
      measure at the boundary surface is not leaving the system.  However,
      since we also have a transversal field enhancement there is now also a
      contribution that is radiating and that is reaching our absorbing
      boundary PML region.\\
      Next, we turn our attention to the case of a larger nanoparticle
      separation of $ d_2=0.5 $ nm with a larger internal dipole.  In this
      case, we drive the plasmonic dimer with the frequency $\omega_2=2.83$ eV
      $(0.104$ a.u.$)$ that corresponds to the dipole localized
      surface-plasmon resonance (D mode). We show in
      Fig.~\ref{fig_Na_297_dimer_0_5_nm_electric_field_fixed_ions} the results
      for this case and use the same ordering and labels for the panels as
      before in Fig.~\ref{fig_Na_297_dimer_0_1_nm_electric_field_fixed_ions}.
      Due to the larger distance between the two sodium nanoparticles, the
      absolute value of the current density is significantly smaller compared
      to the previous case with the smaller distance. In panel b) of
      Fig.~\ref{fig_Na_297_dimer_0_5_nm_electric_field_fixed_ions} it can be
      seen that the induced current densities for the forward coupling F@ED
      case are larger than for the forward and backward coupled FB@ED case.
      Looking at the electric field enhancements at the mid point shown in panel c),
      we find a significantly smaller field enhancement in the FB@ED coupled
      case. This is in contrast to the smaller dimer separation in 
      Fig.~\ref{fig_Na_297_dimer_0_1_nm_electric_field_fixed_ions}, where at least
      for short times the FB@ED coupling has induced larger field enhancements compared to
      the only forward coupled case.
      As before in the case of the $d_1$ separation, we find here
      that the average of the electric field over the detector surface in the
      far field as shown in panel e) is mostly locked to the phase of the incident
      laser.  In the near field at the mid point between the two nanoparticles
      the phase and frequency shifts are again larger for F@ED coupling at short times and the phase turns
      even to the opposite sign compared to FB@ED coupling for intermediate times.
      As before we have also performed here a Helmholtz-decomposition and also
      for the larger nanoparticle separation the longitudinal component of the electric field is leaking much more
      into the far field in the FB@ED coupled case than in the F@ED coupled case.\\
      To illustrate the relative magnitude of longitudinal and transverse components of
      the electric field quantitatively, we show in
      Figures Fig.~\ref{fig_Na_297_dimer_0_1_nm_electric_field_fixed_ions_decomposition}
      and Fig.~\ref{fig_Na_297_dimer_0_5_nm_electric_field_fixed_ions_decomposition}
      the temporal evolution of both field components
      in z-direction at point $ \vec{r}_{\mathrm{mp}} $ and $ \vec{r}_{\mathrm{ffpx}} $ respectively.
      As reference, we also plot the total field with solid lines. Besides the fact that
      the longitudinal enhancement is about one order of magnitude larger for the $ d_1 $ distance
      and even two orders of magnitude larger for the $ d_2 $ distance, we see some phase and frequency shifts between 
      the different fields. The phase shift between the longitudinal and the total field
      in Fig.~\ref{fig_Na_297_dimer_0_1_nm_electric_field_fixed_ions_decomposition} a) and
      Fig.~\ref{fig_Na_297_dimer_0_5_nm_electric_field_fixed_ions_decomposition} a) is very
      small for both distances. Although the phase shift to the transverse field is rather large,
      its small amplitude leads only to a minor contribution for the total field. 
      This behavior differs at the detector surface point $ \vec{r}_{\mathrm{ffpx}} $ illustrated in
      Fig.~\ref{fig_Na_297_dimer_0_1_nm_electric_field_fixed_ions_decomposition} c) and 
      Fig.~\ref{fig_Na_297_dimer_0_5_nm_electric_field_fixed_ions_decomposition} c).
      Here, both the longitudinal and the transverse field have almost the same magnitude and show 
      a clear phase shift.
      The behaviour in the far field 
      in Fig.~\ref{fig_Na_297_dimer_0_1_nm_electric_field_fixed_ions_decomposition} d)
      and Fig.~\ref{fig_Na_297_dimer_0_5_nm_electric_field_fixed_ions_decomposition} d)
      exhibits besides a phase shift also a slight frequency shift. Consequently, the incident laser pulse
      interferes with the induced transverse field wich results in a frequency modification of the outgoing
      laser. Since the transverse field, which reaches the far-field detector region, propagates freely,
      our detector point measures this frequency shift. \\
      Up to now, we have looked at the electric field enhancement as function of time.
      In Fig.~\ref{fig_electric_field_enhancement_contour}, we show contour plots
      of the transversal electric field enhancement as function of space in the x-z plane of 
      the nanoplasmonic dimer. The two plots in the top row correspond to the point
      in time where the incoming laser pulse reaches its maximum, whereas the two
      plots in the bottom row correspond to the point in time where the electric
      field enhancement reaches its maximum. The two plots in the left column have
      been computed with light-matter forward coupling only and in the two plots
      in the right column we have used self-consistent forward-backward coupling.
      As can be seen, the forward coupled cases show a rather uniform electric field
      in the plane which is due to the dipole approximation and the fact that the
      incident wavelength is rather large on the scale of the dimer. On the other
      hand for the fully coupled case on the right hand side local field effects
      are clearly visible. In particular in the plot on the bottom right it can be
      seen that at the maximum of the field enhancement the transversal field
      contribution in fact counter acts the longitudinal contribution since it
      has turned to a negative sign in most regions of space.

    \subsubsection{Next order in multipole coupling and energies}   \label{subsec_matter_maxwell_energies}

      \begin{figure*}[t!]
        \begin{minipage}{1.0 \textwidth}
          \includegraphics[scale=0.5]{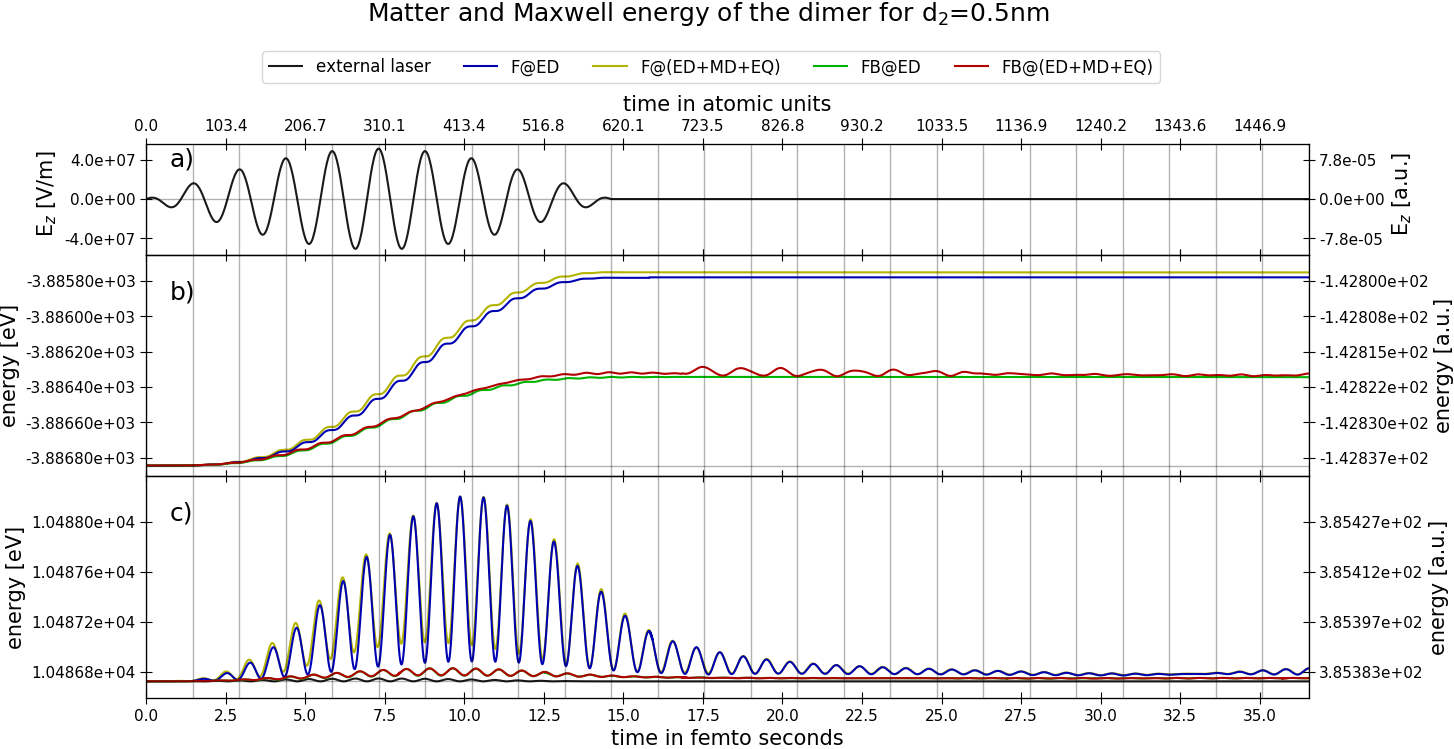}
          \caption{Same as Fig.~\ref{fig_Na_297_dimer_0_1_nm_energy} but here for the larger nanoparticle separation of $ d_1 = 0.5 $ nm. }
          \label{fig_Na_297_dimer_0_5_nm_energy}
        \end{minipage}
      \end{figure*}

      So far we have focussed only on the electric dipole approximation. In the
      following, we include also the next order in the multipole expansion in
      the Kohn-Sham Hamiltonian, namely the electric quadrupole (EQ) term and
      the magnetic dipole (MD) term from
      Eqs.~(\ref{eq_electric_quadrupole_hamiltonian}) and
      (\ref{eq_magnetic_dipole_hamiltonian}).  As before, we consider
      light-matter forward coupling as well as self-consistent forward-backward
      coupling.  This leads to the theory levels F@(ED+MD+EQ) and FB@(ED+MD+EQ)
      respectively.  In Fig.~\ref{fig_Na_297_dimer_0_1_nm_energy} we show for
      the $ d_1=0.1 $ nm configuration as before in panel a) the electric field
      amplitude of the incoming laser. In panel b) we plot the electronic
      energy of the Kohn-Sham system.
%
%
%
      The energies of the four different coupling theory levels F@ED,
      F@(ED+MD+EQ), FB@ED, and FB@(ED+MD+EQ) split up significantly in time and
      stay almost constant when the laser has passed the box.  In all cases the
      matter absorbs energy during the time when the external laser propagates
      through the box.  Similar to the electric fields enhancements, which
      exhibit a clear delay of reaction to the initial laser, also the energy
      gain for the matter is slightly delayed with respect to the external
      laser pulse.  The blue curve shows the conventional F@ED in dipole
      approximation without back reaction of the matter to the field. If we add
      the second order multipole coupling terms, the corresponding F@(ED+MD+EQ)
      run (yellow curve) gains more energy than in dipole approximation. This
      can be understood by the fact that we drive with the frequency of the
      incoming laser in this case the Q mode of the nanoplasmonic dimer which
      has a quadrupole nature. The EQ and MD coupling terms in the Hamiltonian
      can efficiently couple to this mode which results in a larger transfer of
      energy to the electrons.\\ 
      Switching on the backward coupling reduces the energy absorption of the
      matter. Both, the FB@ED and FB@(ED+MD+EQ) cases remain energetically below
      the reference F@ED run.  Again, the additional multipole terms in the
      FB@(ED+MD+EQ) run increase the energy curve compared to the FB@ED
      run. As before, this shift can be attributed to the more efficient energy
      transfer mediated by the additional MD+EQ coupling term.\\
      In panel c) we show the total Maxwell energy, which corresponds to an
      integration of the Maxwell energy density over the whole simulation box.
      First, we note that all Maxwell energies oscillate with twice the
      frequency of the initial laser.  This simply arises due to the squared
      electric and squared magnetic fields in the expression for the energy
      density. On the other hand, the peak positions depend on the respective
      phase shift of the electric and magnetic fields. 
      As we already noticed for the electric field enhancement, the dominant part of the total electric 
      field is given by the longitudinal component, so that also for the Maxwell
      energy the largest contribution originates from the longitudinal part of the 
      electric field. The transverse electric and the magnetic field have only a
      small contribution to the Maxwell energy. This can be seen by comparing the
      scale of the black curve, which corresponds to the energy of the purely transversal
      incoming laser pulse, with the blue curve which shows the Maxwell energy 
      for the forward coupled case in dipole approximation. By also adding here
      the electric quadrupole and magnetic dipole terms to the Hamiltonian, we find
      a substantial gain in electromagnetic energy which exceeds the scale of the
      energy of the incoming laser by more than a factor of five.
%
%
      Again most of this energy is stored in the longitudinal component of the electric field.\\
      When we look at the two forward-backward coupled cases, FB@ED and FB@(ED+MD+EQ), we find
      similar to the energy transfer to the electrons also a smaller transfer of energy
      to the Maxwell fields. Overall, the energy oscillations are about six times weaker and
      clearly phase shifted compared to the forward-coupled case.
      Additionally, we note that in the FB@ED and FB@(ED+MD+EQ) cases extra energy is stored
      inside the electromagnetic fields once the incoming laser has passed the system. This is
      evident from the energy plateau in both cases after about 20 fs, which is above the
      energy value at the initial time.\\
      Finally, we turn our attention to the second dimer configuration with $
      d_2 = 0.5 $ nm.  The matter and Maxwell energies for this case are shown
      in Fig.~(\ref{fig_Na_297_dimer_0_5_nm_energy}). At this point is is
      useful to recall that we drive the nanoplasmonic dimer for the $d_2$ distance
      with a frequency that is resonant to the D mode which has dipole
      character. While overall a similar situation emerges as for the smaller $ d_1
      $ distance,  a notable difference is here that the inclusion of the
      higher order multipole terms, MD and EQ, has little effect on the energy
      gains for the electrons and the electromagnetic fields. The F@ED and
      F@(ED+MD+EQ) energies, as well as the FB@ED and FB@(ED+MD+EQ) energies
      are almost on top of each other. In other words, since we couple to a
      dipole mode of the system, the MD and ED coupling terms have almost no
      effect. In a perturbative analysis this can be understood from the selection
      rules of the involved states for the MD and EQ coupling Hamiltonians.
      From this we can conclude that it depends on the symmetry of the excited
      modes if higher order multipole terms become important.\\
      \begin{figure*}[ht!]
        \begin{minipage}{1.0 \textwidth}
          \includegraphics[scale=0.5]{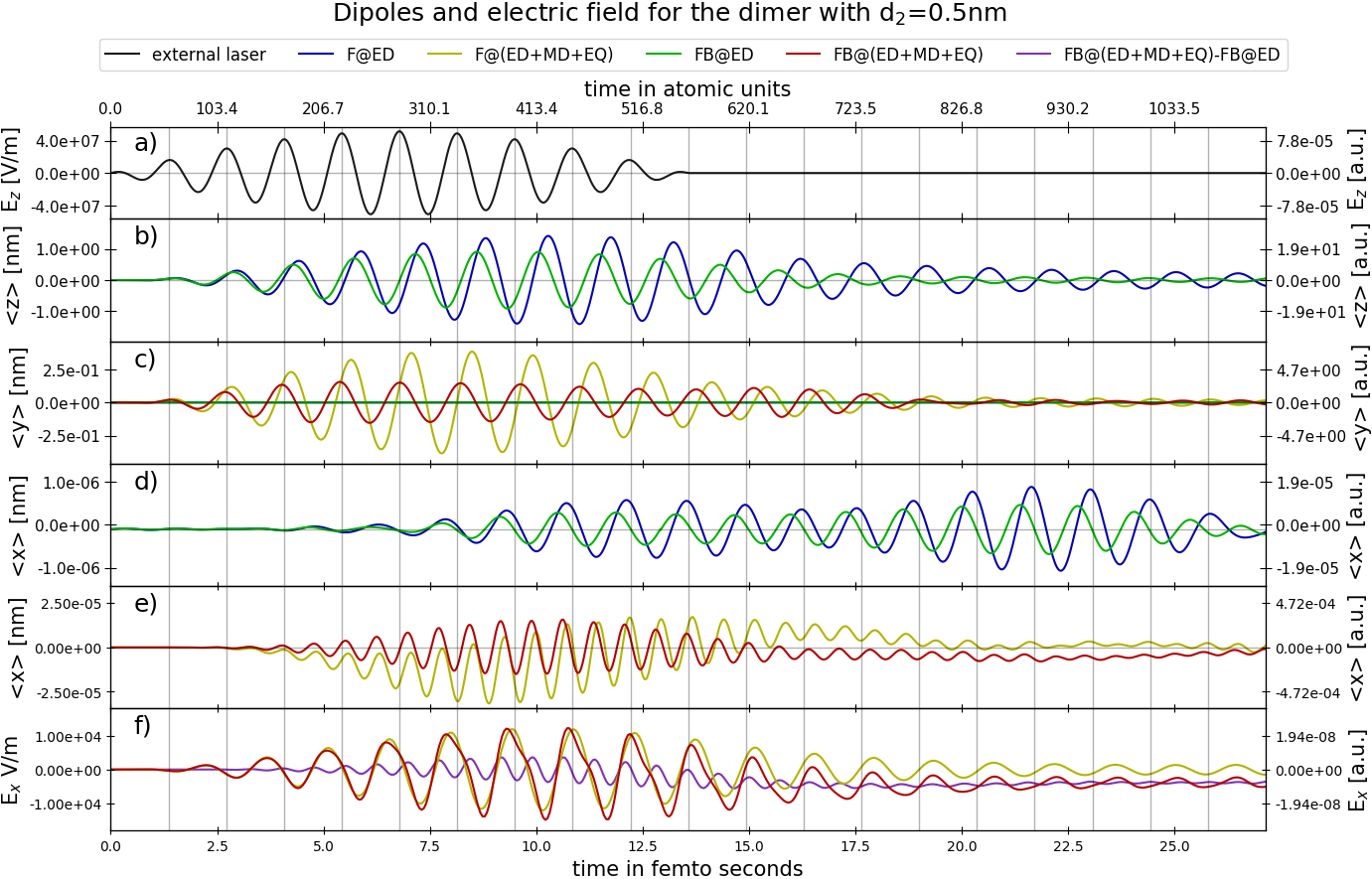}
          \caption{ Position expectation values $ \langle x \rangle $, $ \langle y \rangle $, and $ \langle z \rangle $ 
                    of the nanoplasmonic dimer with $ d_1 = 0.1 $ nm. Panel a) shows the initial laser pulse and panels b) - e)
                    the dipoles of the dimer. Driving the system with a coupling beyond dipole approximation
                    induces higher-order harmonics along the x-axis. This is directly detected
                    by the $ E_x $ field component at point $ \vec{r}_{\mathrm{ffpy}} $. }
          \label{fig_Na_297_dimer_0_5_nm_multipoles}
        \end{minipage}
      \end{figure*}
      The common fact, that for both distances $d_1$ and $d_2$ the forward- and backward
      coupling matter energies remain always below the forward
      coupling runs demonstrates that the matter absorbs less energy if the
      back-reaction is taken into account. In addition to the larger
      absorption of energy, the forward coupling causes larger Maxwell
      energy amplitudes inside the simulation box. This is remarkable since we
      observed in 
      Fig.~\ref{fig_Na_297_dimer_0_1_nm_electric_field_fixed_ions} and
      Fig.~\ref{fig_Na_297_dimer_0_5_nm_electric_field_fixed_ions} that the
      self-consistent forward-backward coupling yields a larger enhancement of
      the field. As consequence, in some regions of the dimer large field
      enhancements occur, but the mean amplification is clearly weaker than for
      the only forward coupled cases. Furthermore, the forward coupling runs
      break energy conservation, since the laser pumps the matter system
      without any loss.  In the forward-backward coupled simulations this is
      not as severe anymore, and explains that the energy absorption and the
      mean Maxwell field enhancement is always smaller compared to the forward
      coupling runs. The situation would be entirely different if we could
      enclose the laser pulse completely in the Maxwell box. Then the pulse
      would not be external anymore and in the forward-backward case full
      energy conservation holds. For optical wavelengths this requires enormous 
      Maxwell simulation boxes if atomic scale grid spacings are used. But it
      becomes feasible for hard x-rays, where much smaller Maxwell grids are
      needed due to the shorter wavelength.

      \begin{figure}[ht!]
        \begin{minipage}{0.46 \textwidth}
          \includegraphics[scale=0.5]{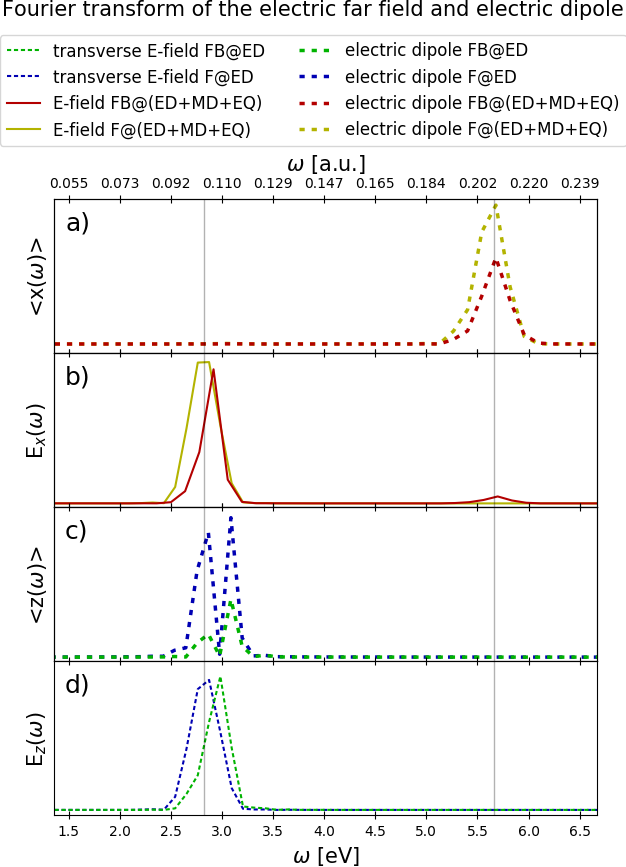}
          \caption{Panel a) and b) show the Fourier transforms of the dipole and the electric field at the far-field point $ \vec{r}_{\mathrm{ffpy}} $
                   from panels e) and f) in Fig.~\ref{fig_Na_297_dimer_0_5_nm_multipoles}
                   respectively. While in the matter observable in a) only the second harmonic peak is visible,
                   the Fourier spectrum of the electric field in b) contains the fundamental laser frequency and the second harmonic.
                   In panel d) we show the Fourier transform of the electric field at the far-field point $ \vec{r}_{\mathrm{ffpx}} $ along the laser propagation axis
                   in dipole approximation. The field is shifted in frequency when self-consistent forward-backward coupling
                   is used. The matter dipoles for this case are shown in panel c). In the forward coupled case a spurious
                   peak appears at the energy of the incoming laser (2.83 eV). This peak is surpressed in the forward-backward coupled
                   case and the dipole spectrum in this case also matches better the actual emitted radition field in panel d).
                   }
          \label{fig_second_harmonic_peak}
        \end{minipage}
      \end{figure}

      \begin{figure*}[ht!]
         \begin{minipage}{1.0 \textwidth}
           \includegraphics[scale=0.5]{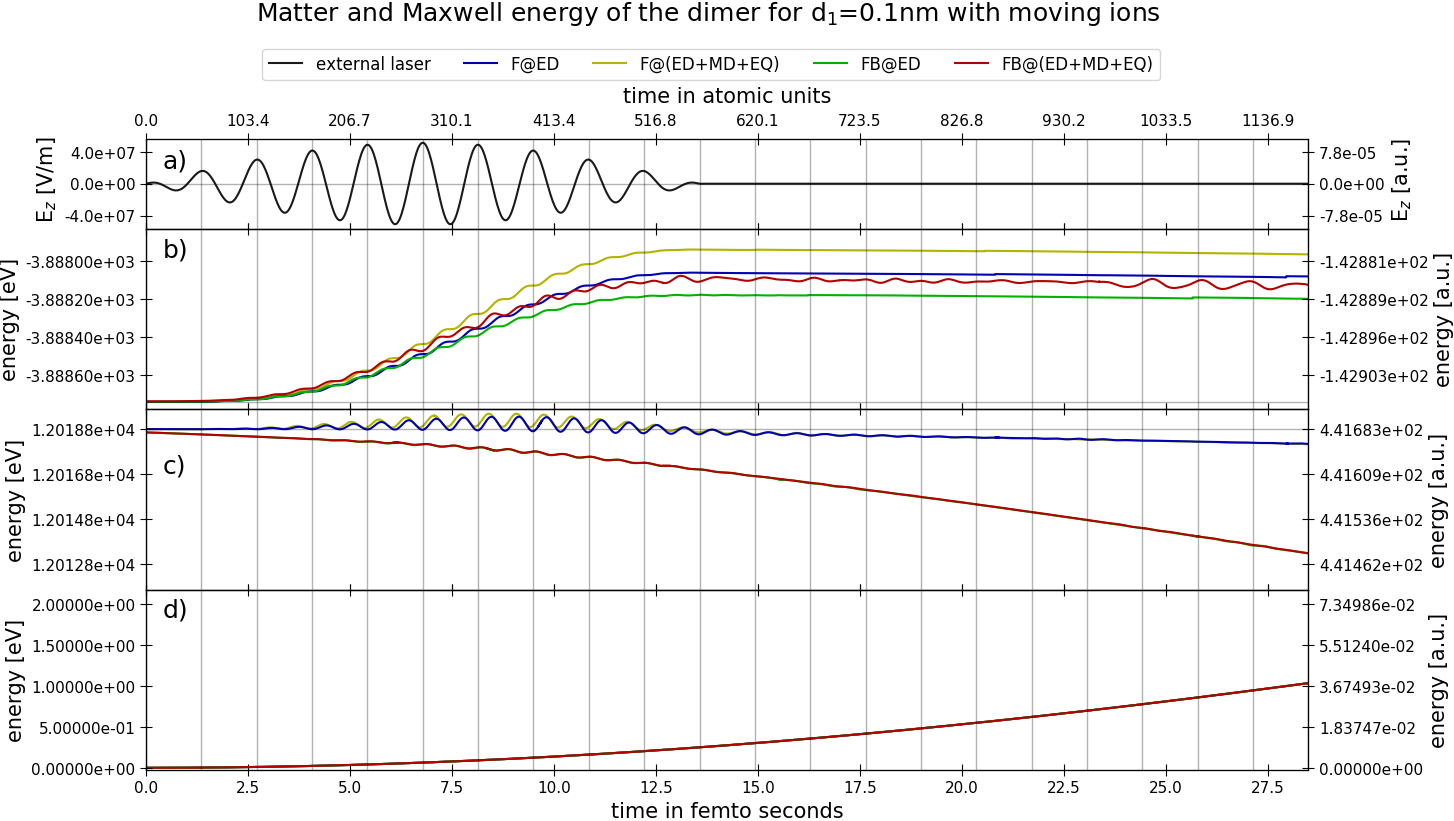}
          \caption{Same as Fig.~\ref{fig_Na_297_dimer_0_1_nm_energy} but now including the motion of the ions in classical
          Ehrenfest approximation and the classical Lorentz forces on the nuclei with local electromagnetic fields taken directly from
          the Maxwell grid.}
           \label{fig_Na_297_dimer_0_1_nm_moved_ions_energy}
         \end{minipage}
      \end{figure*}


      
      \begin{figure*}[ht!]
        \begin{minipage}{1.0 \textwidth}
          \center
          \includegraphics[scale=0.5]{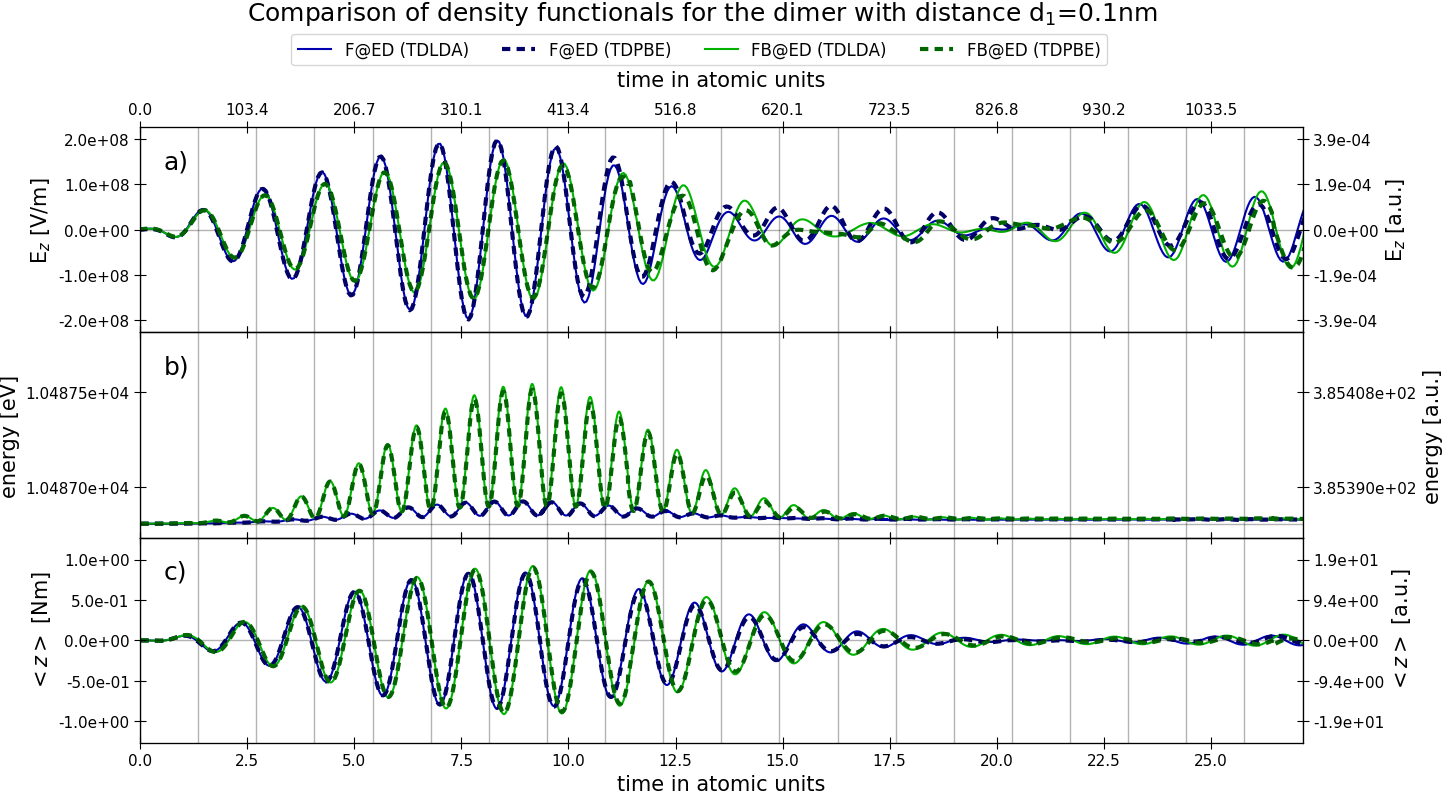}
          \caption{
	  In the figure we show in panel a) the electric field at the origin,
          in panel b) the integrated Maxwell energy in the simulation box, and in panel
          c) the dipole expectation value in z-direction.  We compare TDPBE results
          (dashed lines) with the TDLDA results (solid lines).  The difference between
          TDPBE and TDLDA (dashed vs.~solid lines) is much smaller than the difference
          between only forward coupling and self-consistent forward-backward coupling
          (blue vs.~green lines). In particular, for forward-backward coupling a clear
          frequency shift is visible already after a short time.   
                   }
          \label{fig_density_functionals}
        \end{minipage}
      \end{figure*}

    \subsubsection{Electromagnetic detectors and harmonic generation}

      It is common practice in most quantum simulations to use matter
      expectation values to approximate optical spectra. A prime example is the
      dipole expectation value. The Fourier transform of the dipole is
      routinely used to compute absorption spectra in the linear case or
      high-harmonic spectra in the non-linear case.  With a Maxwell propagator
      at hand it becomes now feasible to directly look at the emitted
      radiation. In other words, it is not necessary anymore to take a detour
      and to approximate the emitted radiation from matter observables.  Rather, we
      can look directly at the emitted radiation.  This provides a paradigm
      shift, since we now can essentially define "electromagnetic detectors" in
      the far-field close to the box boundaries of the Maxwell simulation box
      which allow to perform numerical simulations that very closely resemble
      the experimental situation.\\
      To showcase this new paradigm, we plot in
      Fig.~\ref{fig_Na_297_dimer_0_5_nm_multipoles} in panels b)-e) dipole
      expectation values for our nanoplasmonic dimer.  In panel f) we show the
      x-component of the electric field in the far field, which corresponds to
      our electromagnetic detector in this case. Furthermore, in
      Fig.~\ref{fig_second_harmonic_peak}, we show the Fourier transforms of
      the last two panels from Fig.~\ref{fig_Na_297_dimer_0_5_nm_multipoles}.
      Panel a) from Fig.~\ref{fig_second_harmonic_peak} is the Fourier transform
      of panel e) in Fig.~\ref{fig_Na_297_dimer_0_5_nm_multipoles} and panel b)
      is the Fourier transform of panel f).\\
      Several observations can be deduced from these results. First, comparing
      the forward coupling cases, F@ED and F@(ED+MD+EQ), with the fully coupled
      dipole signals, FB@ED and FB@(ED+MD+EQ), shows a damping in the amplitude
      which is consistent with the observations we already made for the
      electric field enhancements and the energies.  Second, adding MD and EQ coupling terms to the Hamiltonian
      produces in both, the only forward coupled case as well as in the
      forward-backward coupled case dipole signals which oscillate with twice
      the frequency of the incoming laser. In other words, second harmonic
      generation is only found in this case if we go beyond the dipole approximation.  And
      finally, looking at the electric field in the far-field we also see the
      second harmonic signal. This signal only emerges in the fully
      forward-backward coupled case, since in the forward coupled case the
      matter can not influence the electromagnetic fields.  This is a prime
      example that a self-consistent forward-backward coupling to Maxwell's equations is needed to achieve a
      comprehensive physical picture.

    \subsubsection{Ion motion}

      In all the simulations that we have considered so far, we have used the
      Born-Oppenheimer approximation and have clamped the classical nuclei at
      the optimized ground state geometry. In this section we now release this
      constraint and we also allow the nuclei to move according to the
      Ehrenfest equations of motion \cite{andrade_2009} and the classical Lorentz forces that we
      introduced in Sec.~\ref{subsec_classical_nuclei}. For the Lorentz forces
      on the nuclei we can take directly the electromagnetic fields that we
      propagate on our Riemann-Silberstein grid. This allows to capture nuclear
      forces due to local field effects. For all the following cases, we take
      as initial condition for the Ehrenfest equations the atomic positions of
      the optimized ground state and set the initial velocities to zero. This
      effectively corresponds to a rather "cold" nuclear subsystem. More
      sophisticated velocity distributions could be used, e.g. thermalized
      velocity distributions from molecular dynamics runs coupled to a
      thermostat, but we leave such temperature studies for the future.\\ 
      In Fig.~\ref{fig_Na_297_dimer_0_1_nm_moved_ions_energy} we show again
      matter and Maxwell energies for the nanoparticle distance $d_1$, but now
      for moving ions. As before, panel a) shows the incoming laser, panel b)
      the matter energies and panel c) the Maxwell energies. In addition we now
      also add panel d) which shows the sum of the kinetic energy of all nuclei
      as function of time.\\ 
      The additional
      ionic motion causes on this rather short time scale of about 30 fs some
      additional fluctuations in the matter energy evolution, but the main
      behavior is very similar to the case with fixed ions. However, looking
      at the Maxwell energies in panel c) reveals a small
      overall decrease of the Maxwell energy in the forward coupled and a rather
      strong decrease in the self-consistent forward-backward case. Since the
      electronic energy remains almost identical to the case of clamped ions,
      the losses in the Maxwell energy are directly transferred to the nuclei.
      This can be seen in panel d) where the kinetic energy of the nuclei
      grows. It is quite remarkable that this increase is rather fast (30 fs)
      as nuclear motion is typically attributed to take place on a pico-second
      time scale. 
      

    \subsubsection{Comparison of different density functionals}

     All of the result shown so far have been computed exclusively with TDLDA
     as choice for the approximate exchange-correlation functional for the
     longitudinal part of the light-matter interaction. To assess the relative
     importance of exchange-correlation effects versus self-consistent
     light-matter interaction, we have performed also simulations with the PBE
     functional.  In Fig.~\ref{fig_density_functionals} we show in panel a) the
     electric field at the origin, in panel b) the electromagnetic energy, and
     in panel c) the electric dipole in z-direction. In all panels, we compare
     the TDPBE results (dashed lines) with the TDLDA results (solid lines).
     From the results it is apparent that the difference between TDPBE and
     TDLDA (dashed vs.~solid lines) is much smaller than the difference between
     only forward coupling and self-consistent forward-backward coupling (blue
     vs.~green lines).  While the change in amplitude is minor for the electric
     field and the dipole, a clear frequency shift for forward-backward coupling is visible already after
     short time.  This shift has also an effect on the electromagnetic energy
     which shows a larger deviation also in amplitude.  In other words, for the
     present example it is more important to switch on self-consistent
     forward-backward light-matter interactions than to include further exchange-correlation contributions to the effective Kohn-Sham potentials. This
     supports the need for a self-consistent coupling to Maxwell's equations to
     achieve a comprehensive description of light-matter interactions.







\section{Summary and conclusions} \label{sec_summary}

In the present work we provide a comprehensive derivation of a formally exact
density-functional approach for fully self-consistent light-matter interactions
of photons, electrons, and effective nuclei.  This approach corresponds to an
exact reformulation of the full quantum problem of non-relativistic QED in
terms of non-linear quantum Navier-Stokes equations. We also introduce a
Kohn-Sham construction, which allows to map the dynamics of interacting
particles to the effective dynamics of non-interacting particles with the same
densities. This turns the approach into a computationally feasible method. In
the mean-field approximation for the effective nuclei, our
approach corresponds to coupled Ehrenfest-Maxwell-Pauli-Kohn-Sham equations. \\ 
Using the Riemann-Silberstein formulation of classical electrodynamics, we then
show how to couple the Ehrenfest equations for the nuclei and the
Pauli-Kohn-Sham equations for the electrons self-consistently to the
time-evolution of the electromagnetic fields. The reformulation of Maxwell's
equations in a Schr\"odinger-like form allows us to use time-evolution
techniques that have originally been developed for the dynamics of
wavefunctions in quantum mechanics. With help of the Riemann-Silberstein vector
we can seamlessly combine fully-microscopic electrodynamics with macroscopic
electrodynamics for linear media.  This allows us to easily incorporate
macroscopic optical elements like lenses or mirrors in a microscopic
description and opens the path to many applications, such as, e.g., molecules in
optical cavities or close to boundaries.  For the coupling between light and
matter, we employ in practice a multipole expansion based on the Power-Zienau-Woolley
transformation.  We introduce a predictor-corrector scheme for a
self-consistent coupling of light and matter.  To address the different time
and length-scales which arise for the coupling of molecular or nano-scale
systems to light with optical wavelengths, we introduce a multi-scale approach
in space and time. For an efficient absorption of outgoing electromagnetic
waves, we present a perfectly matched layer for the Riemann-Silberstein vector.
We have implemented our approach in the real-time real-space code Octopus and
illustrate the approach with an example of light-matter interactions for a
nanoplasmonic sodium dimer. The results from the example show that a
self-consistent forward-backward coupling of light and matter is necessary for
a comprehensive physical description of the system.  Including the propagation
of the microscopic electrodynamical fields in the real-time simulation allows
us to define electromagnetic detectors in the far-field of the simulation box.
In this way, it is not necessary anymore to take a detour and to approximate the
emitted radiation from matter observables. Rather, we can look directly at the
emitted radiation. This is a distinguished feature of our implementation and
has direct application for many spectroscopies.\\
For the nanoplasmonic system that we have selected here as an example, we find that the
difference for observables computed with the local density approximation and
gradient-corrected functionals is minor compared to the difference of only
forward coupling and self-consistent forward-backward light-matter interactions.  Very
often in literature the discrepancies between theory and experiment are
attributed to missing exchange-correlation effects. From the present results,
we can at least say that self-consistent forward-backward light-matter
couplings should always be included and that their contribution can easily
match the magnitude of longitudinal exchange-correlation contributions or even
exceed them.  \\
While the Ehrenfest-Maxwell-Pauli-Kohn-Sham limit is the level of
the theory for which we present all the results in the present work, the exact
density-functional formulation of the Pauli-Fierz field theory provides a sound
framework to go beyond this mean-field limit. Further steps along these lines will
be the subject of future investigations.\\


\section{Outlook} \label{sec_outlook}

Although we have touched in this work many aspects that enter fully
self-consistent light-matter coupling, there are still several tasks that
remain to be addressed in the future. As most important task, functionals for
the exchange-correlation terms that we introduced in the density-functional
framework for non-relativistic QED have to be constructed. Including such exchange-correlation
terms will allow to go beyond the classical mean-field description for photons
(and likewise for the effective nuclei). In particular, this allows to include
vacuum fluctuations of the electromagnetic field in the description beyond the physical-mass approximation. \\
Further, we have considered in the implemenetation so far only the lowest orders of a multipole
expansion based on the Power-Zienau-Woolley transformation. Clearly,
full-minimal coupling is required for a complete description of
non-relativistic QED.  In our implementation, we already compute the full
vector potential with spatial resolution on the grid. The next step is to plug
this potential back into the Hamiltonian and to treat full minimal coupling
with full spatial resolution. It is then valuable to have both, the full
minimal-coupling, as well as the multipole expansion available in the same
implementation. When certain effects appear for a given application in full
minimal-coupling, it is then possible to go back to the multipole expansion and
to check in which order the effects start to emerge. One such example is the
second-harmonic generation that we discussed which occurred in our nanoplasmonic system
only by including magnetic dipole and electric quadrupole coupling.\\
For the back-coupling from matter to the electromagnetic
fields, we have used in the present work the standard paramagnetic current
density. We still need to include the diamagnetic current and magnetization
current contributions of the electrons, and also the ionic current of the
nuclei for a complete description of the back-coupling.\\
Another line of research that is required in the future is the dependence of
the pseudo-potentials on the self-consistent light matter coupling. In the
present work we have used the conventional field free pseudo-potentials.
However, when constructing the pseudo-potentials the proper starting point
should be a light-matter coupled all-electron Hamiltonian which introduces a
dependence on the vector potential in the pseudo-potential.  All the
computations that we have shown start from the field-free ground-state of the
matter. It is clear that also the ground state should be found when
light-matter coupling is already switched on.  This requires to solve
stationary Maxwell equations self-consistently coupled to the stationary
Kohn-Sham equations. Temperature effects can be included in our framework by
starting from an equilibrated thermal initial state for the ions and initiating
the non-equilibrium dynamics afterwards. Finally, we have focussed here on
finite quantum systems. Extending the present formulation to periodic systems
opens a whole new class of possibilities, ranging from 1D and 2D to 3D periodic
structures.\\ 
Having all these additional aspects taken into account, allows to perform
full-featured ab-initio simulations for a vast set of physical applications.
Examples include systems studied in nano-optics and nano-plasmonics, (photo)
electrocatalysis, light-matter coupling in 2D materials, cases where laser
pulses carry orbital angular momentum, or light-tailored chemical reactions in
optical cavities to name only a few.

\begin{acknowledgments}
We acknowledge useful discussions with Nicolas Tancogne-Dejean, Aaron Kelly,
Johannes Flick, Christian Sch{\"a}fer, Norah Hoffman, Alberto Castro, and Hardy
Gross.  We thank Michele Compostella for developing the visualization software
that has been used for the present work. Angel Rubio acknowledges financial
support by the European Research Council  (ERC-2015-AdG-694097)  and  Grupos
Consolidados (IT578-13).  The Flatiron Institute is a division of the Simons
Foundation.
\end{acknowledgments}


\bibliography{qedft_empks} 

\end{document}